\documentclass[%
 preprint,
 amsmath,amssymb,
 aps,
]{revtex4-1}

\usepackage{graphicx}
\usepackage{dcolumn}
\usepackage{bm}
\usepackage{xcolor}
\usepackage[version=4]{mhchem}
\usepackage{braket}
\usepackage{soul}
\usepackage{hyperref}

\begin{document}

\preprint{}

\title{Quantum Embedding Theories to Simulate Condensed Systems on Quantum Computers}

\author{Christian Vorwerk}
\thanks{These two authors contributed equally}
\affiliation{Pritzker School of Molecular Engineering, University of Chicago, Chicago IL 60637, USA}%

\author{Nan Sheng}
\thanks{These two authors contributed equally}
\affiliation{Department of Chemistry, University of Chicago, IL 60637, USA}

\author{Marco Govoni}
 \email{mgovoni@anl.gov}
\affiliation{Materials Science Division and Center for Molecular Engineering, Argonne National Laboratory, Lemont IL 60439, USA}
\affiliation{Pritzker School of Molecular Engineering, University of Chicago, IL 60637, USA}%

\author{Benchen Huang}
\affiliation{Department of Chemistry, University of Chicago, IL 60637, USA~}

\author{Giulia Galli}
\email{gagalli@uchicago.edu}
\affiliation{Pritzker School of Molecular Engineering and Department of Chemistry, University of Chicago, Chicago IL 60637, USA}
\affiliation{Materials Science Division and Center for Molecular Engineering, Argonne National Laboratory, Lemont IL 60439, USA}

\date{\today}

\begin{abstract}
Quantum computers hold promise to  improve the efficiency of quantum
  simulations of materials and to enable the investigation of systems and
  properties more complex than tractable at present on classical architectures.
  Here, we discuss computational frameworks to carry out electronic structure
  calculations of solids on noisy intermediate scale quantum computers using
  embedding theories, and we give examples for a 
  specific class of materials, i.e., spin defects in solids. These are promising
  systems to build future quantum technologies, e.g., computers, sensors and
  devices for quantum communications. Although quantum simulations on quantum
  architectures are in their infancy, promising results for realistic systems
  appear to be within reach.
\end{abstract}

\maketitle

\section{Introduction}
Quantum simulations of the physical properties of molecules and solids on
classical computers are routinely used to tackle many problems in materials
science and
chemistry~\cite{jones2015,krylov2018,schleder2019,maurer_advances_2019,
bogojeski_quantum_2020,mcardle2020}.
These simulations are aimed at understanding a variety of complex 
systems, in diverse fields such as
catalysis~\cite{bell_quantum_2011,xu_theoretical_2019} and quantum information
science~\cite{wolfowicz2021,dreyer2018,weber2010}, as well as at generating data
for  computations based on machine learning and artificial
intelligence~\cite{agrawal2016,himanen2019,s.dong2021}. The use of quantum computers
promises to improve the efficiency of quantum simulations and to enable the
adoption of high level theories for systems more complex than tractable at
present~\cite{yuan2020,elfving2020,vonburg2021,liu2021,ollitrault2021}.

A fundamental step in the calculation of the electronic structure of molecules and solids at
the quantum-mechanical level of theory is the solution of the time-independent
Schr\"odinger equation describing interacting electrons in the field of the
nuclei; such solution provides the basis for the evaluation of numerous
properties, including total  and excitation energies and optimized
geometries. A common strategy adopted on the quantum hardware available at present, namely noisy intermediate scale quantum computers (NISQ), is that of writing the electronic structure problem in terms of
second-quantized Hamiltonians, whose parameters  are determined on a classical
computer from quantum chemistry~\cite{helgaker2014} or density functional theory (DFT)
and many-body perturbation theory~\cite{martin2020,martin2016} (MBPT) calculations.
The Hamiltonian is then represented in terms of qubits and quantum gates, using, for instance, 
the Jordan-Wigner transform~\cite{jordan1934}, the parity encoding~\cite{bravyi2002,seeley_bravyi-kitaev_2012}, the Verstraete-Cirac mapping
\cite{verstraete2005} or the Bravyi-Kitaev transform~\cite{bravyi2002}; these mappings 
are implemented in several open-source codes (e.g., Qiskit
\cite{aleksandrowicz2019qiskit}, OpenFermion~\cite{mcclean2020}). Finally, the lowest eigenvalues of the Hamiltonian are computed with an 
algorithm 
compatible with the number of available qubits and circuit depths of the
hardware. 
Topics of active research include the study of the scaling of the
algorithms used in the  diagonalization of the Hamiltonian, for example variational quantum eigensolver (VQE)
\cite{peruzzo2014,mcclean2016} and quantum phase estimation~\cite{nielsen2010}, with specific attention to the dimensionality of the problem and whether shallow
or deep circuits are employed~\cite{bravyi2020}.

Progress has been reported in the past decade in solving the Schr\"odinger
equation for molecular systems on quantum computers
\cite{aspuru-guzik2005,lanyon2010,li2011,peruzzo2014,shen2017,omalley2016,
santagati2018,kandala2017,hempel2018,colless2018,kandala2019,ryabinkin2018,
li2019,nam2020,mccaskey2019,gao2021,smart2019,sagastizabal2019,higgott2019,
google2020,metcalf2020,rossmannek2021,kawashima2021,teplukhin2021,kirsopp2021,jones2021} ranging from simple ions,
including \ce{H3+} and \ce{HHe+} to molecules such as water, alkali hydrides and
hydrogen chains with up to 12 atoms, for which Hartree-Fock solutions have been
recently published~\cite{google2020}. An interesting summary of calculations of
molecular systems on quantum computers is given in Ref.~\cite{mcardle2020},
which also discusses the algorithms used to solve the molecular Hamiltonian in
second quantization. 

Recently  progress has also been reported in quantum computations of model Hamiltonians, such as the
Hubbard~\cite{kivlichan2020,cruz2020,montanaro2020,uvarov2020,motta2020} and
Heisenberg Hamiltonians~\cite{li2011,motta2020,mei2020}, infinite hydrogen chains~\cite{mizuta2021,liu2020}, and a two-dimensional electron gas in a strong magnetic field (Hamiltonian describing quantized Landau levels)~\cite{kaicher2020,rahmani2020}. Most of these calculations have been carried out on a quantum simulator (i.e., the quantum computer is simulated on a classical one),  with just few examples~\cite{cruz2020,montanaro2020,mei2020,motta2020} using quantum hardware.  In addition, periodic Hubbard
models have been solved on quantum simulators~\cite{kreula2016,kreula_non-linear_nodate,jaderberg2020,lupo_maximally_2021,bauer2016,rubin2016,mineh2021,li_toward_2021} either within dynamical mean field theory
(DMFT)\cite{georges1992,georges1996,georges2004,anisimov1997,kotliar2006} or density
matrix embedding theory (DMET)~\cite{wouters2016,knizia2012,knizia2013,pham2020,hermes2019,pham2018}, with three recent examples of  DMFT~\cite{rungger2020,keen2020,yao2021} and DMET~\cite{kawashima2021,tilly2021} calculations  reported on quantum computers as well. DMFT and DMET are powerful approaches~\cite{georges1992,georges1996,georges2004,anisimov1997,kotliar2006} to map the strongly-correlated states of a solid onto a self-consistent quantum impurity
problem~\cite{lupo_maximally_2021}. A review of algorithms used for DMFT and DMET
calculations of strongly correlated systems is presented in Ref.~\cite{bassman2021}. In addition, quantum
simulations have been applied 
to solve tight-binding Hamiltonians of weakly
correlated solids~\cite{cerasoli2020,sureshbabu2021,choudhary2021} on quantum simulators and quantum computers.


For most molecular and solid-state systems, the solution of the electronic structure problem using the 
full many-body Hamiltonian is still out of reach on both classical and near term quantum architectures, due to the large size of the Hilbert space.  Interestingly, there are  myriad of important problems in condensed matter physics, materials science and chemistry  that
can naturally be formulated in terms of active regions surrounded by a host medium,  for example point defects in materials, active site of
catalysts, molecular adsorbates on surfaces and nanostructures embedded in
condensed systems, including matrices or solvents, to name a few. In addition strongly correlated states arising from $d$ and $f$ orbitals in oxides and other solids may be described by an active space embedded in a condensed medium. Hence, all of these problems may be addressed using embedding theories~\cite{libisch2014,wesolowski2015,jacob2014,wouters2016,knizia2012,knizia2013,
pham2020,hermes2019,pham2018,ma2020,ma2020a,ma2021,lan2017,zgid2017,rusakov2019,
biermann2003,biermann2014,boehnke2016,choi2016,nilsson2017,sun2002,
lichtenstein1998,anisimov1997,dhawan2021,otten2021}
which separate the electronic structure problem of an active region from that of the periodic, host environment. In particular, 
using embedding theories, one may define second-quantized Hamiltonians
and formulate the electronic structure problem of active regions in solids in a fashion similar to that of
molecular systems. 

\begin{figure}[t!]
\includegraphics[width=\textwidth]{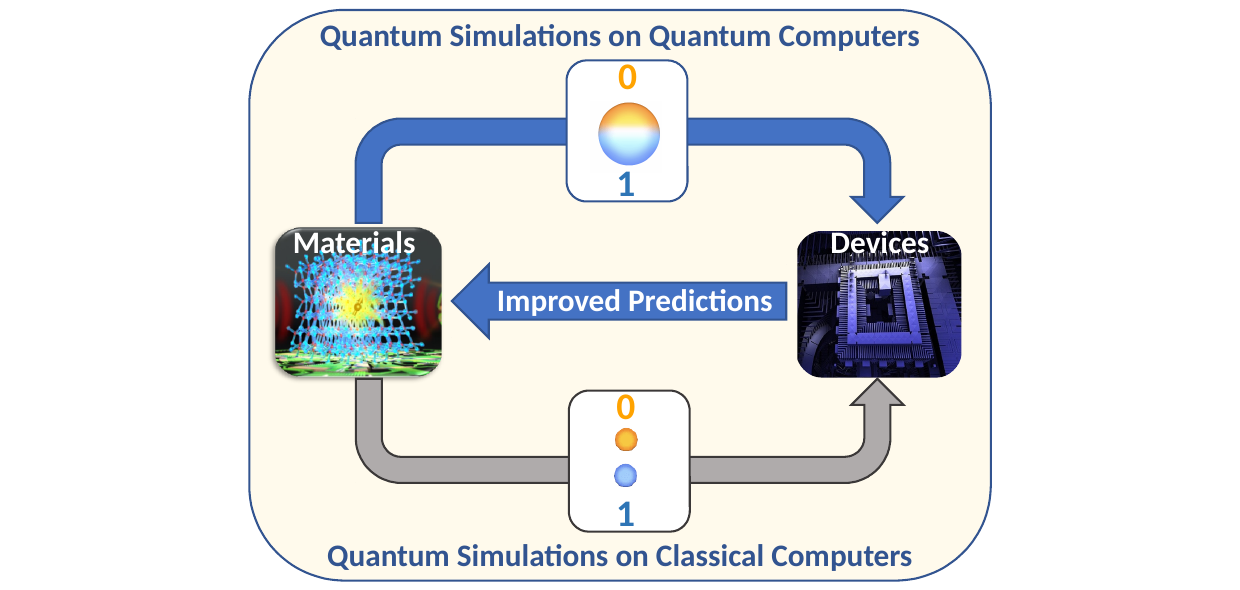}
\caption{\label{fig:intro} From materials to devices and back again: the bottom
  and top arrows indicate quantum simulations of materials carried out on a
  classical or  quantum computer, respectively, leading to the prediction and
  design of components for quantum architectures; the latter may then be used to
  perform quantum simulations (as indicated by the middle arrow) and in turn
  improve predictive capabilities for  materials and devices.}
\end{figure}

In this perspective we discuss frameworks to carry out quantum-mechanical
calculations for solids on near-term quantum computers using embedding theories,
with examples on a specific class of materials, i.e.,  spin-defects in solids (see
Fig.~\ref{fig:intro}), for which calculations of ground and excited state properties on a quantum computer have been recently reported.  A spin-defect, i.e. a point defect with specific spin properties~\cite{wolfowicz2021}, is a promising system to realize a qubit and hence to build future quantum
technologies, e.g., computers, sensors and novel devices for quantum
communications~\cite{weber2010,seo2016,seo2017,ivady2018,dreyer2018,smart2019,anderson2019}.
We discuss hybrid classical/quantum calculations of the electronic structure of spin defects and we envision a general feedback loop, where
quantum simulations of materials properties on a quantum device may lead to the
prediction of new materials and properties for the design of improved quantum
devices, which will in turn lead to enhanced property predictions (see
Fig.~\ref{fig:intro}). We conclude with a discussion of current open challenges.

\section{Quantum Embedding Theories}

\begin{figure}[t!]
\includegraphics[width=\textwidth]{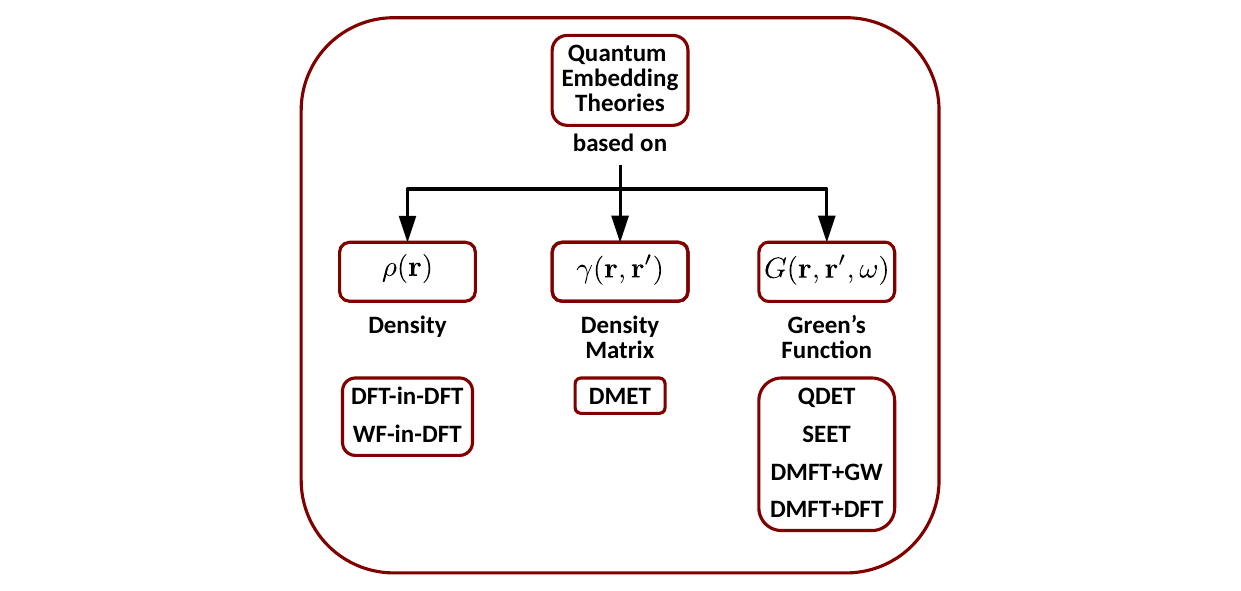}
\caption{\label{fig:overview} A summary of quantum embedding theories used in condensed-matter physics and quantum chemistry. Density-based theories
  encompass density-functional-theory-in-density-functional-theory (DFT-in-DFT)
  and  wavefunction-in-DFT (WF-in-DFT)
  embedding~\cite{libisch2014,wesolowski2015,jacob2014} schemes. Density-matrix
  embedding theory
  (DMET)~\cite{wouters2016,knizia2012,knizia2013,pham2020,hermes2019,pham2018}
  employs the density matrix ($\gamma$) to define an   embedding protocol. Green's ($G$)
  function-based methods include quantum defect embedding theory
  (QDET)~\cite{ma2020,ma2020a,ma2021}, self-energy embedding theory
  (SEET)~\cite{lan2017,zgid2017,rusakov2019}, dynamical mean-field theory (DMFT)+$GW$~\cite{biermann2003,biermann2014,boehnke2016,choi2016,nilsson2017,sun2002}, and DMFT+DFT~\cite{lichtenstein1998,anisimov1997} embedding.}
\end{figure}

As mentioned in the introduction, quantum embedding theories are frameworks to solve the
time-independent Schr\"odinger equation for a system of electrons by separating
the problem into the calculation of the energy levels or density of a so-called
\textit{active space} (or ``fragment'', ``impurity'') and  those of the
remaining \textit{environment}. Each  part of the system is described at the
quantum-mechanical level~\cite{sun2016,jones2020}, with the active space being
treated with a more accurate and computationally more expensive theoretical
method than the environment~\cite{sun2016}. In contrast, quantum
embedding {\it models} describe only the active space with quantum-mechanical
methods, while employing a classical description  for the
environment~\cite{jones2020,lin2006,wang2014,pezeshki2015}. 

The three key ingredients of a quantum embedding theory are the strategy used to
partition the full system into active space and environment, the computational
method adopted to describe the two portions, and the approximation for the
interaction between active space and environment~\cite{he2020}. As illustrated
in Fig.~\ref{fig:overview}, we can classify embedding theories by
identifying the key quantity used to realize the embedding~\cite{sun2016,jones2020}. In density functional embedding theory (DFET),
the  active space is defined by a region in real space and the density of the
system is partitioned into that of the active region and of the environment. DFT
calculations for the environment yield an exchange-correlation embedding
potential ~\cite{libisch2014,wesolowski2015,jacob2014,gujarati2022} which then enters the
Schr\"odinger equation for the active space; such equation is solved with a high-level
quantum-chemical method.  In density matrix embedding
theories~\cite{wouters2016,knizia2012,knizia2013,lau2021,pham2020,hermes2019,pham2018,cui2020,cui2022},
the active space is again defined by selecting a specific region of real space.  However, the
electronic structure of the active space is determined  by solving a
Schr\"odinger equation at a high level of theory  with additional bath
orbitals which  account for the interaction with the environment. This
framework is commonly known as the quantum impurity problem due to its
similarity with the Anderson impurity problem~\cite{anderson1961}. The bath
orbitals are obtained from a low-level calculation of the full system with an
additional one-particle operator; the latter is  constructed to satisfy the condition
that the density matrix at the low- and  high-level of theory be
identical. 

In Green's function embedding theories, such as dynamical
mean-field theory embedding (DMFT+DFT~\cite{lichtenstein1998,anisimov1997} and
DMFT+$GW$~\cite{biermann2003,biermann2014,boehnke2016,choi2016,nilsson2017,sun2002})
or self-energy embedding theory (SEET)~\cite{lan2017,zgid2017,rusakov2019}, the dynamical and non-local self-energy of the active space is expressed as a sum of terms evaluated at a high and low level of theory, with an additional double counting term. DMFT based methods and SEET approaches  differ by the choice of high- and low-level methods and by the technique used to separate the terms of the total self-energy of the system.


Recently, we proposed a Green's function based quantum embedding theory for the calculation of
defect properties in solids~\cite{ma2020,ma2020a,ma2021}, which we call 
quantum defect embedding theory (QDET). 
Note that the term {\it defect} here is not restricted to defects in solids and simply denotes a {\it small guest} region
embedded in a {\it large host} condensed system. Similar to all Green's function based
methods, in QDET the active space is defined by a set of single particle
electronic states. The set includes the states localized in proximity of the
defect or impurity and, in some cases, contains additional single particle orbitals
belonging to  the host material. Within QDET, one constructs an effective Hamiltonian in second quantization
which operates on the active space, using a potential which includes the effect
of the environment through an effective many-body screening term. The
Hamiltonian is solved using a full configuration interaction (FCI) approach,
thus including correlation effects between the electronic states of the active
space. 

The vast majority of calculations for model solids or materials carried out on quantum simulators or quantum computers, have adopted Green's function embedding theories. Few exceptions include tight-binding Hamiltonians~\cite{cerasoli2020,sureshbabu2021} and  the infinitely coordinated Bethe lattice Hubbard model treated within DMET~\cite{tilly2021}. Therefore it is instructive to compare in detail Green's function based quantum embedding theories before describing quantum computations, and such comparison is presented in the next section.

\subsection{Comparison between Green's function based quantum embedding theories} 

As discussed below, although in principle
QDET,
DMFT+$GW$  and  SEET are related methods,  these three frameworks target vastly different
properties and hence applications, and can handle systems of different sizes.  

Dynamical Mean Field Theory
was originally proposed to solve interacting lattice models, such as the Hubbard
and the periodic Anderson model~\cite{georges1996,werner2007}. Building on the original formulation, the framework commonly known as DMFT+$GW$
aims at  describing correlated bands in periodic solids, e.g., transition-metal
compounds with correlated 3$d$- and lanthanides with correlated
4$f$-bands~\cite{kotliar2006,nilsson2018}, for example
\ce{SrVO3}~\cite{sakuma2013,boehnke2016,petocchi2020,tomczak2017}, \ce{La2CuO4},
and NiO~\cite{choi2016}.  The approach is designed to yield thermodynamical properties
and charged excitations, which can be obtained from the one-body Green’s function. Current DMFT+$GW$ calculations on classical computers can 
tackle cells of the order of  5-20 atoms and dense grids of k-points. 

Within  this approach, the states of a chosen  active space 
 (e.g., the $3d$ bands
in transition metal compounds~\cite{kotliar2006,nilsson2018})  are
mapped onto an effective impurity problem, which by construction reproduces the
Green's function of the active space. The impurity and  the states of the full solid are connected through a so called hybridization self-energy $\Delta$,
which is determined self-consistently by requiring the impurity Green's function
be identical to the local Green's function of the active space. The influence of
the environment on the active space is described within MBPT, using a
diagrammatic expansion of the self-energy ($\Sigma$) in terms of the screened
Coulomb potential $W$. Vertex corrections are usually neglected and hence $\Sigma=
\mathrm{i}GW$~\cite{reining2018,onida2002,hedin1999,aryasetiawan1998,golze2019},
where $G$ is the Green's
function~\cite{choi2016,choi2019,nilsson2017,petocchi2020,tomczak2012,tomczak2017}.
Within DMFT+$GW$, the self-energy of the active space is given by the sum of the
DMFT and $GW$ self-energies plus a double counting term. The latter is necessary to
remove the contribution of diagrams that are contained in both the $GW$  and 
DMFT self-energies. Additionally, the interaction within the active space is
screened by the polarizability of the environment, and evaluated, e.g., by using the constrained random-phase approximation
(cRPA) method~\cite{aryasetiawan2004,aryasetiawan2009,miyake2008}. 

 Various schemes have been proposed in the literature to evaluate the dynamical screening, as well as the charge self-consistency~\cite{hampel2020,bhandary2021}, and double counting terms 
\cite{tomczak2012,nilsson2017,lee2017}.
While early implementations assumed the local interactions to be static within DMFT, dynamical interactions are now included in  DMFT+$GW$~\cite{biermann2014} calculations. These interactions may be computed  using quantum
Monte Carlo (QMC) solvers~\cite{eidelstein2020,seth2016,werner2007,werner2010} or exact diagonalization
(ED)~\cite{medvedeva2017} methods; however, the implementation of  QMC and ED solvers has been  only reported for approximate dynamically screened interactions ~\cite{biermann2014,werner2016,tomczak2017}.
An approximation to DMFT+$GW$ was introduced more than twenty years ago~\cite{anisimov1997,lichtenstein1998}, where the environment is described by DFT instead of many body perturbation theory; the approach is now known as DMFT+DFT. The combination of DMFT for a correlated subspace of orbitals and DFT for the remaining non-correlated states in the solid has become a widespread technique to study strongly correlated systems~\cite{kotliar2006,adler2018}. An exact double counting  scheme has been derived~\cite{haule2015} for DMFT+DFT, although approximate double counting schemes are most often employed in the literature~\cite{anisimov1997,haule2010,haule2014,vanroekeghem2014}. 

While DMFT+$GW$ targets the properties of solids, SEET was originally introduced to describe correlated states in molecular
systems. However, an extension to periodic systems has recently been
proposed~\cite{rusakov2019,yeh_electron_2021,iskakov_ab_2020}, and applied to hydrogen chains, and to unit cells of oxides with 2 to 5 atoms, including SrVO$_3$, SrMO$_3$, MnO and NiO. SEET yields ground state properties at zero temperature, for example
total energies.
In 
(SEET)~\cite{lan2017,zgid2017,kananenka2015,lan2015}   an active
space is  defined by a subspace of the full electronic Hilbert space, and is  used to separate the independent-particle Green's function
$G^0$ into that of the active space and its environment. Using the Dyson
equation $G=G^0+G^0\Sigma G$, where $G$  is the full  Green's function,  one
then defines an impurity problem for the active space,  with an explicit
hybridization term ($\Delta$) with the environment. The interaction within the
active space is given by the bare Coulomb potential. The self-energy of the
active space is obtained as the sum of the impurity self-energy, the self-energy
computed at a low-level of theory (e.g.,
$GW$~\cite{reining2018,onida2002,hedin1999,aryasetiawan1998}) and a
double counting term. The  quantum impurity problem is solved by
FCI~\cite{zgid2017,kananenka2015,lan2017a}.

\begin{figure}[t!]
\includegraphics[width=\textwidth]{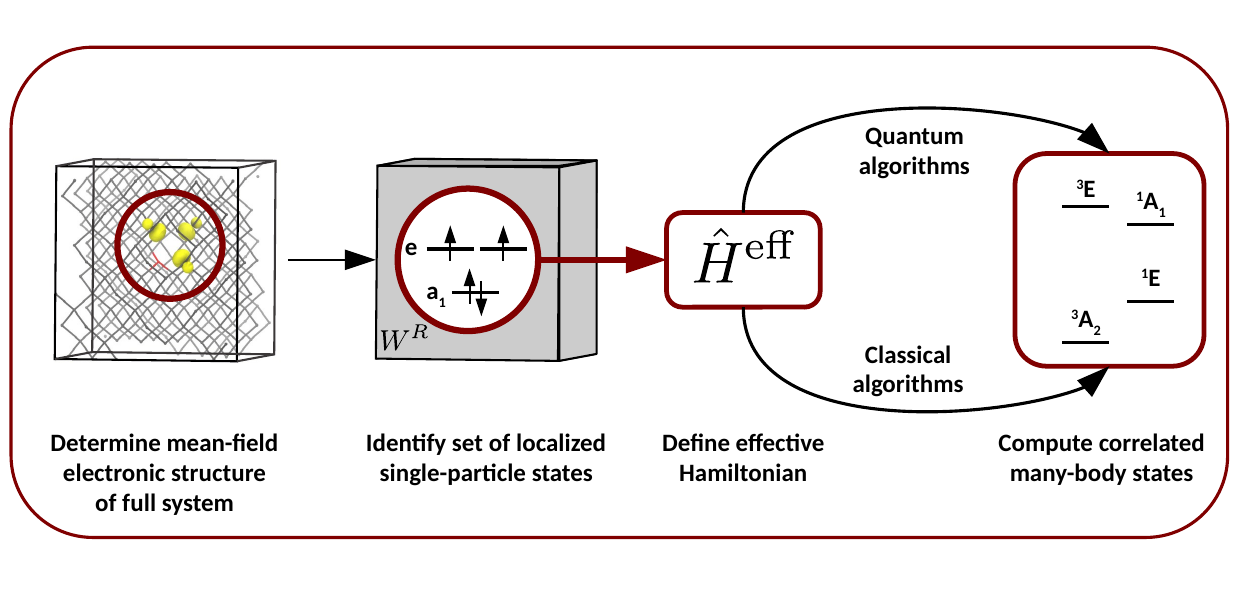}
\caption{\label{fig:workflow} Layout of materials simulations using the quantum
  defect embedding theory (QDET) of Ref.~\cite{ma2020a,ma2020,ma2021} on a
  classical and quantum computer. In the second and fourth panel from the left,
  representative single particle and many body states are shown, respectively.
  The term $W^{\mathrm{E}}$ (see text) denotes the screening  that the environment exerts
  on the active space. The effective Hamiltonian $H^{\mathrm{eff}}$ describes the active
  space (see text).}
\end{figure}

 Finally, QDET is designed to determine correlated electronic states of active regions in periodic solids, e.g., those of point defects embedded in a periodic crystal, and to obtain neutral excitations within the
active space,  for example the excited states of a spin defect with localized single particle states within a host crystal. The approach has recently been applied to defects in diamond and silicon
carbide~\cite{ma2020a,ma2020,ma2021}, as well as hexagonal
BN~\cite{muechler2021}, with periodic cells containing between 215 atoms and 511 atoms.
The specific steps of a QDET calculation are summarized in Fig.~\ref{fig:workflow}).  
After performing a (hybrid) DFT calculation for the full solid using a plane-wave basis set, an active space
is defined by selecting a number of Kohn-Sham states (e.g., by choosing four localized orbitals in the proximity of the
negatively-charged nitrogen vacancy center (\ce{NV-}) in diamond as in Ref.~\cite{ma2020}). 
The reduced polarizability $\chi^R$ and screening $W^R$ are calculated without explicit
summations over empty states (for example using the WEST code
\cite{govoni2015,scherpelz2016,govoni2018,govoni2021}). This is an important feature which makes QDET scalable to large supercells.
Using  the reduced screening as well as a double counting correction, an effective potential and an effective Hamiltonian $H^{\mathrm{eff}}$ are derived, describing the states belonging to the active space in the field of the environment (the rest of the crystal). For a given active space, $H^{\mathrm{eff}}$ is unique.
Note that screening effects may be evaluated either within the RPA or explicitly taking into
account the derivative of the exchange-correlation potential ($f^{\mathrm{xc}}$)~\cite{ma2020}.
We also note that an exact double counting correction within the $G_0W_0$ approximation, which is fully consistent with those employed in DMFT+$GW$, has been rigorously derived, implemented and tested for QDET~\cite{sheng2022}. 

In the following, we clarify the connection between DMFT+$GW$, SEET, and QDET with an instructive example, and we discuss how correlation effects are treated by the different methods. We consider a set of localized electronic states of a defect in an insulator or semiconductor in the dilute limit (i.e. in the limit where there is a single defect within the crystal and hence no interaction between defects is present). This specific case is known as the quantum impurity problem in the atomic
limit~\cite{biermann2014,casula2012,krivenko2015}. 
 Ground and excited states of some of these defects are of interest for the realization of qubits,  that are building blocks of quantum technologies, including quantum sensing, computing, and communication~\cite{weber2010,seo2016,seo2017,ivady2018,dreyer2018,smart2019,anderson2019}. While defect states in semiconductors do not form correlated bands and hence are not commonly studied using DMFT-based frameworks, they represent a useful example to illustrate similarities and differences between methods.

Consider a supercell of  a periodic solid in real space and assume the 
reciprocal space is sampled using only the $\Gamma$=(0,0,0) point.  The mapping of the active space (localized electronic states of the mean-field Hamiltonian)
to the impurity problem is exact and the hybridization vanishes, i.e., $\Delta =
0$. As such, a single, non-self-consistent solution of the impurity problem yields the electronic states of
the active space. Assuming the interactions to be instantaneous, the solution of
the impurity problem can be obtained by exact diagonalization, which is
equivalent to full configuration interaction (FCI). Under these specific
conditions, the DMFT+$GW$  and QDET frameworks are identical. However, the properties that can be computed using these two frameworks and their respective regimes of validity do differ, due to the different approximations adopted in practice when carrying out calculations.  Most notably, the two methods differ in their treatment of the impurity Hamiltonian: DMFT+$GW$ targets correlated bands in crystalline solids, where the onsite repulsion $U$ is the dominant term; hence the effective impurity Hamiltonian is commonly approximated by the Anderson impurity model~\cite{anderson1961} which only includes density-density repulsion terms. Some multi-orbital implementations of DMFT have been proposed in the literature, however they are rather challenging, from a computational standpoint~\cite{nomura2014,mizuno2021}. QDET, on the other hand, is specifically designed to compute neutral excitations of localized correlated states, where the treatment of exchange interactions is crucial (for example in multiplet excitations), and therefore no approximations to the effective Hamiltonian are applied.

 For a defect in a periodic supercell, both QDET and SEET obtain the 
self-energy as the sum of a quantum impurity self-energy, the $GW$
self-energy, and a double counting term. If the active space is chosen to be a
subset of the quasi-particle states obtained from the $GW$ calculation of the
full system, then the hybridization term vanishes; in this case the  SEET+$GW$
framework reduces to the atomic impurity problem for the active space and becomes similar to QDET. The main
difference resides in the screening of the interaction
in the active space which is  the bare Coulomb potential in SEET+$GW$; it is instead 
the screened Coulomb interaction within QDET. This difference also leads to
different expressions for the double counting correction to the active-space
self-energy~\cite{zgid2017}. 

In summary in the atomic limit,  that is in the case of a single defect  embedded in a surrounding host with which it does {\it not} hybridize, the DMFT+$GW$ and QDET methods coincide. SEET and QDET also turn out to be similar frameworks in the case of zero hybridization, however they differ by the type of interactions (bare or screened) entering the total potential acting on the active space.


\section{Quantum embedding electronic structure calculations on a quantum computer}
%
\begin{figure}[t!]
\includegraphics[width=\textwidth]{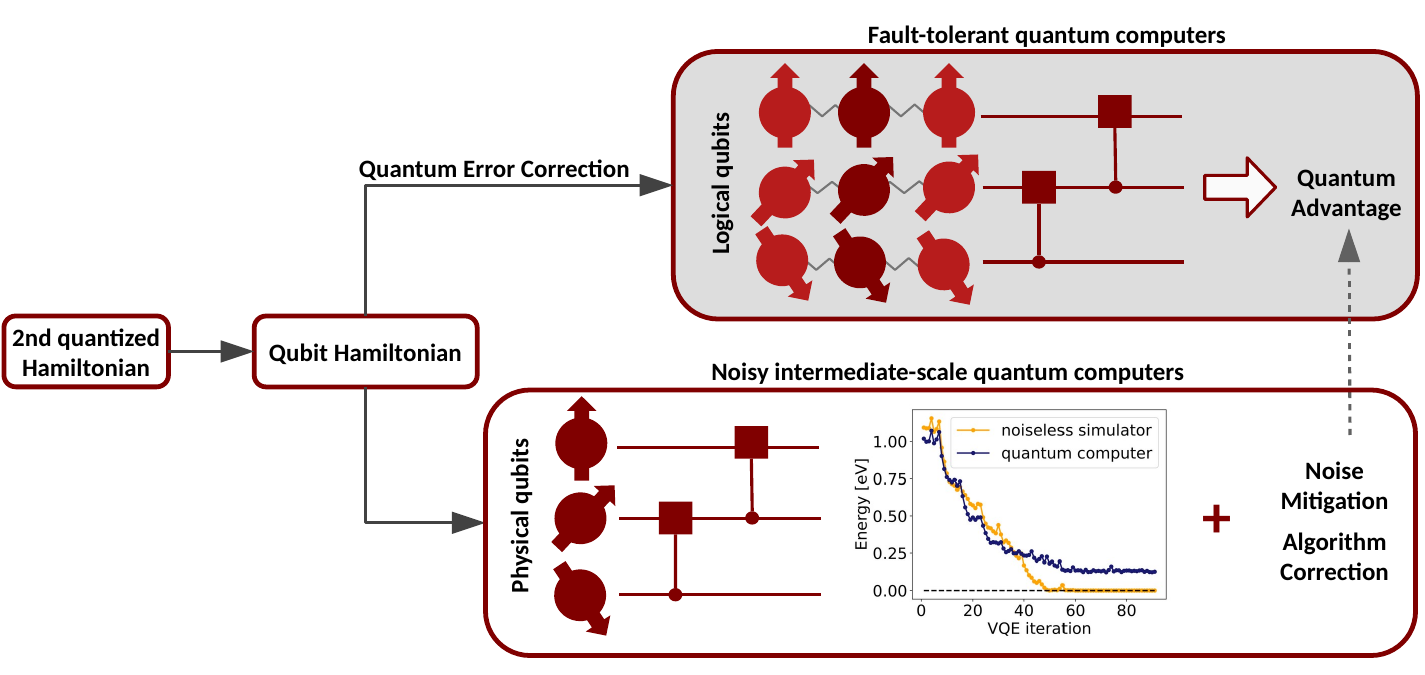}
\caption{\label{fig:quantumflow}Schematic representation of a quantum simulation: A
  second-quantized Hamiltonian is mapped to a spin Hamiltonian with  a qubit
  representation. In the current NISQ era, the calculation with physical qubits
  requires  noise mitigation and algorithm corrections (see text). In the future,
  quantum error correction  is expected to lead to a logical qubit representation, and in turn to  quantum advantage.}
\end{figure}

By reducing the complexity of electronic structure calculations of solids, quantum embedding theories lend themselves to possible implementations on quantum computers, which hold promises to reduce the exponential scaling of FCI  or ED calculations. 
Although quantum advantage for useful chemistry and physics problems has not yet been demonstrated, the use of quantum hardware is expected to eventually pave the  way to implement algorithms with a polynomial instead of exponential scaling as a function of the system size for the  solution of the time independent Schr\"{o}edinger equation. Indeed, in a classical computer $N$ classical bits can
represent one $N$-bit number, but $N$ quantum bits (qubits) can represent $2^N$ bit numbers on a quantum computer, providing a powerful memory scaling. Hence, it is interesting to explore how to obtain ground and excited states of second-quantized Hamiltonians and how to evaluate spectral functions on a quantum computer (see Fig.~\ref{fig:quantumflow}).

The computational complexity of quantum-embedding calculations based on Green's functions is determined by the size of
the active space.
In DMFT for example, the large
Hilbert space  of correlated $4f$ or $5f$ orbitals (14 partially occupied states) makes the study of lanthanide
and actinide compounds numerically challenging. Similar challenges arise when
the correlated orbitals of several atomic sites are included in the active space
in cellular DMFT~\cite{kotliar2001,deleo2008} (CDMFT). In CDMFT calculations reported so far on classical hardware, typically one uses supercells or clusters with up to 100 sites~\cite{gull_submatrix_2011,leblanc_solutions_2015}, and the maximum number of sites strongly depends on the temperature of the system.  In QDET and SEET, large Hilbert spaces may
arise from systems with a large number of localized defect states or from active regions containing, for example, multiple defects.


At present, SEET, DMFT+$GW$ or DMFT+DFT calculations of solids have not yet been implemented on quantum computers. Recently, first-principles DMFT calculations for \ce{La2CuO4} have carried out on a quantum emulator~\cite{jamet2021}. However, model Hamiltonians derived within DMFT have been diagonalized both on quantum simulators~\cite{kreula2016,bauer2016,kreula_non-linear_nodate,lupo_maximally_2021} and quantum hardware~\cite{rungger2020,keen2020,yao2021}. Note that the Hubbard and Heisenberg model Hamiltonians considered so far for quantum computations are finite-site models with typically 2 sites or orbitals, which can be directly encoded on quantum hardware; instead, the infinite Hubbard lattice cannot be directly encoded and DMFT has been used to map the infinite lattice to a finite-size problem. DMFT calculations on quantum computers have so far been performed only for the single-band Hubbard model on the Bethe lattice~\cite{rungger2020,keen2020,yao2021}. These calculations, using either the variational quantum eigensolver (VQE)~\cite{rungger2020,yao2021} or explicit time evolution techniques~\cite{keen2020} to obtain the impurity Green's function, have shown that quantum computing can yield results in good agreement with known analytical limits. However, the quantum noise of NISQ-era hardware leads to erroneous contributions to the self-energy $\Sigma$, which may prevent the DMFT calculation from converging~\cite{rungger2020,keen2020}. Thus, regularization by applying sum rules~\cite{rungger2020} or by approximations to $\Sigma(\omega)$~\cite{keen2020} are currently required. The noise can introduce unphysical poles in the spectral function, which in turn prevents  DMFT calculations  from reaching self-consistency.
~\cite{wecker2015,bauer2016,rubin2016,rungger2020,keen2020,jaderberg2020,yao2021,kreula2016}. 

Seminal applications of quantum embedding theories to realistic materials  on a quantum computer have been carried out using  QDET. Ma et al.~\cite{ma2020a} reported calculations of the strongly correlated states of the \ce{NV-} in diamond on a NISQ computer using a minimum model, and they obtained results consistent with those  of quantum emulators, the latter coinciding with the results of classical calculations. However,  the energies obtained on a quantum  computer  turned out to be slightly higher than those computed with a quantum emulator, due to hardware noise.

Recently, Huang \textit{et al.} calculated the excitations of the \ce{NV-} defect in diamond and the neutral divacancy (\ce{VV^0}) in 4H-SiC~\cite{huang2022} on a quantum computer within QDET using a combination of  variational quantum eigensolver (VQE)~\cite{peruzzo2014} and
quantum subspace expansion (QSE)~\cite{mcclean2017} algorithms. They found that hardware noise leads to poor conservation of the number of electrons, thus violating the variational principle. A post-selection of results enforcing particle conservation  was shown to significantly improve the accuracy of the calculations. An additional error reduction was obtained through a zero noise extrapolation scheme (ZNE)~\cite{endo2021,kandala2019}, an error mitigation scheme that doesn't require additional qubits.

Calculations on quantum computers using quantum embedding methods have been performed on a variety of quantum hardware, including IBM superconducting qubits~\cite{rungger2020,keen2020,ma2020,huang2022}, ion trap quantum qubits at the University of Maryland~\cite{rungger2020}, and the quantum cloud service by Rigetti~\cite{yao2021}. Generally, the calculations are performed using 4 qubits~\cite{rungger2020,keen2020,ma2020,huang2022}, with the exception of Ref.~\cite{yao2021} which employed 2 qubits.

 In spite of encouraging results on quantum computers, establishing which algorithms are better suited to obtain many-body energies of electronic states of solids on NISQ hardware remains an
open problem~\cite{huang2022,bauer2020}. For example, recent papers have proposed alternative methods to find the
eigenstates of a Fermionic Hamiltonian that are not based on the variational principle and
therefore do not involve an optimization procedure~\cite{motta2020}.
In particular, Ref.~\cite{korol2021} proposed an algorithm to prepare approximate ground states with
shallow circuit and, interestingly, just one parameter to define trial wave-functions.

\section{Outlook}

In conclusion, we have discussed  several quantum embedding
theories\cite{libisch2014,wesolowski2015,jacob2014,wouters2016,knizia2012,
knizia2013,pham2020,hermes2019,pham2018,ma2020,ma2020a,ma2021,lan2017,zgid2017,
rusakov2019,biermann2003,biermann2014,boehnke2016,choi2016,nilsson2017,sun2002,
lichtenstein1998,anisimov1997} which are promising frameworks for quantum
simulations of heterogeneous solids on near term quantum computers.
Although limited  to few systems (spin defects in semiconductors), the results obtained so far on quantum hardware  indicate that quantum simulations of strongly-correlated sites in
periodic systems are within reach for NISQ quantum computers.  Spin qubits~\cite{wolfowicz2021} represent just one of the possible applications
of quantum embedding theories, which may in principle be applied to a variety of
localized highly-correlated states, including those found in solvated ions and
nanostructures, adsorbates on surfaces and catalytic sites at surfaces and
interfaces. 

The verdict is not yet out, on whether quantum computers will
substantially improve the scaling of algorithms in use in classical computers to
diagonalize the Hamiltonian of active sites or to compute spectral functions of realistic materials beyond model Hamiltonians.
 Several estimates~\cite{peruzzo2014,aspuru-guzik2005,jaderberg2020,rungger2020,wecker2014,mcardle2020}  point at a few hundreds {\it logical} qubits as the requirement to reach quantum
advantage in terms of algorithmic scaling.
Although the relatively small number of logical qubits is encouraging, all algorithms in use today
require large gate counts of at least millions of error-corrected
gates~\cite{wecker2014,bauer2016}, going beyond the capability of current NISQ hardware~\cite{preskill2018} for the implementation of hundred logical qubits. Further development of improved quantum algorithms
and error-mitigation will therefore be crucial to demonstrate improved scaling on NISQ
machines.

In addition, since the lowest achievable scaling 
of quantum computational algorithms has not yet been fully determined,
the size of the active space for which quantum simulations become advantageous
over classical ones is still unknown. Fault-tolerant error correction
schemes\cite[and references therein]{lebreuilly_autonomous_2021}, necessary to
apply quantum simulations to realistic materials and to obtain full quantum advantage, are an active and yet
relatively new field of research and questions regarding which kind of schemes
might be needed to achieve chemical accuracy are still open questions. Furthermore, for all hybrid simulations where data are moved from classical to
quantum computers, communication schemes need to be carefully engineered in
order to avoid transfer  bottlenecks.

Finally, we note that here we just
briefly addressed the problem of solving the time-independent Schr\"odinger
equation for electrons in solids, namely a basic electronic structure
problem at fixed ionic coordinates. The calculations of additional properties of
materials, including structural stability~\cite{fedorov2021}, electron-phonon
interaction~\cite{macridin_electron-phonon_2018} and finite-temperature
properties\ \cite{powers_exploring_2021,wu_variational_2019} remain, as of yet, largely unexplored on quantum computers.

\section*{Author Contributions}
G.G. conceived this perspective and formulated the final content with all
authors. All authors contributed to the writing of the manuscript.

\begin{acknowledgments}
We thank Emanuel Gull and He Ma for fruitful discussions. This work was supported by MICCoM, as part of the Computational
  Materials Sciences Program funded by the U.S. Department of Energy. This
  research used resources of the National Energy Research Scientific Computing
  Center (NERSC), a DOE Office of Science User Facility supported by the Office
  of Science of the US Department of Energy under Contract No.
  DE-AC02-05CH11231, resources of the Argonne Leadership Computing Facility,
  which is a DOE Office of Science User Facility supported under Contract
  DE-AC02-06CH11357, and resources of the Oak Ridge Leadership
Computing Facility at the Oak Ridge National Laboratory, which is supported by the Office
of Science of the U.S. Department of Energy under Contract No. DE-AC05-00OR22725. We acknowledge the use of
  IBM Quantum services for this work. The views expressed are those of the
  authors, and do not reflect the official policy or position of IBM or the IBM
  Quantum team.
\end{acknowledgments}

\bibliography{bibliography}

\begin{thebibliography}{188}%
\makeatletter
\providecommand \@ifxundefined [1]{%
 \@ifx{#1\undefined}
}%
\providecommand \@ifnum [1]{%
 \ifnum #1\expandafter \@firstoftwo
 \else \expandafter \@secondoftwo
 \fi
}%
\providecommand \@ifx [1]{%
 \ifx #1\expandafter \@firstoftwo
 \else \expandafter \@secondoftwo
 \fi
}%
\providecommand \natexlab [1]{#1}%
\providecommand \enquote  [1]{``#1''}%
\providecommand \bibnamefont  [1]{#1}%
\providecommand \bibfnamefont [1]{#1}%
\providecommand \citenamefont [1]{#1}%
\providecommand \href@noop [0]{\@secondoftwo}%
\providecommand \href [0]{\begingroup \@sanitize@url \@href}%
\providecommand \@href[1]{\@@startlink{#1}\@@href}%
\providecommand \@@href[1]{\endgroup#1\@@endlink}%
\providecommand \@sanitize@url [0]{\catcode `\\12\catcode `\$12\catcode
  `\&12\catcode `\#12\catcode `\^12\catcode `\_12\catcode `\%12\relax}%
\providecommand \@@startlink[1]{}%
\providecommand \@@endlink[0]{}%
\providecommand \url  [0]{\begingroup\@sanitize@url \@url }%
\providecommand \@url [1]{\endgroup\@href {#1}{\urlprefix }}%
\providecommand \urlprefix  [0]{URL }%
\providecommand \Eprint [0]{\href }%
\providecommand \doibase [0]{http://dx.doi.org/}%
\providecommand \selectlanguage [0]{\@gobble}%
\providecommand \bibinfo  [0]{\@secondoftwo}%
\providecommand \bibfield  [0]{\@secondoftwo}%
\providecommand \translation [1]{[#1]}%
\providecommand \BibitemOpen [0]{}%
\providecommand \bibitemStop [0]{}%
\providecommand \bibitemNoStop [0]{.\EOS\space}%
\providecommand \EOS [0]{\spacefactor3000\relax}%
\providecommand \BibitemShut  [1]{\csname bibitem#1\endcsname}%
\let\auto@bib@innerbib\@empty
\bibitem [{\citenamefont {Jones}(2015)}]{jones2015}%
  \BibitemOpen
  \bibfield  {author} {\bibinfo {author} {\bibfnamefont {R.~O.}\ \bibnamefont
  {Jones}},\ }\href {\doibase 10.1103/RevModPhys.87.897} {\bibfield  {journal}
  {\bibinfo  {journal} {Reviews of Modern Physics}\ }\textbf {\bibinfo {volume}
  {87}},\ \bibinfo {pages} {897} (\bibinfo {year} {2015})}\BibitemShut
  {NoStop}%
\bibitem [{\citenamefont {Krylov}\ \emph {et~al.}(2018)\citenamefont {Krylov},
  \citenamefont {Windus}, \citenamefont {Barnes}, \citenamefont
  {{Marin-Rimoldi}}, \citenamefont {Nash}, \citenamefont {Pritchard},
  \citenamefont {Smith}, \citenamefont {Altarawy}, \citenamefont {Saxe},
  \citenamefont {Clementi}, \citenamefont {Crawford}, \citenamefont {Harrison},
  \citenamefont {Jha}, \citenamefont {Pande},\ and\ \citenamefont
  {{Head-Gordon}}}]{krylov2018}%
  \BibitemOpen
  \bibfield  {author} {\bibinfo {author} {\bibfnamefont {A.}~\bibnamefont
  {Krylov}}, \bibinfo {author} {\bibfnamefont {T.~L.}\ \bibnamefont {Windus}},
  \bibinfo {author} {\bibfnamefont {T.}~\bibnamefont {Barnes}}, \bibinfo
  {author} {\bibfnamefont {E.}~\bibnamefont {{Marin-Rimoldi}}}, \bibinfo
  {author} {\bibfnamefont {J.~A.}\ \bibnamefont {Nash}}, \bibinfo {author}
  {\bibfnamefont {B.}~\bibnamefont {Pritchard}}, \bibinfo {author}
  {\bibfnamefont {D.~G.~A.}\ \bibnamefont {Smith}}, \bibinfo {author}
  {\bibfnamefont {D.}~\bibnamefont {Altarawy}}, \bibinfo {author}
  {\bibfnamefont {P.}~\bibnamefont {Saxe}}, \bibinfo {author} {\bibfnamefont
  {C.}~\bibnamefont {Clementi}}, \bibinfo {author} {\bibfnamefont {T.~D.}\
  \bibnamefont {Crawford}}, \bibinfo {author} {\bibfnamefont {R.~J.}\
  \bibnamefont {Harrison}}, \bibinfo {author} {\bibfnamefont {S.}~\bibnamefont
  {Jha}}, \bibinfo {author} {\bibfnamefont {V.~S.}\ \bibnamefont {Pande}}, \
  and\ \bibinfo {author} {\bibfnamefont {T.}~\bibnamefont {{Head-Gordon}}},\
  }\href {\doibase 10.1063/1.5052551} {\bibfield  {journal} {\bibinfo
  {journal} {The Journal of Chemical Physics}\ }\textbf {\bibinfo {volume}
  {149}},\ \bibinfo {pages} {180901} (\bibinfo {year} {2018})}\BibitemShut
  {NoStop}%
\bibitem [{\citenamefont {Schleder}\ \emph {et~al.}(2019)\citenamefont
  {Schleder}, \citenamefont {Padilha}, \citenamefont {Acosta}, \citenamefont
  {Costa},\ and\ \citenamefont {Fazzio}}]{schleder2019}%
  \BibitemOpen
  \bibfield  {author} {\bibinfo {author} {\bibfnamefont {G.~R.}\ \bibnamefont
  {Schleder}}, \bibinfo {author} {\bibfnamefont {A.~C.~M.}\ \bibnamefont
  {Padilha}}, \bibinfo {author} {\bibfnamefont {C.~M.}\ \bibnamefont {Acosta}},
  \bibinfo {author} {\bibfnamefont {M.}~\bibnamefont {Costa}}, \ and\ \bibinfo
  {author} {\bibfnamefont {A.}~\bibnamefont {Fazzio}},\ }\href {\doibase
  10.1088/2515-7639/ab084b} {\bibfield  {journal} {\bibinfo  {journal} {Journal
  of Physics: Materials}\ }\textbf {\bibinfo {volume} {2}},\ \bibinfo {pages}
  {032001} (\bibinfo {year} {2019})}\BibitemShut {NoStop}%
\bibitem [{\citenamefont {Maurer}\ \emph {et~al.}(2019)\citenamefont {Maurer},
  \citenamefont {Freysoldt}, \citenamefont {Reilly}, \citenamefont
  {Brandenburg}, \citenamefont {Hofmann}, \citenamefont {Björkman},
  \citenamefont {Lebègue},\ and\ \citenamefont
  {Tkatchenko}}]{maurer_advances_2019}%
  \BibitemOpen
  \bibfield  {author} {\bibinfo {author} {\bibfnamefont {R.~J.}\ \bibnamefont
  {Maurer}}, \bibinfo {author} {\bibfnamefont {C.}~\bibnamefont {Freysoldt}},
  \bibinfo {author} {\bibfnamefont {A.~M.}\ \bibnamefont {Reilly}}, \bibinfo
  {author} {\bibfnamefont {J.~G.}\ \bibnamefont {Brandenburg}}, \bibinfo
  {author} {\bibfnamefont {O.~T.}\ \bibnamefont {Hofmann}}, \bibinfo {author}
  {\bibfnamefont {T.}~\bibnamefont {Björkman}}, \bibinfo {author}
  {\bibfnamefont {S.}~\bibnamefont {Lebègue}}, \ and\ \bibinfo {author}
  {\bibfnamefont {A.}~\bibnamefont {Tkatchenko}},\ }\href {\doibase
  10.1146/annurev-matsci-070218-010143} {\bibfield  {journal} {\bibinfo
  {journal} {Annual Review of Materials Research}\ }\textbf {\bibinfo {volume}
  {49}},\ \bibinfo {pages} {1} (\bibinfo {year} {2019})}\BibitemShut {NoStop}%
\bibitem [{\citenamefont {Bogojeski}\ \emph {et~al.}(2020)\citenamefont
  {Bogojeski}, \citenamefont {Vogt-Maranto}, \citenamefont {Tuckerman},
  \citenamefont {Müller},\ and\ \citenamefont
  {Burke}}]{bogojeski_quantum_2020}%
  \BibitemOpen
  \bibfield  {author} {\bibinfo {author} {\bibfnamefont {M.}~\bibnamefont
  {Bogojeski}}, \bibinfo {author} {\bibfnamefont {L.}~\bibnamefont
  {Vogt-Maranto}}, \bibinfo {author} {\bibfnamefont {M.~E.}\ \bibnamefont
  {Tuckerman}}, \bibinfo {author} {\bibfnamefont {K.-R.}\ \bibnamefont
  {Müller}}, \ and\ \bibinfo {author} {\bibfnamefont {K.}~\bibnamefont
  {Burke}},\ }\href {\doibase 10.1038/s41467-020-19093-1} {\bibfield  {journal}
  {\bibinfo  {journal} {Nature Communications}\ }\textbf {\bibinfo {volume}
  {11}},\ \bibinfo {pages} {5223} (\bibinfo {year} {2020})}\BibitemShut
  {NoStop}%
\bibitem [{\citenamefont {McArdle}\ \emph {et~al.}(2020)\citenamefont
  {McArdle}, \citenamefont {Endo}, \citenamefont {{Aspuru-Guzik}},
  \citenamefont {Benjamin},\ and\ \citenamefont {Yuan}}]{mcardle2020}%
  \BibitemOpen
  \bibfield  {author} {\bibinfo {author} {\bibfnamefont {S.}~\bibnamefont
  {McArdle}}, \bibinfo {author} {\bibfnamefont {S.}~\bibnamefont {Endo}},
  \bibinfo {author} {\bibfnamefont {A.}~\bibnamefont {{Aspuru-Guzik}}},
  \bibinfo {author} {\bibfnamefont {S.~C.}\ \bibnamefont {Benjamin}}, \ and\
  \bibinfo {author} {\bibfnamefont {X.}~\bibnamefont {Yuan}},\ }\href {\doibase
  10.1103/RevModPhys.92.015003} {\bibfield  {journal} {\bibinfo  {journal}
  {Reviews of Modern Physics}\ }\textbf {\bibinfo {volume} {92}},\ \bibinfo
  {pages} {015003} (\bibinfo {year} {2020})}\BibitemShut {NoStop}%
\bibitem [{\citenamefont {Bell}\ and\ \citenamefont
  {Head-Gordon}(2011)}]{bell_quantum_2011}%
  \BibitemOpen
  \bibfield  {author} {\bibinfo {author} {\bibfnamefont {A.~T.}\ \bibnamefont
  {Bell}}\ and\ \bibinfo {author} {\bibfnamefont {M.}~\bibnamefont
  {Head-Gordon}},\ }\href {\doibase 10.1146/annurev-chembioeng-061010-114108}
  {\bibfield  {journal} {\bibinfo  {journal} {Annual Review of Chemical and
  Biomolecular Engineering}\ }\textbf {\bibinfo {volume} {2}},\ \bibinfo
  {pages} {453} (\bibinfo {year} {2011})}\BibitemShut {NoStop}%
\bibitem [{\citenamefont {Xu}\ and\ \citenamefont
  {Carter}(2019)}]{xu_theoretical_2019}%
  \BibitemOpen
  \bibfield  {author} {\bibinfo {author} {\bibfnamefont {S.}~\bibnamefont
  {Xu}}\ and\ \bibinfo {author} {\bibfnamefont {E.~A.}\ \bibnamefont
  {Carter}},\ }\href {\doibase 10.1021/acs.chemrev.8b00481} {\bibfield
  {journal} {\bibinfo  {journal} {Chemical Reviews}\ }\textbf {\bibinfo
  {volume} {119}},\ \bibinfo {pages} {6631} (\bibinfo {year}
  {2019})}\BibitemShut {NoStop}%
\bibitem [{\citenamefont {Wolfowicz}\ \emph {et~al.}(2021)\citenamefont
  {Wolfowicz}, \citenamefont {Heremans}, \citenamefont {Anderson},
  \citenamefont {Kanai}, \citenamefont {Seo}, \citenamefont {Gali},
  \citenamefont {Galli},\ and\ \citenamefont {Awschalom}}]{wolfowicz2021}%
  \BibitemOpen
  \bibfield  {author} {\bibinfo {author} {\bibfnamefont {G.}~\bibnamefont
  {Wolfowicz}}, \bibinfo {author} {\bibfnamefont {F.~J.}\ \bibnamefont
  {Heremans}}, \bibinfo {author} {\bibfnamefont {C.~P.}\ \bibnamefont
  {Anderson}}, \bibinfo {author} {\bibfnamefont {S.}~\bibnamefont {Kanai}},
  \bibinfo {author} {\bibfnamefont {H.}~\bibnamefont {Seo}}, \bibinfo {author}
  {\bibfnamefont {A.}~\bibnamefont {Gali}}, \bibinfo {author} {\bibfnamefont
  {G.}~\bibnamefont {Galli}}, \ and\ \bibinfo {author} {\bibfnamefont {D.~D.}\
  \bibnamefont {Awschalom}},\ }\href {\doibase 10.1038/s41578-021-00306-y}
  {\bibfield  {journal} {\bibinfo  {journal} {Nature Reviews Materials}\ ,\
  \bibinfo {pages} {1}} (\bibinfo {year} {2021})}\BibitemShut {NoStop}%
\bibitem [{\citenamefont {Dreyer}\ \emph {et~al.}(2018)\citenamefont {Dreyer},
  \citenamefont {Alkauskas}, \citenamefont {Lyons}, \citenamefont {Janotti},\
  and\ \citenamefont {Van~de Walle}}]{dreyer2018}%
  \BibitemOpen
  \bibfield  {author} {\bibinfo {author} {\bibfnamefont {C.~E.}\ \bibnamefont
  {Dreyer}}, \bibinfo {author} {\bibfnamefont {A.}~\bibnamefont {Alkauskas}},
  \bibinfo {author} {\bibfnamefont {J.~L.}\ \bibnamefont {Lyons}}, \bibinfo
  {author} {\bibfnamefont {A.}~\bibnamefont {Janotti}}, \ and\ \bibinfo
  {author} {\bibfnamefont {C.~G.}\ \bibnamefont {Van~de Walle}},\ }\href
  {\doibase 10.1146/annurev-matsci-070317-124453} {\bibfield  {journal}
  {\bibinfo  {journal} {Annual Review of Materials Research}\ }\textbf
  {\bibinfo {volume} {48}},\ \bibinfo {pages} {1} (\bibinfo {year}
  {2018})}\BibitemShut {NoStop}%
\bibitem [{\citenamefont {Weber}\ \emph {et~al.}(2010)\citenamefont {Weber},
  \citenamefont {Koehl}, \citenamefont {Varley}, \citenamefont {Janotti},
  \citenamefont {Buckley}, \citenamefont {de~Walle},\ and\ \citenamefont
  {Awschalom}}]{weber2010}%
  \BibitemOpen
  \bibfield  {author} {\bibinfo {author} {\bibfnamefont {J.~R.}\ \bibnamefont
  {Weber}}, \bibinfo {author} {\bibfnamefont {W.~F.}\ \bibnamefont {Koehl}},
  \bibinfo {author} {\bibfnamefont {J.~B.}\ \bibnamefont {Varley}}, \bibinfo
  {author} {\bibfnamefont {A.}~\bibnamefont {Janotti}}, \bibinfo {author}
  {\bibfnamefont {B.~B.}\ \bibnamefont {Buckley}}, \bibinfo {author}
  {\bibfnamefont {C.~G.~V.}\ \bibnamefont {de~Walle}}, \ and\ \bibinfo {author}
  {\bibfnamefont {D.~D.}\ \bibnamefont {Awschalom}},\ }\href {\doibase
  10.1073/pnas.1003052107} {\bibfield  {journal} {\bibinfo  {journal}
  {Proceedings of the National Academy of Sciences}\ }\textbf {\bibinfo
  {volume} {107}},\ \bibinfo {pages} {8513} (\bibinfo {year}
  {2010})}\BibitemShut {NoStop}%
\bibitem [{\citenamefont {Agrawal}\ and\ \citenamefont
  {Choudhary}(2016)}]{agrawal2016}%
  \BibitemOpen
  \bibfield  {author} {\bibinfo {author} {\bibfnamefont {A.}~\bibnamefont
  {Agrawal}}\ and\ \bibinfo {author} {\bibfnamefont {A.}~\bibnamefont
  {Choudhary}},\ }\href {\doibase 10.1063/1.4946894} {\bibfield  {journal}
  {\bibinfo  {journal} {APL Materials}\ }\textbf {\bibinfo {volume} {4}},\
  \bibinfo {pages} {053208} (\bibinfo {year} {2016})}\BibitemShut {NoStop}%
\bibitem [{\citenamefont {Himanen}\ \emph {et~al.}(2019)\citenamefont
  {Himanen}, \citenamefont {Geurts}, \citenamefont {Foster},\ and\
  \citenamefont {Rinke}}]{himanen2019}%
  \BibitemOpen
  \bibfield  {author} {\bibinfo {author} {\bibfnamefont {L.}~\bibnamefont
  {Himanen}}, \bibinfo {author} {\bibfnamefont {A.}~\bibnamefont {Geurts}},
  \bibinfo {author} {\bibfnamefont {A.~S.}\ \bibnamefont {Foster}}, \ and\
  \bibinfo {author} {\bibfnamefont {P.}~\bibnamefont {Rinke}},\ }\href
  {\doibase 10.1002/advs.201900808} {\bibfield  {journal} {\bibinfo  {journal}
  {Advanced Science}\ }\textbf {\bibinfo {volume} {6}},\ \bibinfo {pages}
  {1900808} (\bibinfo {year} {2019})}\BibitemShut {NoStop}%
\bibitem [{\citenamefont {S.~Dong}\ \emph {et~al.}(2021)\citenamefont
  {S.~Dong}, \citenamefont {Govoni},\ and\ \citenamefont {Galli}}]{s.dong2021}%
  \BibitemOpen
  \bibfield  {author} {\bibinfo {author} {\bibfnamefont {S.}~\bibnamefont
  {S.~Dong}}, \bibinfo {author} {\bibfnamefont {M.}~\bibnamefont {Govoni}}, \
  and\ \bibinfo {author} {\bibfnamefont {G.}~\bibnamefont {Galli}},\ }\href
  {\doibase 10.1039/D1SC00503K} {\bibfield  {journal} {\bibinfo  {journal}
  {Chemical Science}\ }\textbf {\bibinfo {volume} {12}},\ \bibinfo {pages}
  {4970} (\bibinfo {year} {2021})}\BibitemShut {NoStop}%
\bibitem [{\citenamefont {Yuan}(2020)}]{yuan2020}%
  \BibitemOpen
  \bibfield  {author} {\bibinfo {author} {\bibfnamefont {X.}~\bibnamefont
  {Yuan}},\ }\href {\doibase 10.1126/science.abd3880} {\bibfield  {journal}
  {\bibinfo  {journal} {Science}\ }\textbf {\bibinfo {volume} {369}},\ \bibinfo
  {pages} {1054} (\bibinfo {year} {2020})}\BibitemShut {NoStop}%
\bibitem [{\citenamefont {Elfving}\ \emph {et~al.}(2020)\citenamefont
  {Elfving}, \citenamefont {Broer}, \citenamefont {Webber}, \citenamefont
  {Gavartin}, \citenamefont {Halls}, \citenamefont {Lorton},\ and\
  \citenamefont {Bochevarov}}]{elfving2020}%
  \BibitemOpen
  \bibfield  {author} {\bibinfo {author} {\bibfnamefont {V.~E.}\ \bibnamefont
  {Elfving}}, \bibinfo {author} {\bibfnamefont {B.~W.}\ \bibnamefont {Broer}},
  \bibinfo {author} {\bibfnamefont {M.}~\bibnamefont {Webber}}, \bibinfo
  {author} {\bibfnamefont {J.}~\bibnamefont {Gavartin}}, \bibinfo {author}
  {\bibfnamefont {M.~D.}\ \bibnamefont {Halls}}, \bibinfo {author}
  {\bibfnamefont {K.~P.}\ \bibnamefont {Lorton}}, \ and\ \bibinfo {author}
  {\bibfnamefont {A.}~\bibnamefont {Bochevarov}},\ }\href
  {http://arxiv.org/abs/2009.12472} {\bibfield  {journal} {\bibinfo  {journal}
  {arXiv:2009.12472 [physics, physics:quant-ph]}\ } (\bibinfo {year}
  {2020})}\BibitemShut {NoStop}%
\bibitem [{\citenamefont {{von Burg}}\ \emph {et~al.}(2021)\citenamefont {{von
  Burg}}, \citenamefont {Low}, \citenamefont {H{\"a}ner}, \citenamefont
  {Steiger}, \citenamefont {Reiher}, \citenamefont {Roetteler},\ and\
  \citenamefont {Troyer}}]{vonburg2021}%
  \BibitemOpen
  \bibfield  {author} {\bibinfo {author} {\bibfnamefont {V.}~\bibnamefont {{von
  Burg}}}, \bibinfo {author} {\bibfnamefont {G.~H.}\ \bibnamefont {Low}},
  \bibinfo {author} {\bibfnamefont {T.}~\bibnamefont {H{\"a}ner}}, \bibinfo
  {author} {\bibfnamefont {D.~S.}\ \bibnamefont {Steiger}}, \bibinfo {author}
  {\bibfnamefont {M.}~\bibnamefont {Reiher}}, \bibinfo {author} {\bibfnamefont
  {M.}~\bibnamefont {Roetteler}}, \ and\ \bibinfo {author} {\bibfnamefont
  {M.}~\bibnamefont {Troyer}},\ }\href {http://arxiv.org/abs/2007.14460}
  {\bibfield  {journal} {\bibinfo  {journal} {arXiv:2007.14460 [physics,
  physics:quant-ph]}\ } (\bibinfo {year} {2021})}\BibitemShut {NoStop}%
\bibitem [{\citenamefont {Liu}\ \emph {et~al.}(2021)\citenamefont {Liu},
  \citenamefont {Low}, \citenamefont {Steiger}, \citenamefont {H{\"a}ner},
  \citenamefont {Reiher},\ and\ \citenamefont {Troyer}}]{liu2021}%
  \BibitemOpen
  \bibfield  {author} {\bibinfo {author} {\bibfnamefont {H.}~\bibnamefont
  {Liu}}, \bibinfo {author} {\bibfnamefont {G.~H.}\ \bibnamefont {Low}},
  \bibinfo {author} {\bibfnamefont {D.~S.}\ \bibnamefont {Steiger}}, \bibinfo
  {author} {\bibfnamefont {T.}~\bibnamefont {H{\"a}ner}}, \bibinfo {author}
  {\bibfnamefont {M.}~\bibnamefont {Reiher}}, \ and\ \bibinfo {author}
  {\bibfnamefont {M.}~\bibnamefont {Troyer}},\ }\href
  {http://arxiv.org/abs/2102.10081} {\bibfield  {journal} {\bibinfo  {journal}
  {arXiv:2102.10081 [quant-ph]}\ } (\bibinfo {year} {2021})}\BibitemShut
  {NoStop}%
\bibitem [{\citenamefont {Ollitrault}\ \emph {et~al.}(2021)\citenamefont
  {Ollitrault}, \citenamefont {Miessen},\ and\ \citenamefont
  {Tavernelli}}]{ollitrault2021}%
  \BibitemOpen
  \bibfield  {author} {\bibinfo {author} {\bibfnamefont {P.~J.}\ \bibnamefont
  {Ollitrault}}, \bibinfo {author} {\bibfnamefont {A.}~\bibnamefont {Miessen}},
  \ and\ \bibinfo {author} {\bibfnamefont {I.}~\bibnamefont {Tavernelli}},\
  }\href {\doibase 10.1021/acs.accounts.1c00514} {\bibfield  {journal}
  {\bibinfo  {journal} {Accounts of Chemical Research}\ } (\bibinfo {year}
  {2021}),\ 10.1021/acs.accounts.1c00514}\BibitemShut {NoStop}%
\bibitem [{\citenamefont {Helgaker}\ \emph {et~al.}(2014)\citenamefont
  {Helgaker}, \citenamefont {Jorgensen},\ and\ \citenamefont
  {Olsen}}]{helgaker2014}%
  \BibitemOpen
  \bibfield  {author} {\bibinfo {author} {\bibfnamefont {T.}~\bibnamefont
  {Helgaker}}, \bibinfo {author} {\bibfnamefont {P.}~\bibnamefont {Jorgensen}},
  \ and\ \bibinfo {author} {\bibfnamefont {J.}~\bibnamefont {Olsen}},\
  }\href@noop {} {\emph {\bibinfo {title} {Molecular electronic-structure
  theory}}}\ (\bibinfo  {publisher} {John Wiley \& Sons},\ \bibinfo {year}
  {2014})\BibitemShut {NoStop}%
\bibitem [{\citenamefont {Martin}(2020)}]{martin2020}%
  \BibitemOpen
  \bibfield  {author} {\bibinfo {author} {\bibfnamefont {R.~M.}\ \bibnamefont
  {Martin}},\ }\href@noop {} {\emph {\bibinfo {title} {Electronic structure:
  basic theory and practical methods}}}\ (\bibinfo  {publisher} {Cambridge
  university press},\ \bibinfo {year} {2020})\BibitemShut {NoStop}%
\bibitem [{\citenamefont {Martin}\ \emph {et~al.}(2016)\citenamefont {Martin},
  \citenamefont {Reining},\ and\ \citenamefont {Ceperley}}]{martin2016}%
  \BibitemOpen
  \bibfield  {author} {\bibinfo {author} {\bibfnamefont {R.~M.}\ \bibnamefont
  {Martin}}, \bibinfo {author} {\bibfnamefont {L.}~\bibnamefont {Reining}}, \
  and\ \bibinfo {author} {\bibfnamefont {D.~M.}\ \bibnamefont {Ceperley}},\
  }\href@noop {} {\emph {\bibinfo {title} {Interacting electrons}}}\ (\bibinfo
  {publisher} {Cambridge University Press},\ \bibinfo {year}
  {2016})\BibitemShut {NoStop}%
\bibitem [{\citenamefont {Jordan}\ \emph {et~al.}(1934)\citenamefont {Jordan},
  \citenamefont {v.~Neumann},\ and\ \citenamefont {Wigner}}]{jordan1934}%
  \BibitemOpen
  \bibfield  {author} {\bibinfo {author} {\bibfnamefont {P.}~\bibnamefont
  {Jordan}}, \bibinfo {author} {\bibfnamefont {J.}~\bibnamefont {v.~Neumann}},
  \ and\ \bibinfo {author} {\bibfnamefont {E.}~\bibnamefont {Wigner}},\ }\href
  {\doibase 10.2307/1968117} {\bibfield  {journal} {\bibinfo  {journal} {Annals
  of Mathematics}\ }\textbf {\bibinfo {volume} {35}},\ \bibinfo {pages} {29}
  (\bibinfo {year} {1934})}\BibitemShut {NoStop}%
\bibitem [{\citenamefont {Bravyi}\ and\ \citenamefont
  {Kitaev}(2002)}]{bravyi2002}%
  \BibitemOpen
  \bibfield  {author} {\bibinfo {author} {\bibfnamefont {S.~B.}\ \bibnamefont
  {Bravyi}}\ and\ \bibinfo {author} {\bibfnamefont {A.~Y.}\ \bibnamefont
  {Kitaev}},\ }\href {\doibase 10.1006/aphy.2002.6254} {\bibfield  {journal}
  {\bibinfo  {journal} {Annals of Physics}\ }\textbf {\bibinfo {volume}
  {298}},\ \bibinfo {pages} {210} (\bibinfo {year} {2002})}\BibitemShut
  {NoStop}%
\bibitem [{\citenamefont {Seeley}\ \emph {et~al.}(2012)\citenamefont {Seeley},
  \citenamefont {Richard},\ and\ \citenamefont
  {Love}}]{seeley_bravyi-kitaev_2012}%
  \BibitemOpen
  \bibfield  {author} {\bibinfo {author} {\bibfnamefont {J.~T.}\ \bibnamefont
  {Seeley}}, \bibinfo {author} {\bibfnamefont {M.~J.}\ \bibnamefont {Richard}},
  \ and\ \bibinfo {author} {\bibfnamefont {P.~J.}\ \bibnamefont {Love}},\
  }\href {\doibase 10.1063/1.4768229} {\bibfield  {journal} {\bibinfo
  {journal} {The Journal of Chemical Physics}\ }\textbf {\bibinfo {volume}
  {137}},\ \bibinfo {pages} {224109} (\bibinfo {year} {2012})}\BibitemShut
  {NoStop}%
\bibitem [{\citenamefont {Verstraete}\ and\ \citenamefont
  {Cirac}(2005)}]{verstraete2005}%
  \BibitemOpen
  \bibfield  {author} {\bibinfo {author} {\bibfnamefont {F.}~\bibnamefont
  {Verstraete}}\ and\ \bibinfo {author} {\bibfnamefont {J.~I.}\ \bibnamefont
  {Cirac}},\ }\href {\doibase 10.1088/1742-5468/2005/09/P09012} {\bibfield
  {journal} {\bibinfo  {journal} {Journal of Statistical Mechanics: Theory and
  Experiment}\ }\textbf {\bibinfo {volume} {2005}},\ \bibinfo {pages} {P09012}
  (\bibinfo {year} {2005})}\BibitemShut {NoStop}%
\bibitem [{\citenamefont {Aleksandrowicz}\ \emph {et~al.}(2019)\citenamefont
  {Aleksandrowicz}, \citenamefont {Alexander}, \citenamefont {Barkoutsos},
  \citenamefont {Bello}, \citenamefont {Ben-Haim}, \citenamefont {Bucher},
  \citenamefont {Cabrera-Hern{\'a}ndez}, \citenamefont {Carballo-Franquis},
  \citenamefont {Chen}, \citenamefont {Chen} \emph
  {et~al.}}]{aleksandrowicz2019qiskit}%
  \BibitemOpen
  \bibfield  {author} {\bibinfo {author} {\bibfnamefont {G.}~\bibnamefont
  {Aleksandrowicz}}, \bibinfo {author} {\bibfnamefont {T.}~\bibnamefont
  {Alexander}}, \bibinfo {author} {\bibfnamefont {P.}~\bibnamefont
  {Barkoutsos}}, \bibinfo {author} {\bibfnamefont {L.}~\bibnamefont {Bello}},
  \bibinfo {author} {\bibfnamefont {Y.}~\bibnamefont {Ben-Haim}}, \bibinfo
  {author} {\bibfnamefont {D.}~\bibnamefont {Bucher}}, \bibinfo {author}
  {\bibfnamefont {F.}~\bibnamefont {Cabrera-Hern{\'a}ndez}}, \bibinfo {author}
  {\bibfnamefont {J.}~\bibnamefont {Carballo-Franquis}}, \bibinfo {author}
  {\bibfnamefont {A.}~\bibnamefont {Chen}}, \bibinfo {author} {\bibfnamefont
  {C.}~\bibnamefont {Chen}},  \emph {et~al.},\ }\href@noop {} {\bibfield
  {journal} {\bibinfo  {journal} {Accessed on: Mar}\ }\textbf {\bibinfo
  {volume} {16}} (\bibinfo {year} {2019})}\BibitemShut {NoStop}%
\bibitem [{\citenamefont {McClean}\ \emph {et~al.}(2020)\citenamefont
  {McClean}, \citenamefont {Rubin}, \citenamefont {Sung}, \citenamefont
  {Kivlichan}, \citenamefont {{Bonet-Monroig}}, \citenamefont {Cao},
  \citenamefont {Dai}, \citenamefont {Fried}, \citenamefont {Gidney},
  \citenamefont {Gimby}, \citenamefont {Gokhale}, \citenamefont {H{\"a}ner},
  \citenamefont {Hardikar}, \citenamefont {Havl{\'i}{\v c}ek}, \citenamefont
  {Higgott}, \citenamefont {Huang}, \citenamefont {Izaac}, \citenamefont
  {Jiang}, \citenamefont {Liu}, \citenamefont {McArdle}, \citenamefont
  {Neeley}, \citenamefont {O'Brien}, \citenamefont {O'Gorman}, \citenamefont
  {Ozfidan}, \citenamefont {Radin}, \citenamefont {Romero}, \citenamefont
  {Sawaya}, \citenamefont {Senjean}, \citenamefont {Setia}, \citenamefont
  {Sim}, \citenamefont {Steiger}, \citenamefont {Steudtner}, \citenamefont
  {Sun}, \citenamefont {Sun}, \citenamefont {Wang}, \citenamefont {Zhang},\
  and\ \citenamefont {Babbush}}]{mcclean2020}%
  \BibitemOpen
  \bibfield  {author} {\bibinfo {author} {\bibfnamefont {J.~R.}\ \bibnamefont
  {McClean}}, \bibinfo {author} {\bibfnamefont {N.~C.}\ \bibnamefont {Rubin}},
  \bibinfo {author} {\bibfnamefont {K.~J.}\ \bibnamefont {Sung}}, \bibinfo
  {author} {\bibfnamefont {I.~D.}\ \bibnamefont {Kivlichan}}, \bibinfo {author}
  {\bibfnamefont {X.}~\bibnamefont {{Bonet-Monroig}}}, \bibinfo {author}
  {\bibfnamefont {Y.}~\bibnamefont {Cao}}, \bibinfo {author} {\bibfnamefont
  {C.}~\bibnamefont {Dai}}, \bibinfo {author} {\bibfnamefont {E.~S.}\
  \bibnamefont {Fried}}, \bibinfo {author} {\bibfnamefont {C.}~\bibnamefont
  {Gidney}}, \bibinfo {author} {\bibfnamefont {B.}~\bibnamefont {Gimby}},
  \bibinfo {author} {\bibfnamefont {P.}~\bibnamefont {Gokhale}}, \bibinfo
  {author} {\bibfnamefont {T.}~\bibnamefont {H{\"a}ner}}, \bibinfo {author}
  {\bibfnamefont {T.}~\bibnamefont {Hardikar}}, \bibinfo {author}
  {\bibfnamefont {V.}~\bibnamefont {Havl{\'i}{\v c}ek}}, \bibinfo {author}
  {\bibfnamefont {O.}~\bibnamefont {Higgott}}, \bibinfo {author} {\bibfnamefont
  {C.}~\bibnamefont {Huang}}, \bibinfo {author} {\bibfnamefont
  {J.}~\bibnamefont {Izaac}}, \bibinfo {author} {\bibfnamefont
  {Z.}~\bibnamefont {Jiang}}, \bibinfo {author} {\bibfnamefont
  {X.}~\bibnamefont {Liu}}, \bibinfo {author} {\bibfnamefont {S.}~\bibnamefont
  {McArdle}}, \bibinfo {author} {\bibfnamefont {M.}~\bibnamefont {Neeley}},
  \bibinfo {author} {\bibfnamefont {T.}~\bibnamefont {O'Brien}}, \bibinfo
  {author} {\bibfnamefont {B.}~\bibnamefont {O'Gorman}}, \bibinfo {author}
  {\bibfnamefont {I.}~\bibnamefont {Ozfidan}}, \bibinfo {author} {\bibfnamefont
  {M.~D.}\ \bibnamefont {Radin}}, \bibinfo {author} {\bibfnamefont
  {J.}~\bibnamefont {Romero}}, \bibinfo {author} {\bibfnamefont {N.~P.~D.}\
  \bibnamefont {Sawaya}}, \bibinfo {author} {\bibfnamefont {B.}~\bibnamefont
  {Senjean}}, \bibinfo {author} {\bibfnamefont {K.}~\bibnamefont {Setia}},
  \bibinfo {author} {\bibfnamefont {S.}~\bibnamefont {Sim}}, \bibinfo {author}
  {\bibfnamefont {D.~S.}\ \bibnamefont {Steiger}}, \bibinfo {author}
  {\bibfnamefont {M.}~\bibnamefont {Steudtner}}, \bibinfo {author}
  {\bibfnamefont {Q.}~\bibnamefont {Sun}}, \bibinfo {author} {\bibfnamefont
  {W.}~\bibnamefont {Sun}}, \bibinfo {author} {\bibfnamefont {D.}~\bibnamefont
  {Wang}}, \bibinfo {author} {\bibfnamefont {F.}~\bibnamefont {Zhang}}, \ and\
  \bibinfo {author} {\bibfnamefont {R.}~\bibnamefont {Babbush}},\ }\href
  {\doibase 10.1088/2058-9565/ab8ebc} {\bibfield  {journal} {\bibinfo
  {journal} {Quantum Science and Technology}\ }\textbf {\bibinfo {volume}
  {5}},\ \bibinfo {pages} {034014} (\bibinfo {year} {2020})}\BibitemShut
  {NoStop}%
\bibitem [{\citenamefont {Peruzzo}\ \emph {et~al.}(2014)\citenamefont
  {Peruzzo}, \citenamefont {McClean}, \citenamefont {Shadbolt}, \citenamefont
  {Yung}, \citenamefont {Zhou}, \citenamefont {Love}, \citenamefont
  {{Aspuru-Guzik}},\ and\ \citenamefont {O'Brien}}]{peruzzo2014}%
  \BibitemOpen
  \bibfield  {author} {\bibinfo {author} {\bibfnamefont {A.}~\bibnamefont
  {Peruzzo}}, \bibinfo {author} {\bibfnamefont {J.}~\bibnamefont {McClean}},
  \bibinfo {author} {\bibfnamefont {P.}~\bibnamefont {Shadbolt}}, \bibinfo
  {author} {\bibfnamefont {M.-H.}\ \bibnamefont {Yung}}, \bibinfo {author}
  {\bibfnamefont {X.-Q.}\ \bibnamefont {Zhou}}, \bibinfo {author}
  {\bibfnamefont {P.~J.}\ \bibnamefont {Love}}, \bibinfo {author}
  {\bibfnamefont {A.}~\bibnamefont {{Aspuru-Guzik}}}, \ and\ \bibinfo {author}
  {\bibfnamefont {J.~L.}\ \bibnamefont {O'Brien}},\ }\href {\doibase
  10.1038/ncomms5213} {\bibfield  {journal} {\bibinfo  {journal} {Nature
  Communications}\ }\textbf {\bibinfo {volume} {5}},\ \bibinfo {pages} {4213}
  (\bibinfo {year} {2014})}\BibitemShut {NoStop}%
\bibitem [{\citenamefont {McClean}\ \emph {et~al.}(2016)\citenamefont
  {McClean}, \citenamefont {Romero}, \citenamefont {Babbush},\ and\
  \citenamefont {{Aspuru-Guzik}}}]{mcclean2016}%
  \BibitemOpen
  \bibfield  {author} {\bibinfo {author} {\bibfnamefont {J.~R.}\ \bibnamefont
  {McClean}}, \bibinfo {author} {\bibfnamefont {J.}~\bibnamefont {Romero}},
  \bibinfo {author} {\bibfnamefont {R.}~\bibnamefont {Babbush}}, \ and\
  \bibinfo {author} {\bibfnamefont {A.}~\bibnamefont {{Aspuru-Guzik}}},\ }\href
  {\doibase 10.1088/1367-2630/18/2/023023} {\bibfield  {journal} {\bibinfo
  {journal} {New Journal of Physics}\ }\textbf {\bibinfo {volume} {18}},\
  \bibinfo {pages} {023023} (\bibinfo {year} {2016})}\BibitemShut {NoStop}%
\bibitem [{\citenamefont {Nielsen}\ and\ \citenamefont
  {Chuang}(2010)}]{nielsen2010}%
  \BibitemOpen
  \bibfield  {author} {\bibinfo {author} {\bibfnamefont {M.~A.}\ \bibnamefont
  {Nielsen}}\ and\ \bibinfo {author} {\bibfnamefont {I.~L.}\ \bibnamefont
  {Chuang}},\ }\href {\doibase 10.1017/CBO9780511976667} {\emph {\bibinfo
  {title} {Quantum Computation and Quantum Information: 10th {{Anniversary}}
  Edition}}}\ (\bibinfo  {publisher} {{Cambridge University Press}},\ \bibinfo
  {address} {{Cambridge}},\ \bibinfo {year} {2010})\BibitemShut {NoStop}%
\bibitem [{\citenamefont {Bravyi}\ \emph {et~al.}(2020)\citenamefont {Bravyi},
  \citenamefont {Gosset}, \citenamefont {K{\"o}nig},\ and\ \citenamefont
  {Tomamichel}}]{bravyi2020}%
  \BibitemOpen
  \bibfield  {author} {\bibinfo {author} {\bibfnamefont {S.}~\bibnamefont
  {Bravyi}}, \bibinfo {author} {\bibfnamefont {D.}~\bibnamefont {Gosset}},
  \bibinfo {author} {\bibfnamefont {R.}~\bibnamefont {K{\"o}nig}}, \ and\
  \bibinfo {author} {\bibfnamefont {M.}~\bibnamefont {Tomamichel}},\ }\href
  {\doibase 10.1038/s41567-020-0948-z} {\bibfield  {journal} {\bibinfo
  {journal} {Nature Physics}\ }\textbf {\bibinfo {volume} {16}},\ \bibinfo
  {pages} {1040} (\bibinfo {year} {2020})}\BibitemShut {NoStop}%
\bibitem [{\citenamefont {{Aspuru-Guzik}}\ \emph {et~al.}(2005)\citenamefont
  {{Aspuru-Guzik}}, \citenamefont {Dutoi}, \citenamefont {Love},\ and\
  \citenamefont {{Head-Gordon}}}]{aspuru-guzik2005}%
  \BibitemOpen
  \bibfield  {author} {\bibinfo {author} {\bibfnamefont {A.}~\bibnamefont
  {{Aspuru-Guzik}}}, \bibinfo {author} {\bibfnamefont {A.~D.}\ \bibnamefont
  {Dutoi}}, \bibinfo {author} {\bibfnamefont {P.~J.}\ \bibnamefont {Love}}, \
  and\ \bibinfo {author} {\bibfnamefont {M.}~\bibnamefont {{Head-Gordon}}},\
  }\href {\doibase 10.1126/science.1113479} {\bibfield  {journal} {\bibinfo
  {journal} {Science}\ }\textbf {\bibinfo {volume} {309}},\ \bibinfo {pages}
  {1704} (\bibinfo {year} {2005})}\BibitemShut {NoStop}%
\bibitem [{\citenamefont {Lanyon}\ \emph {et~al.}(2010)\citenamefont {Lanyon},
  \citenamefont {Whitfield}, \citenamefont {Gillett}, \citenamefont {Goggin},
  \citenamefont {Almeida}, \citenamefont {Kassal}, \citenamefont {Biamonte},
  \citenamefont {Mohseni}, \citenamefont {Powell}, \citenamefont {Barbieri},
  \citenamefont {{Aspuru-Guzik}},\ and\ \citenamefont {White}}]{lanyon2010}%
  \BibitemOpen
  \bibfield  {author} {\bibinfo {author} {\bibfnamefont {B.~P.}\ \bibnamefont
  {Lanyon}}, \bibinfo {author} {\bibfnamefont {J.~D.}\ \bibnamefont
  {Whitfield}}, \bibinfo {author} {\bibfnamefont {G.~G.}\ \bibnamefont
  {Gillett}}, \bibinfo {author} {\bibfnamefont {M.~E.}\ \bibnamefont {Goggin}},
  \bibinfo {author} {\bibfnamefont {M.~P.}\ \bibnamefont {Almeida}}, \bibinfo
  {author} {\bibfnamefont {I.}~\bibnamefont {Kassal}}, \bibinfo {author}
  {\bibfnamefont {J.~D.}\ \bibnamefont {Biamonte}}, \bibinfo {author}
  {\bibfnamefont {M.}~\bibnamefont {Mohseni}}, \bibinfo {author} {\bibfnamefont
  {B.~J.}\ \bibnamefont {Powell}}, \bibinfo {author} {\bibfnamefont
  {M.}~\bibnamefont {Barbieri}}, \bibinfo {author} {\bibfnamefont
  {A.}~\bibnamefont {{Aspuru-Guzik}}}, \ and\ \bibinfo {author} {\bibfnamefont
  {A.~G.}\ \bibnamefont {White}},\ }\href {\doibase 10.1038/nchem.483}
  {\bibfield  {journal} {\bibinfo  {journal} {Nature Chemistry}\ }\textbf
  {\bibinfo {volume} {2}},\ \bibinfo {pages} {106} (\bibinfo {year}
  {2010})}\BibitemShut {NoStop}%
\bibitem [{\citenamefont {Li}\ \emph {et~al.}(2011)\citenamefont {Li},
  \citenamefont {Yung}, \citenamefont {Chen}, \citenamefont {Lu}, \citenamefont
  {Whitfield}, \citenamefont {Peng}, \citenamefont {{Aspuru-Guzik}},\ and\
  \citenamefont {Du}}]{li2011}%
  \BibitemOpen
  \bibfield  {author} {\bibinfo {author} {\bibfnamefont {Z.}~\bibnamefont
  {Li}}, \bibinfo {author} {\bibfnamefont {M.-H.}\ \bibnamefont {Yung}},
  \bibinfo {author} {\bibfnamefont {H.}~\bibnamefont {Chen}}, \bibinfo {author}
  {\bibfnamefont {D.}~\bibnamefont {Lu}}, \bibinfo {author} {\bibfnamefont
  {J.~D.}\ \bibnamefont {Whitfield}}, \bibinfo {author} {\bibfnamefont
  {X.}~\bibnamefont {Peng}}, \bibinfo {author} {\bibfnamefont {A.}~\bibnamefont
  {{Aspuru-Guzik}}}, \ and\ \bibinfo {author} {\bibfnamefont {J.}~\bibnamefont
  {Du}},\ }\href {\doibase 10.1038/srep00088} {\bibfield  {journal} {\bibinfo
  {journal} {Scientific Reports}\ }\textbf {\bibinfo {volume} {1}},\ \bibinfo
  {pages} {88} (\bibinfo {year} {2011})}\BibitemShut {NoStop}%
\bibitem [{\citenamefont {Shen}\ \emph {et~al.}(2017)\citenamefont {Shen},
  \citenamefont {Zhang}, \citenamefont {Zhang}, \citenamefont {Zhang},
  \citenamefont {Yung},\ and\ \citenamefont {Kim}}]{shen2017}%
  \BibitemOpen
  \bibfield  {author} {\bibinfo {author} {\bibfnamefont {Y.}~\bibnamefont
  {Shen}}, \bibinfo {author} {\bibfnamefont {X.}~\bibnamefont {Zhang}},
  \bibinfo {author} {\bibfnamefont {S.}~\bibnamefont {Zhang}}, \bibinfo
  {author} {\bibfnamefont {J.-N.}\ \bibnamefont {Zhang}}, \bibinfo {author}
  {\bibfnamefont {M.-H.}\ \bibnamefont {Yung}}, \ and\ \bibinfo {author}
  {\bibfnamefont {K.}~\bibnamefont {Kim}},\ }\href {\doibase
  10.1103/PhysRevA.95.020501} {\bibfield  {journal} {\bibinfo  {journal}
  {Physical Review A}\ }\textbf {\bibinfo {volume} {95}},\ \bibinfo {pages}
  {020501} (\bibinfo {year} {2017})}\BibitemShut {NoStop}%
\bibitem [{\citenamefont {O'Malley}\ \emph {et~al.}(2016)\citenamefont
  {O'Malley}, \citenamefont {Babbush}, \citenamefont {Kivlichan}, \citenamefont
  {Romero}, \citenamefont {McClean}, \citenamefont {Barends}, \citenamefont
  {Kelly}, \citenamefont {Roushan}, \citenamefont {Tranter}, \citenamefont
  {Ding}, \citenamefont {Campbell}, \citenamefont {Chen}, \citenamefont {Chen},
  \citenamefont {Chiaro}, \citenamefont {Dunsworth}, \citenamefont {Fowler},
  \citenamefont {Jeffrey}, \citenamefont {Lucero}, \citenamefont {Megrant},
  \citenamefont {Mutus}, \citenamefont {Neeley}, \citenamefont {Neill},
  \citenamefont {Quintana}, \citenamefont {Sank}, \citenamefont {Vainsencher},
  \citenamefont {Wenner}, \citenamefont {White}, \citenamefont {Coveney},
  \citenamefont {Love}, \citenamefont {Neven}, \citenamefont {{Aspuru-Guzik}},\
  and\ \citenamefont {Martinis}}]{omalley2016}%
  \BibitemOpen
  \bibfield  {author} {\bibinfo {author} {\bibfnamefont {P.~J.~J.}\
  \bibnamefont {O'Malley}}, \bibinfo {author} {\bibfnamefont {R.}~\bibnamefont
  {Babbush}}, \bibinfo {author} {\bibfnamefont {I.~D.}\ \bibnamefont
  {Kivlichan}}, \bibinfo {author} {\bibfnamefont {J.}~\bibnamefont {Romero}},
  \bibinfo {author} {\bibfnamefont {J.~R.}\ \bibnamefont {McClean}}, \bibinfo
  {author} {\bibfnamefont {R.}~\bibnamefont {Barends}}, \bibinfo {author}
  {\bibfnamefont {J.}~\bibnamefont {Kelly}}, \bibinfo {author} {\bibfnamefont
  {P.}~\bibnamefont {Roushan}}, \bibinfo {author} {\bibfnamefont
  {A.}~\bibnamefont {Tranter}}, \bibinfo {author} {\bibfnamefont
  {N.}~\bibnamefont {Ding}}, \bibinfo {author} {\bibfnamefont {B.}~\bibnamefont
  {Campbell}}, \bibinfo {author} {\bibfnamefont {Y.}~\bibnamefont {Chen}},
  \bibinfo {author} {\bibfnamefont {Z.}~\bibnamefont {Chen}}, \bibinfo {author}
  {\bibfnamefont {B.}~\bibnamefont {Chiaro}}, \bibinfo {author} {\bibfnamefont
  {A.}~\bibnamefont {Dunsworth}}, \bibinfo {author} {\bibfnamefont {A.~G.}\
  \bibnamefont {Fowler}}, \bibinfo {author} {\bibfnamefont {E.}~\bibnamefont
  {Jeffrey}}, \bibinfo {author} {\bibfnamefont {E.}~\bibnamefont {Lucero}},
  \bibinfo {author} {\bibfnamefont {A.}~\bibnamefont {Megrant}}, \bibinfo
  {author} {\bibfnamefont {J.~Y.}\ \bibnamefont {Mutus}}, \bibinfo {author}
  {\bibfnamefont {M.}~\bibnamefont {Neeley}}, \bibinfo {author} {\bibfnamefont
  {C.}~\bibnamefont {Neill}}, \bibinfo {author} {\bibfnamefont
  {C.}~\bibnamefont {Quintana}}, \bibinfo {author} {\bibfnamefont
  {D.}~\bibnamefont {Sank}}, \bibinfo {author} {\bibfnamefont {A.}~\bibnamefont
  {Vainsencher}}, \bibinfo {author} {\bibfnamefont {J.}~\bibnamefont {Wenner}},
  \bibinfo {author} {\bibfnamefont {T.~C.}\ \bibnamefont {White}}, \bibinfo
  {author} {\bibfnamefont {P.~V.}\ \bibnamefont {Coveney}}, \bibinfo {author}
  {\bibfnamefont {P.~J.}\ \bibnamefont {Love}}, \bibinfo {author}
  {\bibfnamefont {H.}~\bibnamefont {Neven}}, \bibinfo {author} {\bibfnamefont
  {A.}~\bibnamefont {{Aspuru-Guzik}}}, \ and\ \bibinfo {author} {\bibfnamefont
  {J.~M.}\ \bibnamefont {Martinis}},\ }\href {\doibase
  10.1103/PhysRevX.6.031007} {\bibfield  {journal} {\bibinfo  {journal}
  {Physical Review X}\ }\textbf {\bibinfo {volume} {6}},\ \bibinfo {pages}
  {031007} (\bibinfo {year} {2016})}\BibitemShut {NoStop}%
\bibitem [{\citenamefont {Santagati}\ \emph {et~al.}(2018)\citenamefont
  {Santagati}, \citenamefont {Wang}, \citenamefont {Gentile}, \citenamefont
  {Paesani}, \citenamefont {Wiebe}, \citenamefont {McClean}, \citenamefont
  {{Morley-Short}}, \citenamefont {Shadbolt}, \citenamefont {Bonneau},
  \citenamefont {Silverstone}, \citenamefont {Tew}, \citenamefont {Zhou},
  \citenamefont {O'Brien},\ and\ \citenamefont {Thompson}}]{santagati2018}%
  \BibitemOpen
  \bibfield  {author} {\bibinfo {author} {\bibfnamefont {R.}~\bibnamefont
  {Santagati}}, \bibinfo {author} {\bibfnamefont {J.}~\bibnamefont {Wang}},
  \bibinfo {author} {\bibfnamefont {A.~A.}\ \bibnamefont {Gentile}}, \bibinfo
  {author} {\bibfnamefont {S.}~\bibnamefont {Paesani}}, \bibinfo {author}
  {\bibfnamefont {N.}~\bibnamefont {Wiebe}}, \bibinfo {author} {\bibfnamefont
  {J.~R.}\ \bibnamefont {McClean}}, \bibinfo {author} {\bibfnamefont
  {S.}~\bibnamefont {{Morley-Short}}}, \bibinfo {author} {\bibfnamefont
  {P.~J.}\ \bibnamefont {Shadbolt}}, \bibinfo {author} {\bibfnamefont
  {D.}~\bibnamefont {Bonneau}}, \bibinfo {author} {\bibfnamefont {J.~W.}\
  \bibnamefont {Silverstone}}, \bibinfo {author} {\bibfnamefont {D.~P.}\
  \bibnamefont {Tew}}, \bibinfo {author} {\bibfnamefont {X.}~\bibnamefont
  {Zhou}}, \bibinfo {author} {\bibfnamefont {J.~L.}\ \bibnamefont {O'Brien}}, \
  and\ \bibinfo {author} {\bibfnamefont {M.~G.}\ \bibnamefont {Thompson}},\
  }\href {\doibase 10.1126/sciadv.aap9646} {\bibfield  {journal} {\bibinfo
  {journal} {Science Advances}\ }\textbf {\bibinfo {volume} {4}},\ \bibinfo
  {pages} {eaap9646} (\bibinfo {year} {2018})}\BibitemShut {NoStop}%
\bibitem [{\citenamefont {Kandala}\ \emph {et~al.}(2017)\citenamefont
  {Kandala}, \citenamefont {Mezzacapo}, \citenamefont {Temme}, \citenamefont
  {Takita}, \citenamefont {Brink}, \citenamefont {Chow},\ and\ \citenamefont
  {Gambetta}}]{kandala2017}%
  \BibitemOpen
  \bibfield  {author} {\bibinfo {author} {\bibfnamefont {A.}~\bibnamefont
  {Kandala}}, \bibinfo {author} {\bibfnamefont {A.}~\bibnamefont {Mezzacapo}},
  \bibinfo {author} {\bibfnamefont {K.}~\bibnamefont {Temme}}, \bibinfo
  {author} {\bibfnamefont {M.}~\bibnamefont {Takita}}, \bibinfo {author}
  {\bibfnamefont {M.}~\bibnamefont {Brink}}, \bibinfo {author} {\bibfnamefont
  {J.~M.}\ \bibnamefont {Chow}}, \ and\ \bibinfo {author} {\bibfnamefont
  {J.~M.}\ \bibnamefont {Gambetta}},\ }\href {\doibase 10.1038/nature23879}
  {\bibfield  {journal} {\bibinfo  {journal} {Nature}\ }\textbf {\bibinfo
  {volume} {549}},\ \bibinfo {pages} {242} (\bibinfo {year}
  {2017})}\BibitemShut {NoStop}%
\bibitem [{\citenamefont {Hempel}\ \emph {et~al.}(2018)\citenamefont {Hempel},
  \citenamefont {Maier}, \citenamefont {Romero}, \citenamefont {McClean},
  \citenamefont {Monz}, \citenamefont {Shen}, \citenamefont {Jurcevic},
  \citenamefont {Lanyon}, \citenamefont {Love}, \citenamefont {Babbush},
  \citenamefont {{Aspuru-Guzik}}, \citenamefont {Blatt},\ and\ \citenamefont
  {Roos}}]{hempel2018}%
  \BibitemOpen
  \bibfield  {author} {\bibinfo {author} {\bibfnamefont {C.}~\bibnamefont
  {Hempel}}, \bibinfo {author} {\bibfnamefont {C.}~\bibnamefont {Maier}},
  \bibinfo {author} {\bibfnamefont {J.}~\bibnamefont {Romero}}, \bibinfo
  {author} {\bibfnamefont {J.}~\bibnamefont {McClean}}, \bibinfo {author}
  {\bibfnamefont {T.}~\bibnamefont {Monz}}, \bibinfo {author} {\bibfnamefont
  {H.}~\bibnamefont {Shen}}, \bibinfo {author} {\bibfnamefont {P.}~\bibnamefont
  {Jurcevic}}, \bibinfo {author} {\bibfnamefont {B.~P.}\ \bibnamefont
  {Lanyon}}, \bibinfo {author} {\bibfnamefont {P.}~\bibnamefont {Love}},
  \bibinfo {author} {\bibfnamefont {R.}~\bibnamefont {Babbush}}, \bibinfo
  {author} {\bibfnamefont {A.}~\bibnamefont {{Aspuru-Guzik}}}, \bibinfo
  {author} {\bibfnamefont {R.}~\bibnamefont {Blatt}}, \ and\ \bibinfo {author}
  {\bibfnamefont {C.~F.}\ \bibnamefont {Roos}},\ }\href {\doibase
  10.1103/PhysRevX.8.031022} {\bibfield  {journal} {\bibinfo  {journal}
  {Physical Review X}\ }\textbf {\bibinfo {volume} {8}},\ \bibinfo {pages}
  {031022} (\bibinfo {year} {2018})}\BibitemShut {NoStop}%
\bibitem [{\citenamefont {Colless}\ \emph {et~al.}(2018)\citenamefont
  {Colless}, \citenamefont {Ramasesh}, \citenamefont {Dahlen}, \citenamefont
  {Blok}, \citenamefont {{Kimchi-Schwartz}}, \citenamefont {McClean},
  \citenamefont {Carter}, \citenamefont {{de Jong}},\ and\ \citenamefont
  {Siddiqi}}]{colless2018}%
  \BibitemOpen
  \bibfield  {author} {\bibinfo {author} {\bibfnamefont {J.~I.}\ \bibnamefont
  {Colless}}, \bibinfo {author} {\bibfnamefont {V.~V.}\ \bibnamefont
  {Ramasesh}}, \bibinfo {author} {\bibfnamefont {D.}~\bibnamefont {Dahlen}},
  \bibinfo {author} {\bibfnamefont {M.~S.}\ \bibnamefont {Blok}}, \bibinfo
  {author} {\bibfnamefont {M.~E.}\ \bibnamefont {{Kimchi-Schwartz}}}, \bibinfo
  {author} {\bibfnamefont {J.~R.}\ \bibnamefont {McClean}}, \bibinfo {author}
  {\bibfnamefont {J.}~\bibnamefont {Carter}}, \bibinfo {author} {\bibfnamefont
  {W.~A.}\ \bibnamefont {{de Jong}}}, \ and\ \bibinfo {author} {\bibfnamefont
  {I.}~\bibnamefont {Siddiqi}},\ }\href {\doibase 10.1103/PhysRevX.8.011021}
  {\bibfield  {journal} {\bibinfo  {journal} {Physical Review X}\ }\textbf
  {\bibinfo {volume} {8}},\ \bibinfo {pages} {011021} (\bibinfo {year}
  {2018})}\BibitemShut {NoStop}%
\bibitem [{\citenamefont {Kandala}\ \emph {et~al.}(2019)\citenamefont
  {Kandala}, \citenamefont {Temme}, \citenamefont {C{\'o}rcoles}, \citenamefont
  {Mezzacapo}, \citenamefont {Chow},\ and\ \citenamefont
  {Gambetta}}]{kandala2019}%
  \BibitemOpen
  \bibfield  {author} {\bibinfo {author} {\bibfnamefont {A.}~\bibnamefont
  {Kandala}}, \bibinfo {author} {\bibfnamefont {K.}~\bibnamefont {Temme}},
  \bibinfo {author} {\bibfnamefont {A.~D.}\ \bibnamefont {C{\'o}rcoles}},
  \bibinfo {author} {\bibfnamefont {A.}~\bibnamefont {Mezzacapo}}, \bibinfo
  {author} {\bibfnamefont {J.~M.}\ \bibnamefont {Chow}}, \ and\ \bibinfo
  {author} {\bibfnamefont {J.~M.}\ \bibnamefont {Gambetta}},\ }\href {\doibase
  10.1038/s41586-019-1040-7} {\bibfield  {journal} {\bibinfo  {journal}
  {Nature}\ }\textbf {\bibinfo {volume} {567}},\ \bibinfo {pages} {491}
  (\bibinfo {year} {2019})}\BibitemShut {NoStop}%
\bibitem [{\citenamefont {Ryabinkin}\ \emph {et~al.}(2018)\citenamefont
  {Ryabinkin}, \citenamefont {Yen}, \citenamefont {Genin},\ and\ \citenamefont
  {Izmaylov}}]{ryabinkin2018}%
  \BibitemOpen
  \bibfield  {author} {\bibinfo {author} {\bibfnamefont {I.~G.}\ \bibnamefont
  {Ryabinkin}}, \bibinfo {author} {\bibfnamefont {T.-C.}\ \bibnamefont {Yen}},
  \bibinfo {author} {\bibfnamefont {S.~N.}\ \bibnamefont {Genin}}, \ and\
  \bibinfo {author} {\bibfnamefont {A.~F.}\ \bibnamefont {Izmaylov}},\ }\href
  {\doibase 10.1021/acs.jctc.8b00932} {\bibfield  {journal} {\bibinfo
  {journal} {Journal of Chemical Theory and Computation}\ }\textbf {\bibinfo
  {volume} {14}},\ \bibinfo {pages} {6317} (\bibinfo {year}
  {2018})}\BibitemShut {NoStop}%
\bibitem [{\citenamefont {Li}\ \emph {et~al.}(2019)\citenamefont {Li},
  \citenamefont {Liu}, \citenamefont {Wang}, \citenamefont {Ashhab},
  \citenamefont {Cui}, \citenamefont {Chen}, \citenamefont {Peng},\ and\
  \citenamefont {Du}}]{li2019}%
  \BibitemOpen
  \bibfield  {author} {\bibinfo {author} {\bibfnamefont {Z.}~\bibnamefont
  {Li}}, \bibinfo {author} {\bibfnamefont {X.}~\bibnamefont {Liu}}, \bibinfo
  {author} {\bibfnamefont {H.}~\bibnamefont {Wang}}, \bibinfo {author}
  {\bibfnamefont {S.}~\bibnamefont {Ashhab}}, \bibinfo {author} {\bibfnamefont
  {J.}~\bibnamefont {Cui}}, \bibinfo {author} {\bibfnamefont {H.}~\bibnamefont
  {Chen}}, \bibinfo {author} {\bibfnamefont {X.}~\bibnamefont {Peng}}, \ and\
  \bibinfo {author} {\bibfnamefont {J.}~\bibnamefont {Du}},\ }\href {\doibase
  10.1103/PhysRevLett.122.090504} {\bibfield  {journal} {\bibinfo  {journal}
  {Physical Review Letters}\ }\textbf {\bibinfo {volume} {122}},\ \bibinfo
  {pages} {090504} (\bibinfo {year} {2019})}\BibitemShut {NoStop}%
\bibitem [{\citenamefont {Nam}\ \emph {et~al.}(2020)\citenamefont {Nam},
  \citenamefont {Chen}, \citenamefont {Pisenti}, \citenamefont {Wright},
  \citenamefont {Delaney}, \citenamefont {Maslov}, \citenamefont {Brown},
  \citenamefont {Allen}, \citenamefont {Amini}, \citenamefont {Apisdorf},
  \citenamefont {Beck}, \citenamefont {Blinov}, \citenamefont {Chaplin},
  \citenamefont {Chmielewski}, \citenamefont {Collins}, \citenamefont
  {Debnath}, \citenamefont {Hudek}, \citenamefont {Ducore}, \citenamefont
  {Keesan}, \citenamefont {Kreikemeier}, \citenamefont {Mizrahi}, \citenamefont
  {Solomon}, \citenamefont {Williams}, \citenamefont {{Wong-Campos}},
  \citenamefont {Moehring}, \citenamefont {Monroe},\ and\ \citenamefont
  {Kim}}]{nam2020}%
  \BibitemOpen
  \bibfield  {author} {\bibinfo {author} {\bibfnamefont {Y.}~\bibnamefont
  {Nam}}, \bibinfo {author} {\bibfnamefont {J.-S.}\ \bibnamefont {Chen}},
  \bibinfo {author} {\bibfnamefont {N.~C.}\ \bibnamefont {Pisenti}}, \bibinfo
  {author} {\bibfnamefont {K.}~\bibnamefont {Wright}}, \bibinfo {author}
  {\bibfnamefont {C.}~\bibnamefont {Delaney}}, \bibinfo {author} {\bibfnamefont
  {D.}~\bibnamefont {Maslov}}, \bibinfo {author} {\bibfnamefont {K.~R.}\
  \bibnamefont {Brown}}, \bibinfo {author} {\bibfnamefont {S.}~\bibnamefont
  {Allen}}, \bibinfo {author} {\bibfnamefont {J.~M.}\ \bibnamefont {Amini}},
  \bibinfo {author} {\bibfnamefont {J.}~\bibnamefont {Apisdorf}}, \bibinfo
  {author} {\bibfnamefont {K.~M.}\ \bibnamefont {Beck}}, \bibinfo {author}
  {\bibfnamefont {A.}~\bibnamefont {Blinov}}, \bibinfo {author} {\bibfnamefont
  {V.}~\bibnamefont {Chaplin}}, \bibinfo {author} {\bibfnamefont
  {M.}~\bibnamefont {Chmielewski}}, \bibinfo {author} {\bibfnamefont
  {C.}~\bibnamefont {Collins}}, \bibinfo {author} {\bibfnamefont
  {S.}~\bibnamefont {Debnath}}, \bibinfo {author} {\bibfnamefont {K.~M.}\
  \bibnamefont {Hudek}}, \bibinfo {author} {\bibfnamefont {A.~M.}\ \bibnamefont
  {Ducore}}, \bibinfo {author} {\bibfnamefont {M.}~\bibnamefont {Keesan}},
  \bibinfo {author} {\bibfnamefont {S.~M.}\ \bibnamefont {Kreikemeier}},
  \bibinfo {author} {\bibfnamefont {J.}~\bibnamefont {Mizrahi}}, \bibinfo
  {author} {\bibfnamefont {P.}~\bibnamefont {Solomon}}, \bibinfo {author}
  {\bibfnamefont {M.}~\bibnamefont {Williams}}, \bibinfo {author}
  {\bibfnamefont {J.~D.}\ \bibnamefont {{Wong-Campos}}}, \bibinfo {author}
  {\bibfnamefont {D.}~\bibnamefont {Moehring}}, \bibinfo {author}
  {\bibfnamefont {C.}~\bibnamefont {Monroe}}, \ and\ \bibinfo {author}
  {\bibfnamefont {J.}~\bibnamefont {Kim}},\ }\href {\doibase
  10.1038/s41534-020-0259-3} {\bibfield  {journal} {\bibinfo  {journal} {npj
  Quantum Information}\ }\textbf {\bibinfo {volume} {6}},\ \bibinfo {pages} {1}
  (\bibinfo {year} {2020})}\BibitemShut {NoStop}%
\bibitem [{\citenamefont {McCaskey}\ \emph {et~al.}(2019)\citenamefont
  {McCaskey}, \citenamefont {Parks}, \citenamefont {Jakowski}, \citenamefont
  {Moore}, \citenamefont {Morris}, \citenamefont {Humble},\ and\ \citenamefont
  {Pooser}}]{mccaskey2019}%
  \BibitemOpen
  \bibfield  {author} {\bibinfo {author} {\bibfnamefont {A.~J.}\ \bibnamefont
  {McCaskey}}, \bibinfo {author} {\bibfnamefont {Z.~P.}\ \bibnamefont {Parks}},
  \bibinfo {author} {\bibfnamefont {J.}~\bibnamefont {Jakowski}}, \bibinfo
  {author} {\bibfnamefont {S.~V.}\ \bibnamefont {Moore}}, \bibinfo {author}
  {\bibfnamefont {T.~D.}\ \bibnamefont {Morris}}, \bibinfo {author}
  {\bibfnamefont {T.~S.}\ \bibnamefont {Humble}}, \ and\ \bibinfo {author}
  {\bibfnamefont {R.~C.}\ \bibnamefont {Pooser}},\ }\href {\doibase
  10.1038/s41534-019-0209-0} {\bibfield  {journal} {\bibinfo  {journal} {npj
  Quantum Information}\ }\textbf {\bibinfo {volume} {5}},\ \bibinfo {pages} {1}
  (\bibinfo {year} {2019})}\BibitemShut {NoStop}%
\bibitem [{\citenamefont {Gao}\ \emph {et~al.}(2021)\citenamefont {Gao},
  \citenamefont {Nakamura}, \citenamefont {Gujarati}, \citenamefont {Jones},
  \citenamefont {Rice}, \citenamefont {Wood}, \citenamefont {Pistoia},
  \citenamefont {Garcia},\ and\ \citenamefont {Yamamoto}}]{gao2021}%
  \BibitemOpen
  \bibfield  {author} {\bibinfo {author} {\bibfnamefont {Q.}~\bibnamefont
  {Gao}}, \bibinfo {author} {\bibfnamefont {H.}~\bibnamefont {Nakamura}},
  \bibinfo {author} {\bibfnamefont {T.~P.}\ \bibnamefont {Gujarati}}, \bibinfo
  {author} {\bibfnamefont {G.~O.}\ \bibnamefont {Jones}}, \bibinfo {author}
  {\bibfnamefont {J.~E.}\ \bibnamefont {Rice}}, \bibinfo {author}
  {\bibfnamefont {S.~P.}\ \bibnamefont {Wood}}, \bibinfo {author}
  {\bibfnamefont {M.}~\bibnamefont {Pistoia}}, \bibinfo {author} {\bibfnamefont
  {J.~M.}\ \bibnamefont {Garcia}}, \ and\ \bibinfo {author} {\bibfnamefont
  {N.}~\bibnamefont {Yamamoto}},\ }\href {\doibase 10.1021/acs.jpca.0c09530}
  {\bibfield  {journal} {\bibinfo  {journal} {The Journal of Physical
  Chemistry. A}\ }\textbf {\bibinfo {volume} {125}},\ \bibinfo {pages} {1827}
  (\bibinfo {year} {2021})}\BibitemShut {NoStop}%
\bibitem [{\citenamefont {Smart}\ and\ \citenamefont
  {Mazziotti}(2019)}]{smart2019}%
  \BibitemOpen
  \bibfield  {author} {\bibinfo {author} {\bibfnamefont {S.~E.}\ \bibnamefont
  {Smart}}\ and\ \bibinfo {author} {\bibfnamefont {D.~A.}\ \bibnamefont
  {Mazziotti}},\ }\href {\doibase 10.1103/PhysRevA.100.022517} {\bibfield
  {journal} {\bibinfo  {journal} {Physical Review A}\ }\textbf {\bibinfo
  {volume} {100}},\ \bibinfo {pages} {022517} (\bibinfo {year}
  {2019})}\BibitemShut {NoStop}%
\bibitem [{\citenamefont {Sagastizabal}\ \emph {et~al.}(2019)\citenamefont
  {Sagastizabal}, \citenamefont {{Bonet-Monroig}}, \citenamefont {Singh},
  \citenamefont {Rol}, \citenamefont {Bultink}, \citenamefont {Fu},
  \citenamefont {Price}, \citenamefont {Ostroukh}, \citenamefont
  {Muthusubramanian}, \citenamefont {Bruno}, \citenamefont {Beekman},
  \citenamefont {Haider}, \citenamefont {O'Brien},\ and\ \citenamefont
  {DiCarlo}}]{sagastizabal2019}%
  \BibitemOpen
  \bibfield  {author} {\bibinfo {author} {\bibfnamefont {R.}~\bibnamefont
  {Sagastizabal}}, \bibinfo {author} {\bibfnamefont {X.}~\bibnamefont
  {{Bonet-Monroig}}}, \bibinfo {author} {\bibfnamefont {M.}~\bibnamefont
  {Singh}}, \bibinfo {author} {\bibfnamefont {M.~A.}\ \bibnamefont {Rol}},
  \bibinfo {author} {\bibfnamefont {C.~C.}\ \bibnamefont {Bultink}}, \bibinfo
  {author} {\bibfnamefont {X.}~\bibnamefont {Fu}}, \bibinfo {author}
  {\bibfnamefont {C.~H.}\ \bibnamefont {Price}}, \bibinfo {author}
  {\bibfnamefont {V.~P.}\ \bibnamefont {Ostroukh}}, \bibinfo {author}
  {\bibfnamefont {N.}~\bibnamefont {Muthusubramanian}}, \bibinfo {author}
  {\bibfnamefont {A.}~\bibnamefont {Bruno}}, \bibinfo {author} {\bibfnamefont
  {M.}~\bibnamefont {Beekman}}, \bibinfo {author} {\bibfnamefont
  {N.}~\bibnamefont {Haider}}, \bibinfo {author} {\bibfnamefont {T.~E.}\
  \bibnamefont {O'Brien}}, \ and\ \bibinfo {author} {\bibfnamefont
  {L.}~\bibnamefont {DiCarlo}},\ }\href {\doibase 10.1103/PhysRevA.100.010302}
  {\bibfield  {journal} {\bibinfo  {journal} {Physical Review A}\ }\textbf
  {\bibinfo {volume} {100}},\ \bibinfo {pages} {010302} (\bibinfo {year}
  {2019})}\BibitemShut {NoStop}%
\bibitem [{\citenamefont {Higgott}\ \emph {et~al.}(2019)\citenamefont
  {Higgott}, \citenamefont {Wang},\ and\ \citenamefont
  {Brierley}}]{higgott2019}%
  \BibitemOpen
  \bibfield  {author} {\bibinfo {author} {\bibfnamefont {O.}~\bibnamefont
  {Higgott}}, \bibinfo {author} {\bibfnamefont {D.}~\bibnamefont {Wang}}, \
  and\ \bibinfo {author} {\bibfnamefont {S.}~\bibnamefont {Brierley}},\ }\href
  {\doibase 10.22331/q-2019-07-01-156} {\bibfield  {journal} {\bibinfo
  {journal} {Quantum}\ }\textbf {\bibinfo {volume} {3}},\ \bibinfo {pages}
  {156} (\bibinfo {year} {2019})}\BibitemShut {NoStop}%
\bibitem [{\citenamefont {{{Google AI Quantum}}}\ \emph
  {et~al.}(2020)\citenamefont {{{Google AI Quantum}}}, \citenamefont {Arute},
  \citenamefont {Arya}, \citenamefont {Babbush}, \citenamefont {Bacon},
  \citenamefont {Bardin}, \citenamefont {Barends}, \citenamefont {Boixo},
  \citenamefont {Broughton}, \citenamefont {Buckley}, \citenamefont {Buell},
  \citenamefont {Burkett}, \citenamefont {Bushnell}, \citenamefont {Chen},
  \citenamefont {Chen}, \citenamefont {Chiaro}, \citenamefont {Collins},
  \citenamefont {Courtney}, \citenamefont {Demura}, \citenamefont {Dunsworth},
  \citenamefont {Farhi}, \citenamefont {Fowler}, \citenamefont {Foxen},
  \citenamefont {Gidney}, \citenamefont {Giustina}, \citenamefont {Graff},
  \citenamefont {Habegger}, \citenamefont {Harrigan}, \citenamefont {Ho},
  \citenamefont {Hong}, \citenamefont {Huang}, \citenamefont {Huggins},
  \citenamefont {Ioffe}, \citenamefont {Isakov}, \citenamefont {Jeffrey},
  \citenamefont {Jiang}, \citenamefont {Jones}, \citenamefont {Kafri},
  \citenamefont {Kechedzhi}, \citenamefont {Kelly}, \citenamefont {Kim},
  \citenamefont {Klimov}, \citenamefont {Korotkov}, \citenamefont {Kostritsa},
  \citenamefont {Landhuis}, \citenamefont {Laptev}, \citenamefont {Lindmark},
  \citenamefont {Lucero}, \citenamefont {Martin}, \citenamefont {Martinis},
  \citenamefont {McClean}, \citenamefont {McEwen}, \citenamefont {Megrant},
  \citenamefont {Mi}, \citenamefont {Mohseni}, \citenamefont {Mruczkiewicz},
  \citenamefont {Mutus}, \citenamefont {Naaman}, \citenamefont {Neeley},
  \citenamefont {Neill}, \citenamefont {Neven}, \citenamefont {Niu},
  \citenamefont {O'Brien}, \citenamefont {Ostby}, \citenamefont {Petukhov},
  \citenamefont {Putterman}, \citenamefont {Quintana}, \citenamefont {Roushan},
  \citenamefont {Rubin}, \citenamefont {Sank}, \citenamefont {Satzinger},
  \citenamefont {Smelyanskiy}, \citenamefont {Strain}, \citenamefont {Sung},
  \citenamefont {Szalay}, \citenamefont {Takeshita}, \citenamefont
  {Vainsencher}, \citenamefont {White}, \citenamefont {Wiebe}, \citenamefont
  {Yao}, \citenamefont {Yeh},\ and\ \citenamefont {Zalcman}}]{google2020}%
  \BibitemOpen
  \bibfield  {author} {\bibinfo {author} {\bibnamefont {{{Google AI
  Quantum}}}}, \bibinfo {author} {\bibfnamefont {F.}~\bibnamefont {Arute}},
  \bibinfo {author} {\bibfnamefont {K.}~\bibnamefont {Arya}}, \bibinfo {author}
  {\bibfnamefont {R.}~\bibnamefont {Babbush}}, \bibinfo {author} {\bibfnamefont
  {D.}~\bibnamefont {Bacon}}, \bibinfo {author} {\bibfnamefont {J.~C.}\
  \bibnamefont {Bardin}}, \bibinfo {author} {\bibfnamefont {R.}~\bibnamefont
  {Barends}}, \bibinfo {author} {\bibfnamefont {S.}~\bibnamefont {Boixo}},
  \bibinfo {author} {\bibfnamefont {M.}~\bibnamefont {Broughton}}, \bibinfo
  {author} {\bibfnamefont {B.~B.}\ \bibnamefont {Buckley}}, \bibinfo {author}
  {\bibfnamefont {D.~A.}\ \bibnamefont {Buell}}, \bibinfo {author}
  {\bibfnamefont {B.}~\bibnamefont {Burkett}}, \bibinfo {author} {\bibfnamefont
  {N.}~\bibnamefont {Bushnell}}, \bibinfo {author} {\bibfnamefont
  {Y.}~\bibnamefont {Chen}}, \bibinfo {author} {\bibfnamefont {Z.}~\bibnamefont
  {Chen}}, \bibinfo {author} {\bibfnamefont {B.}~\bibnamefont {Chiaro}},
  \bibinfo {author} {\bibfnamefont {R.}~\bibnamefont {Collins}}, \bibinfo
  {author} {\bibfnamefont {W.}~\bibnamefont {Courtney}}, \bibinfo {author}
  {\bibfnamefont {S.}~\bibnamefont {Demura}}, \bibinfo {author} {\bibfnamefont
  {A.}~\bibnamefont {Dunsworth}}, \bibinfo {author} {\bibfnamefont
  {E.}~\bibnamefont {Farhi}}, \bibinfo {author} {\bibfnamefont
  {A.}~\bibnamefont {Fowler}}, \bibinfo {author} {\bibfnamefont
  {B.}~\bibnamefont {Foxen}}, \bibinfo {author} {\bibfnamefont
  {C.}~\bibnamefont {Gidney}}, \bibinfo {author} {\bibfnamefont
  {M.}~\bibnamefont {Giustina}}, \bibinfo {author} {\bibfnamefont
  {R.}~\bibnamefont {Graff}}, \bibinfo {author} {\bibfnamefont
  {S.}~\bibnamefont {Habegger}}, \bibinfo {author} {\bibfnamefont {M.~P.}\
  \bibnamefont {Harrigan}}, \bibinfo {author} {\bibfnamefont {A.}~\bibnamefont
  {Ho}}, \bibinfo {author} {\bibfnamefont {S.}~\bibnamefont {Hong}}, \bibinfo
  {author} {\bibfnamefont {T.}~\bibnamefont {Huang}}, \bibinfo {author}
  {\bibfnamefont {W.~J.}\ \bibnamefont {Huggins}}, \bibinfo {author}
  {\bibfnamefont {L.}~\bibnamefont {Ioffe}}, \bibinfo {author} {\bibfnamefont
  {S.~V.}\ \bibnamefont {Isakov}}, \bibinfo {author} {\bibfnamefont
  {E.}~\bibnamefont {Jeffrey}}, \bibinfo {author} {\bibfnamefont
  {Z.}~\bibnamefont {Jiang}}, \bibinfo {author} {\bibfnamefont
  {C.}~\bibnamefont {Jones}}, \bibinfo {author} {\bibfnamefont
  {D.}~\bibnamefont {Kafri}}, \bibinfo {author} {\bibfnamefont
  {K.}~\bibnamefont {Kechedzhi}}, \bibinfo {author} {\bibfnamefont
  {J.}~\bibnamefont {Kelly}}, \bibinfo {author} {\bibfnamefont
  {S.}~\bibnamefont {Kim}}, \bibinfo {author} {\bibfnamefont {P.~V.}\
  \bibnamefont {Klimov}}, \bibinfo {author} {\bibfnamefont {A.}~\bibnamefont
  {Korotkov}}, \bibinfo {author} {\bibfnamefont {F.}~\bibnamefont {Kostritsa}},
  \bibinfo {author} {\bibfnamefont {D.}~\bibnamefont {Landhuis}}, \bibinfo
  {author} {\bibfnamefont {P.}~\bibnamefont {Laptev}}, \bibinfo {author}
  {\bibfnamefont {M.}~\bibnamefont {Lindmark}}, \bibinfo {author}
  {\bibfnamefont {E.}~\bibnamefont {Lucero}}, \bibinfo {author} {\bibfnamefont
  {O.}~\bibnamefont {Martin}}, \bibinfo {author} {\bibfnamefont {J.~M.}\
  \bibnamefont {Martinis}}, \bibinfo {author} {\bibfnamefont {J.~R.}\
  \bibnamefont {McClean}}, \bibinfo {author} {\bibfnamefont {M.}~\bibnamefont
  {McEwen}}, \bibinfo {author} {\bibfnamefont {A.}~\bibnamefont {Megrant}},
  \bibinfo {author} {\bibfnamefont {X.}~\bibnamefont {Mi}}, \bibinfo {author}
  {\bibfnamefont {M.}~\bibnamefont {Mohseni}}, \bibinfo {author} {\bibfnamefont
  {W.}~\bibnamefont {Mruczkiewicz}}, \bibinfo {author} {\bibfnamefont
  {J.}~\bibnamefont {Mutus}}, \bibinfo {author} {\bibfnamefont
  {O.}~\bibnamefont {Naaman}}, \bibinfo {author} {\bibfnamefont
  {M.}~\bibnamefont {Neeley}}, \bibinfo {author} {\bibfnamefont
  {C.}~\bibnamefont {Neill}}, \bibinfo {author} {\bibfnamefont
  {H.}~\bibnamefont {Neven}}, \bibinfo {author} {\bibfnamefont {M.~Y.}\
  \bibnamefont {Niu}}, \bibinfo {author} {\bibfnamefont {T.~E.}\ \bibnamefont
  {O'Brien}}, \bibinfo {author} {\bibfnamefont {E.}~\bibnamefont {Ostby}},
  \bibinfo {author} {\bibfnamefont {A.}~\bibnamefont {Petukhov}}, \bibinfo
  {author} {\bibfnamefont {H.}~\bibnamefont {Putterman}}, \bibinfo {author}
  {\bibfnamefont {C.}~\bibnamefont {Quintana}}, \bibinfo {author}
  {\bibfnamefont {P.}~\bibnamefont {Roushan}}, \bibinfo {author} {\bibfnamefont
  {N.~C.}\ \bibnamefont {Rubin}}, \bibinfo {author} {\bibfnamefont
  {D.}~\bibnamefont {Sank}}, \bibinfo {author} {\bibfnamefont {K.~J.}\
  \bibnamefont {Satzinger}}, \bibinfo {author} {\bibfnamefont {V.}~\bibnamefont
  {Smelyanskiy}}, \bibinfo {author} {\bibfnamefont {D.}~\bibnamefont {Strain}},
  \bibinfo {author} {\bibfnamefont {K.~J.}\ \bibnamefont {Sung}}, \bibinfo
  {author} {\bibfnamefont {M.}~\bibnamefont {Szalay}}, \bibinfo {author}
  {\bibfnamefont {T.~Y.}\ \bibnamefont {Takeshita}}, \bibinfo {author}
  {\bibfnamefont {A.}~\bibnamefont {Vainsencher}}, \bibinfo {author}
  {\bibfnamefont {T.}~\bibnamefont {White}}, \bibinfo {author} {\bibfnamefont
  {N.}~\bibnamefont {Wiebe}}, \bibinfo {author} {\bibfnamefont {Z.~J.}\
  \bibnamefont {Yao}}, \bibinfo {author} {\bibfnamefont {P.}~\bibnamefont
  {Yeh}}, \ and\ \bibinfo {author} {\bibfnamefont {A.}~\bibnamefont
  {Zalcman}},\ }\href {\doibase 10.1126/science.abb9811} {\bibfield  {journal}
  {\bibinfo  {journal} {Science}\ }\textbf {\bibinfo {volume} {369}},\ \bibinfo
  {pages} {1084} (\bibinfo {year} {2020})}\BibitemShut {NoStop}%
\bibitem [{\citenamefont {Metcalf}\ \emph {et~al.}(2020)\citenamefont
  {Metcalf}, \citenamefont {Bauman}, \citenamefont {Kowalski},\ and\
  \citenamefont {{de Jong}}}]{metcalf2020}%
  \BibitemOpen
  \bibfield  {author} {\bibinfo {author} {\bibfnamefont {M.}~\bibnamefont
  {Metcalf}}, \bibinfo {author} {\bibfnamefont {N.~P.}\ \bibnamefont {Bauman}},
  \bibinfo {author} {\bibfnamefont {K.}~\bibnamefont {Kowalski}}, \ and\
  \bibinfo {author} {\bibfnamefont {W.~A.}\ \bibnamefont {{de Jong}}},\ }\href
  {\doibase 10.1021/acs.jctc.0c00421} {\bibfield  {journal} {\bibinfo
  {journal} {Journal of Chemical Theory and Computation}\ }\textbf {\bibinfo
  {volume} {16}},\ \bibinfo {pages} {6165} (\bibinfo {year}
  {2020})}\BibitemShut {NoStop}%
\bibitem [{\citenamefont {Rossmannek}\ \emph {et~al.}(2021)\citenamefont
  {Rossmannek}, \citenamefont {Barkoutsos}, \citenamefont {Ollitrault},\ and\
  \citenamefont {Tavernelli}}]{rossmannek2021}%
  \BibitemOpen
  \bibfield  {author} {\bibinfo {author} {\bibfnamefont {M.}~\bibnamefont
  {Rossmannek}}, \bibinfo {author} {\bibfnamefont {P.~K.}\ \bibnamefont
  {Barkoutsos}}, \bibinfo {author} {\bibfnamefont {P.~J.}\ \bibnamefont
  {Ollitrault}}, \ and\ \bibinfo {author} {\bibfnamefont {I.}~\bibnamefont
  {Tavernelli}},\ }\href {\doibase 10.1063/5.0029536} {\bibfield  {journal}
  {\bibinfo  {journal} {The Journal of Chemical Physics}\ }\textbf {\bibinfo
  {volume} {154}},\ \bibinfo {pages} {114105} (\bibinfo {year}
  {2021})}\BibitemShut {NoStop}%
\bibitem [{\citenamefont {Kawashima}\ \emph {et~al.}(2021)\citenamefont
  {Kawashima}, \citenamefont {Coons}, \citenamefont {Nam}, \citenamefont
  {Lloyd}, \citenamefont {Matsuura}, \citenamefont {Garza}, \citenamefont
  {Johri}, \citenamefont {Huntington}, \citenamefont {Senicourt}, \citenamefont
  {Maksymov}, \citenamefont {Nguyen}, \citenamefont {Kim}, \citenamefont
  {Alidoust}, \citenamefont {Zaribafiyan},\ and\ \citenamefont
  {Yamazaki}}]{kawashima2021}%
  \BibitemOpen
  \bibfield  {author} {\bibinfo {author} {\bibfnamefont {Y.}~\bibnamefont
  {Kawashima}}, \bibinfo {author} {\bibfnamefont {M.~P.}\ \bibnamefont
  {Coons}}, \bibinfo {author} {\bibfnamefont {Y.}~\bibnamefont {Nam}}, \bibinfo
  {author} {\bibfnamefont {E.}~\bibnamefont {Lloyd}}, \bibinfo {author}
  {\bibfnamefont {S.}~\bibnamefont {Matsuura}}, \bibinfo {author}
  {\bibfnamefont {A.~J.}\ \bibnamefont {Garza}}, \bibinfo {author}
  {\bibfnamefont {S.}~\bibnamefont {Johri}}, \bibinfo {author} {\bibfnamefont
  {L.}~\bibnamefont {Huntington}}, \bibinfo {author} {\bibfnamefont
  {V.}~\bibnamefont {Senicourt}}, \bibinfo {author} {\bibfnamefont {A.~O.}\
  \bibnamefont {Maksymov}}, \bibinfo {author} {\bibfnamefont {J.~H.~V.}\
  \bibnamefont {Nguyen}}, \bibinfo {author} {\bibfnamefont {J.}~\bibnamefont
  {Kim}}, \bibinfo {author} {\bibfnamefont {N.}~\bibnamefont {Alidoust}},
  \bibinfo {author} {\bibfnamefont {A.}~\bibnamefont {Zaribafiyan}}, \ and\
  \bibinfo {author} {\bibfnamefont {T.}~\bibnamefont {Yamazaki}},\ }\href
  {http://arxiv.org/abs/2102.07045} {\bibfield  {journal} {\bibinfo  {journal}
  {arXiv:2102.07045 [quant-ph]}\ } (\bibinfo {year} {2021})}\BibitemShut
  {NoStop}%
\bibitem [{\citenamefont {Teplukhin}\ \emph {et~al.}(2021)\citenamefont
  {Teplukhin}, \citenamefont {Kendrick}, \citenamefont {Mniszewski},
  \citenamefont {Zhang}, \citenamefont {Kumar}, \citenamefont {Negre},
  \citenamefont {Anisimov}, \citenamefont {Tretiak},\ and\ \citenamefont
  {Dub}}]{teplukhin2021}%
  \BibitemOpen
  \bibfield  {author} {\bibinfo {author} {\bibfnamefont {A.}~\bibnamefont
  {Teplukhin}}, \bibinfo {author} {\bibfnamefont {B.~K.}\ \bibnamefont
  {Kendrick}}, \bibinfo {author} {\bibfnamefont {S.~M.}\ \bibnamefont
  {Mniszewski}}, \bibinfo {author} {\bibfnamefont {Y.}~\bibnamefont {Zhang}},
  \bibinfo {author} {\bibfnamefont {A.}~\bibnamefont {Kumar}}, \bibinfo
  {author} {\bibfnamefont {C.~F.~A.}\ \bibnamefont {Negre}}, \bibinfo {author}
  {\bibfnamefont {P.~M.}\ \bibnamefont {Anisimov}}, \bibinfo {author}
  {\bibfnamefont {S.}~\bibnamefont {Tretiak}}, \ and\ \bibinfo {author}
  {\bibfnamefont {P.~A.}\ \bibnamefont {Dub}},\ }\href {\doibase
  10.1038/s41598-021-98331-y} {\bibfield  {journal} {\bibinfo  {journal}
  {Scientific Reports}\ }\textbf {\bibinfo {volume} {11}},\ \bibinfo {pages}
  {18796} (\bibinfo {year} {2021})}\BibitemShut {NoStop}%
\bibitem [{\citenamefont {Kirsopp}\ \emph {et~al.}(2021)\citenamefont
  {Kirsopp}, \citenamefont {Di~Paola}, \citenamefont {Manrique}, \citenamefont
  {Krompiec}, \citenamefont {{Greene-Diniz}}, \citenamefont {Guba},
  \citenamefont {Meyder}, \citenamefont {Wolf}, \citenamefont {Strahm},\ and\
  \citenamefont {Ramo}}]{kirsopp2021}%
  \BibitemOpen
  \bibfield  {author} {\bibinfo {author} {\bibfnamefont {J.~J.~M.}\
  \bibnamefont {Kirsopp}}, \bibinfo {author} {\bibfnamefont {C.}~\bibnamefont
  {Di~Paola}}, \bibinfo {author} {\bibfnamefont {D.~Z.}\ \bibnamefont
  {Manrique}}, \bibinfo {author} {\bibfnamefont {M.}~\bibnamefont {Krompiec}},
  \bibinfo {author} {\bibfnamefont {G.}~\bibnamefont {{Greene-Diniz}}},
  \bibinfo {author} {\bibfnamefont {W.}~\bibnamefont {Guba}}, \bibinfo {author}
  {\bibfnamefont {A.}~\bibnamefont {Meyder}}, \bibinfo {author} {\bibfnamefont
  {D.}~\bibnamefont {Wolf}}, \bibinfo {author} {\bibfnamefont {M.}~\bibnamefont
  {Strahm}}, \ and\ \bibinfo {author} {\bibfnamefont {D.~M.}\ \bibnamefont
  {Ramo}},\ }\href@noop {} {\bibfield  {journal} {\bibinfo  {journal}
  {arXiv:2110.08163 [physics, physics:quant-ph]}\ } (\bibinfo {year} {2021})},\
  \Eprint {http://arxiv.org/abs/2110.08163} {arXiv:2110.08163 [physics,
  physics:quant-ph]} \BibitemShut {NoStop}%
\bibitem [{\citenamefont {Jones}\ \emph {et~al.}(2021)\citenamefont {Jones},
  \citenamefont {Vallury}, \citenamefont {Hill},\ and\ \citenamefont
  {Hollenberg}}]{jones2021}%
  \BibitemOpen
  \bibfield  {author} {\bibinfo {author} {\bibfnamefont {M.~A.}\ \bibnamefont
  {Jones}}, \bibinfo {author} {\bibfnamefont {H.~J.}\ \bibnamefont {Vallury}},
  \bibinfo {author} {\bibfnamefont {C.~D.}\ \bibnamefont {Hill}}, \ and\
  \bibinfo {author} {\bibfnamefont {L.~C.~L.}\ \bibnamefont {Hollenberg}},\
  }\href@noop {} {\bibfield  {journal} {\bibinfo  {journal} {arXiv:2111.08132
  [physics, physics:quant-ph]}\ } (\bibinfo {year} {2021})},\ \Eprint
  {http://arxiv.org/abs/2111.08132} {arXiv:2111.08132 [physics,
  physics:quant-ph]} \BibitemShut {NoStop}%
\bibitem [{\citenamefont {Kivlichan}\ \emph {et~al.}(2020)\citenamefont
  {Kivlichan}, \citenamefont {Gidney}, \citenamefont {Berry}, \citenamefont
  {Wiebe}, \citenamefont {McClean}, \citenamefont {Sun}, \citenamefont {Jiang},
  \citenamefont {Rubin}, \citenamefont {Fowler}, \citenamefont
  {{Aspuru-Guzik}}, \citenamefont {Neven},\ and\ \citenamefont
  {Babbush}}]{kivlichan2020}%
  \BibitemOpen
  \bibfield  {author} {\bibinfo {author} {\bibfnamefont {I.~D.}\ \bibnamefont
  {Kivlichan}}, \bibinfo {author} {\bibfnamefont {C.}~\bibnamefont {Gidney}},
  \bibinfo {author} {\bibfnamefont {D.~W.}\ \bibnamefont {Berry}}, \bibinfo
  {author} {\bibfnamefont {N.}~\bibnamefont {Wiebe}}, \bibinfo {author}
  {\bibfnamefont {J.}~\bibnamefont {McClean}}, \bibinfo {author} {\bibfnamefont
  {W.}~\bibnamefont {Sun}}, \bibinfo {author} {\bibfnamefont {Z.}~\bibnamefont
  {Jiang}}, \bibinfo {author} {\bibfnamefont {N.}~\bibnamefont {Rubin}},
  \bibinfo {author} {\bibfnamefont {A.}~\bibnamefont {Fowler}}, \bibinfo
  {author} {\bibfnamefont {A.}~\bibnamefont {{Aspuru-Guzik}}}, \bibinfo
  {author} {\bibfnamefont {H.}~\bibnamefont {Neven}}, \ and\ \bibinfo {author}
  {\bibfnamefont {R.}~\bibnamefont {Babbush}},\ }\href {\doibase
  10.22331/q-2020-07-16-296} {\bibfield  {journal} {\bibinfo  {journal}
  {Quantum}\ }\textbf {\bibinfo {volume} {4}},\ \bibinfo {pages} {296}
  (\bibinfo {year} {2020})}\BibitemShut {NoStop}%
\bibitem [{\citenamefont {Cruz}\ \emph {et~al.}(2020)\citenamefont {Cruz},
  \citenamefont {Catarina}, \citenamefont {Gautier},\ and\ \citenamefont
  {{Fern{\'a}ndez-Rossier}}}]{cruz2020}%
  \BibitemOpen
  \bibfield  {author} {\bibinfo {author} {\bibfnamefont {P.~M.~Q.}\
  \bibnamefont {Cruz}}, \bibinfo {author} {\bibfnamefont {G.}~\bibnamefont
  {Catarina}}, \bibinfo {author} {\bibfnamefont {R.}~\bibnamefont {Gautier}}, \
  and\ \bibinfo {author} {\bibfnamefont {J.}~\bibnamefont
  {{Fern{\'a}ndez-Rossier}}},\ }\href {\doibase 10.1088/2058-9565/abaa2c}
  {\bibfield  {journal} {\bibinfo  {journal} {Quantum Science and Technology}\
  }\textbf {\bibinfo {volume} {5}},\ \bibinfo {pages} {044005} (\bibinfo {year}
  {2020})}\BibitemShut {NoStop}%
\bibitem [{\citenamefont {Montanaro}\ and\ \citenamefont
  {Stanisic}(2020)}]{montanaro2020}%
  \BibitemOpen
  \bibfield  {author} {\bibinfo {author} {\bibfnamefont {A.}~\bibnamefont
  {Montanaro}}\ and\ \bibinfo {author} {\bibfnamefont {S.}~\bibnamefont
  {Stanisic}},\ }\href {http://arxiv.org/abs/2006.01179} {\bibfield  {journal}
  {\bibinfo  {journal} {arXiv:2006.01179 [quant-ph]}\ } (\bibinfo {year}
  {2020})}\BibitemShut {NoStop}%
\bibitem [{\citenamefont {Uvarov}\ \emph {et~al.}(2020)\citenamefont {Uvarov},
  \citenamefont {Biamonte},\ and\ \citenamefont {Yudin}}]{uvarov2020}%
  \BibitemOpen
  \bibfield  {author} {\bibinfo {author} {\bibfnamefont {A.}~\bibnamefont
  {Uvarov}}, \bibinfo {author} {\bibfnamefont {J.~D.}\ \bibnamefont
  {Biamonte}}, \ and\ \bibinfo {author} {\bibfnamefont {D.}~\bibnamefont
  {Yudin}},\ }\href {\doibase 10.1103/PhysRevB.102.075104} {\bibfield
  {journal} {\bibinfo  {journal} {Physical Review B}\ }\textbf {\bibinfo
  {volume} {102}},\ \bibinfo {pages} {075104} (\bibinfo {year}
  {2020})}\BibitemShut {NoStop}%
\bibitem [{\citenamefont {Motta}\ \emph {et~al.}(2020)\citenamefont {Motta},
  \citenamefont {Sun}, \citenamefont {Tan}, \citenamefont {O'Rourke},
  \citenamefont {Ye}, \citenamefont {Minnich}, \citenamefont {Brand{\~a}o},\
  and\ \citenamefont {Chan}}]{motta2020}%
  \BibitemOpen
  \bibfield  {author} {\bibinfo {author} {\bibfnamefont {M.}~\bibnamefont
  {Motta}}, \bibinfo {author} {\bibfnamefont {C.}~\bibnamefont {Sun}}, \bibinfo
  {author} {\bibfnamefont {A.~T.~K.}\ \bibnamefont {Tan}}, \bibinfo {author}
  {\bibfnamefont {M.~J.}\ \bibnamefont {O'Rourke}}, \bibinfo {author}
  {\bibfnamefont {E.}~\bibnamefont {Ye}}, \bibinfo {author} {\bibfnamefont
  {A.~J.}\ \bibnamefont {Minnich}}, \bibinfo {author} {\bibfnamefont {F.~G.
  S.~L.}\ \bibnamefont {Brand{\~a}o}}, \ and\ \bibinfo {author} {\bibfnamefont
  {G.~K.-L.}\ \bibnamefont {Chan}},\ }\href {\doibase
  10.1038/s41567-019-0704-4} {\bibfield  {journal} {\bibinfo  {journal} {Nature
  Physics}\ }\textbf {\bibinfo {volume} {16}},\ \bibinfo {pages} {205}
  (\bibinfo {year} {2020})}\BibitemShut {NoStop}%
\bibitem [{\citenamefont {Mei}\ \emph {et~al.}(2020)\citenamefont {Mei},
  \citenamefont {Guo}, \citenamefont {Yu}, \citenamefont {Xiao}, \citenamefont
  {Zhu},\ and\ \citenamefont {Jia}}]{mei2020}%
  \BibitemOpen
  \bibfield  {author} {\bibinfo {author} {\bibfnamefont {F.}~\bibnamefont
  {Mei}}, \bibinfo {author} {\bibfnamefont {Q.}~\bibnamefont {Guo}}, \bibinfo
  {author} {\bibfnamefont {Y.-F.}\ \bibnamefont {Yu}}, \bibinfo {author}
  {\bibfnamefont {L.}~\bibnamefont {Xiao}}, \bibinfo {author} {\bibfnamefont
  {S.-L.}\ \bibnamefont {Zhu}}, \ and\ \bibinfo {author} {\bibfnamefont
  {S.}~\bibnamefont {Jia}},\ }\href {\doibase 10.1103/PhysRevLett.125.160503}
  {\bibfield  {journal} {\bibinfo  {journal} {Physical Review Letters}\
  }\textbf {\bibinfo {volume} {125}},\ \bibinfo {pages} {160503} (\bibinfo
  {year} {2020})}\BibitemShut {NoStop}%
\bibitem [{\citenamefont {Mizuta}\ \emph {et~al.}(2021)\citenamefont {Mizuta},
  \citenamefont {Fujii}, \citenamefont {Fujii}, \citenamefont {Ichikawa},
  \citenamefont {Imamura}, \citenamefont {Okuno},\ and\ \citenamefont
  {Nakagawa}}]{mizuta2021}%
  \BibitemOpen
  \bibfield  {author} {\bibinfo {author} {\bibfnamefont {K.}~\bibnamefont
  {Mizuta}}, \bibinfo {author} {\bibfnamefont {M.}~\bibnamefont {Fujii}},
  \bibinfo {author} {\bibfnamefont {S.}~\bibnamefont {Fujii}}, \bibinfo
  {author} {\bibfnamefont {K.}~\bibnamefont {Ichikawa}}, \bibinfo {author}
  {\bibfnamefont {Y.}~\bibnamefont {Imamura}}, \bibinfo {author} {\bibfnamefont
  {Y.}~\bibnamefont {Okuno}}, \ and\ \bibinfo {author} {\bibfnamefont {Y.~O.}\
  \bibnamefont {Nakagawa}},\ }\href {http://arxiv.org/abs/2104.00855}
  {\bibfield  {journal} {\bibinfo  {journal} {arXiv:2104.00855 [cond-mat,
  physics:quant-ph]}\ } (\bibinfo {year} {2021})}\BibitemShut {NoStop}%
\bibitem [{\citenamefont {Liu}\ \emph {et~al.}(2020)\citenamefont {Liu},
  \citenamefont {Wan}, \citenamefont {Li},\ and\ \citenamefont
  {Yang}}]{liu2020}%
  \BibitemOpen
  \bibfield  {author} {\bibinfo {author} {\bibfnamefont {J.}~\bibnamefont
  {Liu}}, \bibinfo {author} {\bibfnamefont {L.}~\bibnamefont {Wan}}, \bibinfo
  {author} {\bibfnamefont {Z.}~\bibnamefont {Li}}, \ and\ \bibinfo {author}
  {\bibfnamefont {J.}~\bibnamefont {Yang}},\ }\href {\doibase
  10.1021/acs.jctc.0c00881} {\bibfield  {journal} {\bibinfo  {journal} {Journal
  of Chemical Theory and Computation}\ }\textbf {\bibinfo {volume} {16}},\
  \bibinfo {pages} {6904} (\bibinfo {year} {2020})}\BibitemShut {NoStop}%
\bibitem [{\citenamefont {Kaicher}\ \emph {et~al.}(2020)\citenamefont
  {Kaicher}, \citenamefont {J{\"a}ger}, \citenamefont {{Dallaire-Demers}},\
  and\ \citenamefont {Wilhelm}}]{kaicher2020}%
  \BibitemOpen
  \bibfield  {author} {\bibinfo {author} {\bibfnamefont {M.~P.}\ \bibnamefont
  {Kaicher}}, \bibinfo {author} {\bibfnamefont {S.~B.}\ \bibnamefont
  {J{\"a}ger}}, \bibinfo {author} {\bibfnamefont {P.-L.}\ \bibnamefont
  {{Dallaire-Demers}}}, \ and\ \bibinfo {author} {\bibfnamefont {F.~K.}\
  \bibnamefont {Wilhelm}},\ }\href {\doibase 10.1103/PhysRevA.102.022607}
  {\bibfield  {journal} {\bibinfo  {journal} {Physical Review A}\ }\textbf
  {\bibinfo {volume} {102}},\ \bibinfo {pages} {022607} (\bibinfo {year}
  {2020})}\BibitemShut {NoStop}%
\bibitem [{\citenamefont {Rahmani}\ \emph {et~al.}(2020)\citenamefont
  {Rahmani}, \citenamefont {Sung}, \citenamefont {Putterman}, \citenamefont
  {Roushan}, \citenamefont {Ghaemi},\ and\ \citenamefont
  {Jiang}}]{rahmani2020}%
  \BibitemOpen
  \bibfield  {author} {\bibinfo {author} {\bibfnamefont {A.}~\bibnamefont
  {Rahmani}}, \bibinfo {author} {\bibfnamefont {K.~J.}\ \bibnamefont {Sung}},
  \bibinfo {author} {\bibfnamefont {H.}~\bibnamefont {Putterman}}, \bibinfo
  {author} {\bibfnamefont {P.}~\bibnamefont {Roushan}}, \bibinfo {author}
  {\bibfnamefont {P.}~\bibnamefont {Ghaemi}}, \ and\ \bibinfo {author}
  {\bibfnamefont {Z.}~\bibnamefont {Jiang}},\ }\href {\doibase
  10.1103/PRXQuantum.1.020309} {\bibfield  {journal} {\bibinfo  {journal} {PRX
  Quantum}\ }\textbf {\bibinfo {volume} {1}},\ \bibinfo {pages} {020309}
  (\bibinfo {year} {2020})}\BibitemShut {NoStop}%
\bibitem [{\citenamefont {Kreula}\ \emph {et~al.}(2016)\citenamefont {Kreula},
  \citenamefont {{Garc{\'i}a-{\'A}lvarez}}, \citenamefont {Lamata},
  \citenamefont {Clark}, \citenamefont {Solano},\ and\ \citenamefont
  {Jaksch}}]{kreula2016}%
  \BibitemOpen
  \bibfield  {author} {\bibinfo {author} {\bibfnamefont {J.~M.}\ \bibnamefont
  {Kreula}}, \bibinfo {author} {\bibfnamefont {L.}~\bibnamefont
  {{Garc{\'i}a-{\'A}lvarez}}}, \bibinfo {author} {\bibfnamefont
  {L.}~\bibnamefont {Lamata}}, \bibinfo {author} {\bibfnamefont {S.~R.}\
  \bibnamefont {Clark}}, \bibinfo {author} {\bibfnamefont {E.}~\bibnamefont
  {Solano}}, \ and\ \bibinfo {author} {\bibfnamefont {D.}~\bibnamefont
  {Jaksch}},\ }\href {\doibase 10.1140/epjqt/s40507-016-0049-1} {\bibfield
  {journal} {\bibinfo  {journal} {EPJ Quantum Technology}\ }\textbf {\bibinfo
  {volume} {3}},\ \bibinfo {pages} {1} (\bibinfo {year} {2016})}\BibitemShut
  {NoStop}%
\bibitem [{\citenamefont {Kreula}()}]{kreula_non-linear_nodate}%
  \BibitemOpen
  \bibfield  {author} {\bibinfo {author} {\bibfnamefont {J.~M.}\ \bibnamefont
  {Kreula}},\ }\href@noop {} {\ ,\ \bibinfo {pages} {7}}\BibitemShut {NoStop}%
\bibitem [{\citenamefont {Jaderberg}\ \emph {et~al.}(2020)\citenamefont
  {Jaderberg}, \citenamefont {Agarwal}, \citenamefont {Leonhardt},
  \citenamefont {Kiffner},\ and\ \citenamefont {Jaksch}}]{jaderberg2020}%
  \BibitemOpen
  \bibfield  {author} {\bibinfo {author} {\bibfnamefont {B.}~\bibnamefont
  {Jaderberg}}, \bibinfo {author} {\bibfnamefont {A.}~\bibnamefont {Agarwal}},
  \bibinfo {author} {\bibfnamefont {K.}~\bibnamefont {Leonhardt}}, \bibinfo
  {author} {\bibfnamefont {M.}~\bibnamefont {Kiffner}}, \ and\ \bibinfo
  {author} {\bibfnamefont {D.}~\bibnamefont {Jaksch}},\ }\href {\doibase
  10.1088/2058-9565/ab972b} {\bibfield  {journal} {\bibinfo  {journal} {Quantum
  Science and Technology}\ }\textbf {\bibinfo {volume} {5}},\ \bibinfo {pages}
  {034015} (\bibinfo {year} {2020})}\BibitemShut {NoStop}%
\bibitem [{\citenamefont {Lupo}\ \emph {et~al.}()\citenamefont {Lupo},
  \citenamefont {Jamet}, \citenamefont {Tse}, \citenamefont {Rungger},\ and\
  \citenamefont {Weber}}]{lupo_maximally_2021}%
  \BibitemOpen
  \bibfield  {author} {\bibinfo {author} {\bibfnamefont {C.}~\bibnamefont
  {Lupo}}, \bibinfo {author} {\bibfnamefont {F.}~\bibnamefont {Jamet}},
  \bibinfo {author} {\bibfnamefont {W.~H.~T.}\ \bibnamefont {Tse}}, \bibinfo
  {author} {\bibfnamefont {I.}~\bibnamefont {Rungger}}, \ and\ \bibinfo
  {author} {\bibfnamefont {C.}~\bibnamefont {Weber}},\ }\href {\doibase
  10.1038/s43588-021-00090-3} {\ \textbf {\bibinfo {volume} {1}},\ \bibinfo
  {pages} {410}},\ \bibinfo {note} {bandiera\_abtest: a Cg\_type: Nature
  Research Journals Number: 6 Primary\_atype: Research Publisher: Nature
  Publishing Group Subject\_term: Computational methods;Electronic properties
  and materials;Quantum simulation Subject\_term\_id:
  computational-methods;electronic-properties-and-materials;quantum-simulation}\BibitemShut
  {NoStop}%
\bibitem [{\citenamefont {Bauer}\ \emph {et~al.}(2016)\citenamefont {Bauer},
  \citenamefont {Wecker}, \citenamefont {Millis}, \citenamefont {Hastings},\
  and\ \citenamefont {Troyer}}]{bauer2016}%
  \BibitemOpen
  \bibfield  {author} {\bibinfo {author} {\bibfnamefont {B.}~\bibnamefont
  {Bauer}}, \bibinfo {author} {\bibfnamefont {D.}~\bibnamefont {Wecker}},
  \bibinfo {author} {\bibfnamefont {A.~J.}\ \bibnamefont {Millis}}, \bibinfo
  {author} {\bibfnamefont {M.~B.}\ \bibnamefont {Hastings}}, \ and\ \bibinfo
  {author} {\bibfnamefont {M.}~\bibnamefont {Troyer}},\ }\href {\doibase
  10.1103/PhysRevX.6.031045} {\bibfield  {journal} {\bibinfo  {journal}
  {Physical Review X}\ }\textbf {\bibinfo {volume} {6}},\ \bibinfo {pages}
  {031045} (\bibinfo {year} {2016})}\BibitemShut {NoStop}%
\bibitem [{\citenamefont {Rubin}(2016)}]{rubin2016}%
  \BibitemOpen
  \bibfield  {author} {\bibinfo {author} {\bibfnamefont {N.~C.}\ \bibnamefont
  {Rubin}},\ }\href {http://arxiv.org/abs/1610.06910} {\bibfield  {journal}
  {\bibinfo  {journal} {arXiv:1610.06910 [cond-mat, physics:quant-ph]}\ }
  (\bibinfo {year} {2016})}\BibitemShut {NoStop}%
\bibitem [{\citenamefont {Mineh}\ and\ \citenamefont
  {Montanaro}(2021)}]{mineh2021}%
  \BibitemOpen
  \bibfield  {author} {\bibinfo {author} {\bibfnamefont {L.}~\bibnamefont
  {Mineh}}\ and\ \bibinfo {author} {\bibfnamefont {A.}~\bibnamefont
  {Montanaro}},\ }\href@noop {} {\bibfield  {journal} {\bibinfo  {journal}
  {arXiv:2108.08611 [quant-ph]}\ } (\bibinfo {year} {2021})},\ \Eprint
  {http://arxiv.org/abs/2108.08611} {arXiv:2108.08611 [quant-ph]} \BibitemShut
  {NoStop}%
\bibitem [{\citenamefont {Li}\ \emph {et~al.}()\citenamefont {Li},
  \citenamefont {Huang}, \citenamefont {Cao}, \citenamefont {Huang},
  \citenamefont {Shuai}, \citenamefont {Sun}, \citenamefont {Sun},
  \citenamefont {Yuan},\ and\ \citenamefont {Lv}}]{li_toward_2021}%
  \BibitemOpen
  \bibfield  {author} {\bibinfo {author} {\bibfnamefont {W.}~\bibnamefont
  {Li}}, \bibinfo {author} {\bibfnamefont {Z.}~\bibnamefont {Huang}}, \bibinfo
  {author} {\bibfnamefont {C.}~\bibnamefont {Cao}}, \bibinfo {author}
  {\bibfnamefont {Y.}~\bibnamefont {Huang}}, \bibinfo {author} {\bibfnamefont
  {Z.}~\bibnamefont {Shuai}}, \bibinfo {author} {\bibfnamefont
  {X.}~\bibnamefont {Sun}}, \bibinfo {author} {\bibfnamefont {J.}~\bibnamefont
  {Sun}}, \bibinfo {author} {\bibfnamefont {X.}~\bibnamefont {Yuan}}, \ and\
  \bibinfo {author} {\bibfnamefont {D.}~\bibnamefont {Lv}},\ }\href
  {http://arxiv.org/abs/2109.08062} {\ }\Eprint
  {http://arxiv.org/abs/2109.08062} {2109.08062} \BibitemShut {NoStop}%
\bibitem [{\citenamefont {Georges}\ and\ \citenamefont
  {Kotliar}(1992)}]{georges1992}%
  \BibitemOpen
  \bibfield  {author} {\bibinfo {author} {\bibfnamefont {A.}~\bibnamefont
  {Georges}}\ and\ \bibinfo {author} {\bibfnamefont {G.}~\bibnamefont
  {Kotliar}},\ }\href {\doibase 10.1103/PhysRevB.45.6479} {\bibfield  {journal}
  {\bibinfo  {journal} {Physical Review B}\ }\textbf {\bibinfo {volume} {45}},\
  \bibinfo {pages} {6479} (\bibinfo {year} {1992})}\BibitemShut {NoStop}%
\bibitem [{\citenamefont {Georges}\ \emph {et~al.}(1996)\citenamefont
  {Georges}, \citenamefont {Kotliar}, \citenamefont {Krauth},\ and\
  \citenamefont {Rozenberg}}]{georges1996}%
  \BibitemOpen
  \bibfield  {author} {\bibinfo {author} {\bibfnamefont {A.}~\bibnamefont
  {Georges}}, \bibinfo {author} {\bibfnamefont {G.}~\bibnamefont {Kotliar}},
  \bibinfo {author} {\bibfnamefont {W.}~\bibnamefont {Krauth}}, \ and\ \bibinfo
  {author} {\bibfnamefont {M.~J.}\ \bibnamefont {Rozenberg}},\ }\href {\doibase
  10.1103/RevModPhys.68.13} {\bibfield  {journal} {\bibinfo  {journal} {Reviews
  of Modern Physics}\ }\textbf {\bibinfo {volume} {68}},\ \bibinfo {pages} {13}
  (\bibinfo {year} {1996})}\BibitemShut {NoStop}%
\bibitem [{\citenamefont {Georges}(2004)}]{georges2004}%
  \BibitemOpen
  \bibfield  {author} {\bibinfo {author} {\bibfnamefont {A.}~\bibnamefont
  {Georges}},\ }\href {\doibase 10.1063/1.1800733} {\bibfield  {journal}
  {\bibinfo  {journal} {AIP Conference Proceedings}\ }\textbf {\bibinfo
  {volume} {715}},\ \bibinfo {pages} {3} (\bibinfo {year} {2004})}\BibitemShut
  {NoStop}%
\bibitem [{\citenamefont {Anisimov}\ \emph {et~al.}(1997)\citenamefont
  {Anisimov}, \citenamefont {Poteryaev}, \citenamefont {Korotin}, \citenamefont
  {Anokhin},\ and\ \citenamefont {Kotliar}}]{anisimov1997}%
  \BibitemOpen
  \bibfield  {author} {\bibinfo {author} {\bibfnamefont {V.~I.}\ \bibnamefont
  {Anisimov}}, \bibinfo {author} {\bibfnamefont {A.~I.}\ \bibnamefont
  {Poteryaev}}, \bibinfo {author} {\bibfnamefont {M.~A.}\ \bibnamefont
  {Korotin}}, \bibinfo {author} {\bibfnamefont {A.~O.}\ \bibnamefont
  {Anokhin}}, \ and\ \bibinfo {author} {\bibfnamefont {G.}~\bibnamefont
  {Kotliar}},\ }\href {\doibase 10.1088/0953-8984/9/35/010} {\bibfield
  {journal} {\bibinfo  {journal} {Journal of Physics: Condensed Matter}\
  }\textbf {\bibinfo {volume} {9}},\ \bibinfo {pages} {7359} (\bibinfo {year}
  {1997})}\BibitemShut {NoStop}%
\bibitem [{\citenamefont {Kotliar}\ \emph {et~al.}(2006)\citenamefont
  {Kotliar}, \citenamefont {Savrasov}, \citenamefont {Haule}, \citenamefont
  {Oudovenko}, \citenamefont {Parcollet},\ and\ \citenamefont
  {Marianetti}}]{kotliar2006}%
  \BibitemOpen
  \bibfield  {author} {\bibinfo {author} {\bibfnamefont {G.}~\bibnamefont
  {Kotliar}}, \bibinfo {author} {\bibfnamefont {S.~Y.}\ \bibnamefont
  {Savrasov}}, \bibinfo {author} {\bibfnamefont {K.}~\bibnamefont {Haule}},
  \bibinfo {author} {\bibfnamefont {V.~S.}\ \bibnamefont {Oudovenko}}, \bibinfo
  {author} {\bibfnamefont {O.}~\bibnamefont {Parcollet}}, \ and\ \bibinfo
  {author} {\bibfnamefont {C.~A.}\ \bibnamefont {Marianetti}},\ }\href
  {\doibase 10.1103/RevModPhys.78.865} {\bibfield  {journal} {\bibinfo
  {journal} {Reviews of Modern Physics}\ }\textbf {\bibinfo {volume} {78}},\
  \bibinfo {pages} {865} (\bibinfo {year} {2006})}\BibitemShut {NoStop}%
\bibitem [{\citenamefont {Wouters}\ \emph {et~al.}(2016)\citenamefont
  {Wouters}, \citenamefont {{Jim{\'e}nez-Hoyos}}, \citenamefont {Sun},\ and\
  \citenamefont {Chan}}]{wouters2016}%
  \BibitemOpen
  \bibfield  {author} {\bibinfo {author} {\bibfnamefont {S.}~\bibnamefont
  {Wouters}}, \bibinfo {author} {\bibfnamefont {C.~A.}\ \bibnamefont
  {{Jim{\'e}nez-Hoyos}}}, \bibinfo {author} {\bibfnamefont {Q.}~\bibnamefont
  {Sun}}, \ and\ \bibinfo {author} {\bibfnamefont {G.~K.-L.}\ \bibnamefont
  {Chan}},\ }\href {\doibase 10.1021/acs.jctc.6b00316} {\bibfield  {journal}
  {\bibinfo  {journal} {Journal of Chemical Theory and Computation}\ }\textbf
  {\bibinfo {volume} {12}},\ \bibinfo {pages} {2706} (\bibinfo {year}
  {2016})}\BibitemShut {NoStop}%
\bibitem [{\citenamefont {Knizia}\ and\ \citenamefont
  {Chan}(2012)}]{knizia2012}%
  \BibitemOpen
  \bibfield  {author} {\bibinfo {author} {\bibfnamefont {G.}~\bibnamefont
  {Knizia}}\ and\ \bibinfo {author} {\bibfnamefont {G.~K.-L.}\ \bibnamefont
  {Chan}},\ }\href {\doibase 10.1103/PhysRevLett.109.186404} {\bibfield
  {journal} {\bibinfo  {journal} {Physical Review Letters}\ }\textbf {\bibinfo
  {volume} {109}},\ \bibinfo {pages} {186404} (\bibinfo {year}
  {2012})}\BibitemShut {NoStop}%
\bibitem [{\citenamefont {Knizia}\ and\ \citenamefont
  {Chan}(2013)}]{knizia2013}%
  \BibitemOpen
  \bibfield  {author} {\bibinfo {author} {\bibfnamefont {G.}~\bibnamefont
  {Knizia}}\ and\ \bibinfo {author} {\bibfnamefont {G.~K.-L.}\ \bibnamefont
  {Chan}},\ }\href {\doibase 10.1021/ct301044e} {\bibfield  {journal} {\bibinfo
   {journal} {Journal of Chemical Theory and Computation}\ }\textbf {\bibinfo
  {volume} {9}},\ \bibinfo {pages} {1428} (\bibinfo {year} {2013})}\BibitemShut
  {NoStop}%
\bibitem [{\citenamefont {Pham}\ \emph {et~al.}(2020)\citenamefont {Pham},
  \citenamefont {Hermes},\ and\ \citenamefont {Gagliardi}}]{pham2020}%
  \BibitemOpen
  \bibfield  {author} {\bibinfo {author} {\bibfnamefont {H.~Q.}\ \bibnamefont
  {Pham}}, \bibinfo {author} {\bibfnamefont {M.~R.}\ \bibnamefont {Hermes}}, \
  and\ \bibinfo {author} {\bibfnamefont {L.}~\bibnamefont {Gagliardi}},\ }\href
  {\doibase 10.1021/acs.jctc.9b00939} {\bibfield  {journal} {\bibinfo
  {journal} {Journal of Chemical Theory and Computation}\ }\textbf {\bibinfo
  {volume} {16}},\ \bibinfo {pages} {130} (\bibinfo {year} {2020})}\BibitemShut
  {NoStop}%
\bibitem [{\citenamefont {Hermes}\ and\ \citenamefont
  {Gagliardi}(2019)}]{hermes2019}%
  \BibitemOpen
  \bibfield  {author} {\bibinfo {author} {\bibfnamefont {M.~R.}\ \bibnamefont
  {Hermes}}\ and\ \bibinfo {author} {\bibfnamefont {L.}~\bibnamefont
  {Gagliardi}},\ }\href {\doibase 10.1021/acs.jctc.8b01009} {\bibfield
  {journal} {\bibinfo  {journal} {Journal of Chemical Theory and Computation}\
  }\textbf {\bibinfo {volume} {15}},\ \bibinfo {pages} {972} (\bibinfo {year}
  {2019})}\BibitemShut {NoStop}%
\bibitem [{\citenamefont {Pham}\ \emph {et~al.}(2018)\citenamefont {Pham},
  \citenamefont {Bernales},\ and\ \citenamefont {Gagliardi}}]{pham2018}%
  \BibitemOpen
  \bibfield  {author} {\bibinfo {author} {\bibfnamefont {H.~Q.}\ \bibnamefont
  {Pham}}, \bibinfo {author} {\bibfnamefont {V.}~\bibnamefont {Bernales}}, \
  and\ \bibinfo {author} {\bibfnamefont {L.}~\bibnamefont {Gagliardi}},\ }\href
  {\doibase 10.1021/acs.jctc.7b01248} {\bibfield  {journal} {\bibinfo
  {journal} {Journal of Chemical Theory and Computation}\ }\textbf {\bibinfo
  {volume} {14}},\ \bibinfo {pages} {1960} (\bibinfo {year}
  {2018})}\BibitemShut {NoStop}%
\bibitem [{\citenamefont {Rungger}\ \emph {et~al.}(2020)\citenamefont
  {Rungger}, \citenamefont {Fitzpatrick}, \citenamefont {Chen}, \citenamefont
  {Alderete}, \citenamefont {Apel}, \citenamefont {Cowtan}, \citenamefont
  {Patterson}, \citenamefont {Ramo}, \citenamefont {Zhu}, \citenamefont
  {Nguyen}, \citenamefont {Grant}, \citenamefont {Chretien}, \citenamefont
  {Wossnig}, \citenamefont {Linke},\ and\ \citenamefont
  {Duncan}}]{rungger2020}%
  \BibitemOpen
  \bibfield  {author} {\bibinfo {author} {\bibfnamefont {I.}~\bibnamefont
  {Rungger}}, \bibinfo {author} {\bibfnamefont {N.}~\bibnamefont
  {Fitzpatrick}}, \bibinfo {author} {\bibfnamefont {H.}~\bibnamefont {Chen}},
  \bibinfo {author} {\bibfnamefont {C.~H.}\ \bibnamefont {Alderete}}, \bibinfo
  {author} {\bibfnamefont {H.}~\bibnamefont {Apel}}, \bibinfo {author}
  {\bibfnamefont {A.}~\bibnamefont {Cowtan}}, \bibinfo {author} {\bibfnamefont
  {A.}~\bibnamefont {Patterson}}, \bibinfo {author} {\bibfnamefont {D.~M.}\
  \bibnamefont {Ramo}}, \bibinfo {author} {\bibfnamefont {Y.}~\bibnamefont
  {Zhu}}, \bibinfo {author} {\bibfnamefont {N.~H.}\ \bibnamefont {Nguyen}},
  \bibinfo {author} {\bibfnamefont {E.}~\bibnamefont {Grant}}, \bibinfo
  {author} {\bibfnamefont {S.}~\bibnamefont {Chretien}}, \bibinfo {author}
  {\bibfnamefont {L.}~\bibnamefont {Wossnig}}, \bibinfo {author} {\bibfnamefont
  {N.~M.}\ \bibnamefont {Linke}}, \ and\ \bibinfo {author} {\bibfnamefont
  {R.}~\bibnamefont {Duncan}},\ }\href {http://arxiv.org/abs/1910.04735}
  {\bibfield  {journal} {\bibinfo  {journal} {arXiv:1910.04735 [cond-mat,
  physics:quant-ph]}\ } (\bibinfo {year} {2020})}\BibitemShut {NoStop}%
\bibitem [{\citenamefont {Keen}\ \emph {et~al.}(2020)\citenamefont {Keen},
  \citenamefont {Maier}, \citenamefont {Johnston},\ and\ \citenamefont
  {Lougovski}}]{keen2020}%
  \BibitemOpen
  \bibfield  {author} {\bibinfo {author} {\bibfnamefont {T.}~\bibnamefont
  {Keen}}, \bibinfo {author} {\bibfnamefont {T.}~\bibnamefont {Maier}},
  \bibinfo {author} {\bibfnamefont {S.}~\bibnamefont {Johnston}}, \ and\
  \bibinfo {author} {\bibfnamefont {P.}~\bibnamefont {Lougovski}},\ }\href
  {\doibase 10.1088/2058-9565/ab7d4c} {\bibfield  {journal} {\bibinfo
  {journal} {Quantum Science and Technology}\ }\textbf {\bibinfo {volume}
  {5}},\ \bibinfo {pages} {035001} (\bibinfo {year} {2020})}\BibitemShut
  {NoStop}%
\bibitem [{\citenamefont {Yao}\ \emph {et~al.}(2021)\citenamefont {Yao},
  \citenamefont {Zhang}, \citenamefont {Wang}, \citenamefont {Ho},\ and\
  \citenamefont {Orth}}]{yao2021}%
  \BibitemOpen
  \bibfield  {author} {\bibinfo {author} {\bibfnamefont {Y.}~\bibnamefont
  {Yao}}, \bibinfo {author} {\bibfnamefont {F.}~\bibnamefont {Zhang}}, \bibinfo
  {author} {\bibfnamefont {C.-Z.}\ \bibnamefont {Wang}}, \bibinfo {author}
  {\bibfnamefont {K.-M.}\ \bibnamefont {Ho}}, \ and\ \bibinfo {author}
  {\bibfnamefont {P.~P.}\ \bibnamefont {Orth}},\ }\href {\doibase
  10.1103/PhysRevResearch.3.013184} {\bibfield  {journal} {\bibinfo  {journal}
  {Physical Review Research}\ }\textbf {\bibinfo {volume} {3}},\ \bibinfo
  {pages} {013184} (\bibinfo {year} {2021})}\BibitemShut {NoStop}%
\bibitem [{\citenamefont {Tilly}\ \emph {et~al.}(2021)\citenamefont {Tilly},
  \citenamefont {Sriluckshmy}, \citenamefont {Patel}, \citenamefont {Fontana},
  \citenamefont {Rungger}, \citenamefont {Grant}, \citenamefont {Anderson},
  \citenamefont {Tennyson},\ and\ \citenamefont {Booth}}]{tilly2021}%
  \BibitemOpen
  \bibfield  {author} {\bibinfo {author} {\bibfnamefont {J.}~\bibnamefont
  {Tilly}}, \bibinfo {author} {\bibfnamefont {P.~V.}\ \bibnamefont
  {Sriluckshmy}}, \bibinfo {author} {\bibfnamefont {A.}~\bibnamefont {Patel}},
  \bibinfo {author} {\bibfnamefont {E.}~\bibnamefont {Fontana}}, \bibinfo
  {author} {\bibfnamefont {I.}~\bibnamefont {Rungger}}, \bibinfo {author}
  {\bibfnamefont {E.}~\bibnamefont {Grant}}, \bibinfo {author} {\bibfnamefont
  {R.}~\bibnamefont {Anderson}}, \bibinfo {author} {\bibfnamefont
  {J.}~\bibnamefont {Tennyson}}, \ and\ \bibinfo {author} {\bibfnamefont
  {G.~H.}\ \bibnamefont {Booth}},\ }\href@noop {} {\bibfield  {journal}
  {\bibinfo  {journal} {arXiv:2104.05531 [physics, physics:quant-ph]}\ }
  (\bibinfo {year} {2021})},\ \Eprint {http://arxiv.org/abs/2104.05531}
  {arXiv:2104.05531 [physics, physics:quant-ph]} \BibitemShut {NoStop}%
\bibitem [{\citenamefont {Bassman}\ \emph {et~al.}(2021)\citenamefont
  {Bassman}, \citenamefont {Urbanek}, \citenamefont {Metcalf}, \citenamefont
  {Carter}, \citenamefont {Kemper},\ and\ \citenamefont {{de
  Jong}}}]{bassman2021}%
  \BibitemOpen
  \bibfield  {author} {\bibinfo {author} {\bibfnamefont {L.}~\bibnamefont
  {Bassman}}, \bibinfo {author} {\bibfnamefont {M.}~\bibnamefont {Urbanek}},
  \bibinfo {author} {\bibfnamefont {M.}~\bibnamefont {Metcalf}}, \bibinfo
  {author} {\bibfnamefont {J.}~\bibnamefont {Carter}}, \bibinfo {author}
  {\bibfnamefont {A.~F.}\ \bibnamefont {Kemper}}, \ and\ \bibinfo {author}
  {\bibfnamefont {W.}~\bibnamefont {{de Jong}}},\ }\href
  {http://arxiv.org/abs/2101.08836} {\bibfield  {journal} {\bibinfo  {journal}
  {arXiv:2101.08836 [quant-ph]}\ } (\bibinfo {year} {2021})}\BibitemShut
  {NoStop}%
\bibitem [{\citenamefont {Cerasoli}\ \emph {et~al.}(2020)\citenamefont
  {Cerasoli}, \citenamefont {Sherbert}, \citenamefont {S{\l}awi{\'n}ska},\ and\
  \citenamefont {Nardelli}}]{cerasoli2020}%
  \BibitemOpen
  \bibfield  {author} {\bibinfo {author} {\bibfnamefont {F.~T.}\ \bibnamefont
  {Cerasoli}}, \bibinfo {author} {\bibfnamefont {K.}~\bibnamefont {Sherbert}},
  \bibinfo {author} {\bibfnamefont {J.}~\bibnamefont {S{\l}awi{\'n}ska}}, \
  and\ \bibinfo {author} {\bibfnamefont {M.~B.}\ \bibnamefont {Nardelli}},\
  }\href {\doibase 10.1039/D0CP04008H} {\bibfield  {journal} {\bibinfo
  {journal} {Physical Chemistry Chemical Physics}\ }\textbf {\bibinfo {volume}
  {22}},\ \bibinfo {pages} {21816} (\bibinfo {year} {2020})}\BibitemShut
  {NoStop}%
\bibitem [{\citenamefont {Sureshbabu}\ \emph {et~al.}(2021)\citenamefont
  {Sureshbabu}, \citenamefont {Sajjan}, \citenamefont {Oh},\ and\ \citenamefont
  {Kais}}]{sureshbabu2021}%
  \BibitemOpen
  \bibfield  {author} {\bibinfo {author} {\bibfnamefont {S.~H.}\ \bibnamefont
  {Sureshbabu}}, \bibinfo {author} {\bibfnamefont {M.}~\bibnamefont {Sajjan}},
  \bibinfo {author} {\bibfnamefont {S.}~\bibnamefont {Oh}}, \ and\ \bibinfo
  {author} {\bibfnamefont {S.}~\bibnamefont {Kais}},\ }\href {\doibase
  10.1021/acs.jcim.1c00294} {\bibfield  {journal} {\bibinfo  {journal} {Journal
  of Chemical Information and Modeling}\ } (\bibinfo {year} {2021}),\
  10.1021/acs.jcim.1c00294}\BibitemShut {NoStop}%
\bibitem [{\citenamefont {Choudhary}(2021)}]{choudhary2021}%
  \BibitemOpen
  \bibfield  {author} {\bibinfo {author} {\bibfnamefont {K.}~\bibnamefont
  {Choudhary}},\ }\href {http://arxiv.org/abs/2102.11452} {\bibfield  {journal}
  {\bibinfo  {journal} {arXiv:2102.11452 [cond-mat]}\ } (\bibinfo {year}
  {2021})}\BibitemShut {NoStop}%
\bibitem [{\citenamefont {Libisch}\ \emph {et~al.}(2014)\citenamefont
  {Libisch}, \citenamefont {Huang},\ and\ \citenamefont
  {Carter}}]{libisch2014}%
  \BibitemOpen
  \bibfield  {author} {\bibinfo {author} {\bibfnamefont {F.}~\bibnamefont
  {Libisch}}, \bibinfo {author} {\bibfnamefont {C.}~\bibnamefont {Huang}}, \
  and\ \bibinfo {author} {\bibfnamefont {E.~A.}\ \bibnamefont {Carter}},\
  }\href {\doibase 10.1021/ar500086h} {\bibfield  {journal} {\bibinfo
  {journal} {Accounts of Chemical Research}\ }\textbf {\bibinfo {volume}
  {47}},\ \bibinfo {pages} {2768} (\bibinfo {year} {2014})}\BibitemShut
  {NoStop}%
\bibitem [{\citenamefont {Wesolowski}\ \emph {et~al.}(2015)\citenamefont
  {Wesolowski}, \citenamefont {Shedge},\ and\ \citenamefont
  {Zhou}}]{wesolowski2015}%
  \BibitemOpen
  \bibfield  {author} {\bibinfo {author} {\bibfnamefont {T.~A.}\ \bibnamefont
  {Wesolowski}}, \bibinfo {author} {\bibfnamefont {S.}~\bibnamefont {Shedge}},
  \ and\ \bibinfo {author} {\bibfnamefont {X.}~\bibnamefont {Zhou}},\ }\href
  {\doibase 10.1021/cr500502v} {\bibfield  {journal} {\bibinfo  {journal}
  {{Chemical Reviews}}\ }\textbf {\bibinfo {volume} {115}},\ \bibinfo {pages}
  {5891} (\bibinfo {year} {2015})}\BibitemShut {NoStop}%
\bibitem [{\citenamefont {Jacob}\ and\ \citenamefont
  {Neugebauer}(2014)}]{jacob2014}%
  \BibitemOpen
  \bibfield  {author} {\bibinfo {author} {\bibfnamefont {C.~R.}\ \bibnamefont
  {Jacob}}\ and\ \bibinfo {author} {\bibfnamefont {J.}~\bibnamefont
  {Neugebauer}},\ }\href {\doibase 10.1002/wcms.1175} {\bibfield  {journal}
  {\bibinfo  {journal} {WIREs Computational Molecular Science}\ }\textbf
  {\bibinfo {volume} {4}},\ \bibinfo {pages} {325} (\bibinfo {year}
  {2014})}\BibitemShut {NoStop}%
\bibitem [{\citenamefont {Ma}\ \emph {et~al.}(2020{\natexlab{a}})\citenamefont
  {Ma}, \citenamefont {Sheng}, \citenamefont {Govoni},\ and\ \citenamefont
  {Galli}}]{ma2020}%
  \BibitemOpen
  \bibfield  {author} {\bibinfo {author} {\bibfnamefont {H.}~\bibnamefont
  {Ma}}, \bibinfo {author} {\bibfnamefont {N.}~\bibnamefont {Sheng}}, \bibinfo
  {author} {\bibfnamefont {M.}~\bibnamefont {Govoni}}, \ and\ \bibinfo {author}
  {\bibfnamefont {G.}~\bibnamefont {Galli}},\ }\href {\doibase
  10.1039/D0CP04585C} {\bibfield  {journal} {\bibinfo  {journal} {Physical
  Chemistry Chemical Physics}\ }\textbf {\bibinfo {volume} {22}},\ \bibinfo
  {pages} {25522} (\bibinfo {year} {2020}{\natexlab{a}})}\BibitemShut {NoStop}%
\bibitem [{\citenamefont {Ma}\ \emph {et~al.}(2020{\natexlab{b}})\citenamefont
  {Ma}, \citenamefont {Govoni},\ and\ \citenamefont {Galli}}]{ma2020a}%
  \BibitemOpen
  \bibfield  {author} {\bibinfo {author} {\bibfnamefont {H.}~\bibnamefont
  {Ma}}, \bibinfo {author} {\bibfnamefont {M.}~\bibnamefont {Govoni}}, \ and\
  \bibinfo {author} {\bibfnamefont {G.}~\bibnamefont {Galli}},\ }\href
  {\doibase 10.1038/s41524-020-00353-z} {\bibfield  {journal} {\bibinfo
  {journal} {{npj Computational Materials}}\ }\textbf {\bibinfo {volume} {6}},\
  \bibinfo {pages} {1} (\bibinfo {year} {2020}{\natexlab{b}})}\BibitemShut
  {NoStop}%
\bibitem [{\citenamefont {Ma}\ \emph {et~al.}(2021)\citenamefont {Ma},
  \citenamefont {Sheng}, \citenamefont {Govoni},\ and\ \citenamefont
  {Galli}}]{ma2021}%
  \BibitemOpen
  \bibfield  {author} {\bibinfo {author} {\bibfnamefont {H.}~\bibnamefont
  {Ma}}, \bibinfo {author} {\bibfnamefont {N.}~\bibnamefont {Sheng}}, \bibinfo
  {author} {\bibfnamefont {M.}~\bibnamefont {Govoni}}, \ and\ \bibinfo {author}
  {\bibfnamefont {G.}~\bibnamefont {Galli}},\ }\href {\doibase
  10.1021/acs.jctc.0c01258} {\bibfield  {journal} {\bibinfo  {journal} {Journal
  of Chemical Theory and Computation}\ }\textbf {\bibinfo {volume} {17}},\
  \bibinfo {pages} {2116} (\bibinfo {year} {2021})}\BibitemShut {NoStop}%
\bibitem [{\citenamefont {Lan}\ and\ \citenamefont {Zgid}(2017)}]{lan2017}%
  \BibitemOpen
  \bibfield  {author} {\bibinfo {author} {\bibfnamefont {T.~N.}\ \bibnamefont
  {Lan}}\ and\ \bibinfo {author} {\bibfnamefont {D.}~\bibnamefont {Zgid}},\
  }\href {\doibase 10.1021/acs.jpclett.7b00689} {\bibfield  {journal} {\bibinfo
   {journal} {The Journal of Physical Chemistry Letters}\ }\textbf {\bibinfo
  {volume} {8}},\ \bibinfo {pages} {2200} (\bibinfo {year} {2017})}\BibitemShut
  {NoStop}%
\bibitem [{\citenamefont {Zgid}\ and\ \citenamefont {Gull}(2017)}]{zgid2017}%
  \BibitemOpen
  \bibfield  {author} {\bibinfo {author} {\bibfnamefont {D.}~\bibnamefont
  {Zgid}}\ and\ \bibinfo {author} {\bibfnamefont {E.}~\bibnamefont {Gull}},\
  }\href {\doibase 10.1088/1367-2630/aa5d34} {\bibfield  {journal} {\bibinfo
  {journal} {New Journal of Physics}\ }\textbf {\bibinfo {volume} {19}},\
  \bibinfo {pages} {023047} (\bibinfo {year} {2017})}\BibitemShut {NoStop}%
\bibitem [{\citenamefont {Rusakov}\ \emph {et~al.}(2019)\citenamefont
  {Rusakov}, \citenamefont {Iskakov}, \citenamefont {Tran},\ and\ \citenamefont
  {Zgid}}]{rusakov2019}%
  \BibitemOpen
  \bibfield  {author} {\bibinfo {author} {\bibfnamefont {A.~A.}\ \bibnamefont
  {Rusakov}}, \bibinfo {author} {\bibfnamefont {S.}~\bibnamefont {Iskakov}},
  \bibinfo {author} {\bibfnamefont {L.~N.}\ \bibnamefont {Tran}}, \ and\
  \bibinfo {author} {\bibfnamefont {D.}~\bibnamefont {Zgid}},\ }\href {\doibase
  10.1021/acs.jctc.8b00927} {\bibfield  {journal} {\bibinfo  {journal} {Journal
  of Chemical Theory and Computation}\ }\textbf {\bibinfo {volume} {15}},\
  \bibinfo {pages} {229} (\bibinfo {year} {2019})}\BibitemShut {NoStop}%
\bibitem [{\citenamefont {Biermann}\ \emph {et~al.}(2003)\citenamefont
  {Biermann}, \citenamefont {Aryasetiawan},\ and\ \citenamefont
  {Georges}}]{biermann2003}%
  \BibitemOpen
  \bibfield  {author} {\bibinfo {author} {\bibfnamefont {S.}~\bibnamefont
  {Biermann}}, \bibinfo {author} {\bibfnamefont {F.}~\bibnamefont
  {Aryasetiawan}}, \ and\ \bibinfo {author} {\bibfnamefont {A.}~\bibnamefont
  {Georges}},\ }\href {\doibase 10.1103/PhysRevLett.90.086402} {\bibfield
  {journal} {\bibinfo  {journal} {Physical Review Letters}\ }\textbf {\bibinfo
  {volume} {90}},\ \bibinfo {pages} {086402} (\bibinfo {year}
  {2003})}\BibitemShut {NoStop}%
\bibitem [{\citenamefont {Biermann}(2014)}]{biermann2014}%
  \BibitemOpen
  \bibfield  {author} {\bibinfo {author} {\bibfnamefont {S.}~\bibnamefont
  {Biermann}},\ }\href {\doibase 10.1088/0953-8984/26/17/173202} {\bibfield
  {journal} {\bibinfo  {journal} {Journal of Physics: Condensed Matter}\
  }\textbf {\bibinfo {volume} {26}},\ \bibinfo {pages} {173202} (\bibinfo
  {year} {2014})}\BibitemShut {NoStop}%
\bibitem [{\citenamefont {Boehnke}\ \emph {et~al.}(2016)\citenamefont
  {Boehnke}, \citenamefont {Nilsson}, \citenamefont {Aryasetiawan},\ and\
  \citenamefont {Werner}}]{boehnke2016}%
  \BibitemOpen
  \bibfield  {author} {\bibinfo {author} {\bibfnamefont {L.}~\bibnamefont
  {Boehnke}}, \bibinfo {author} {\bibfnamefont {F.}~\bibnamefont {Nilsson}},
  \bibinfo {author} {\bibfnamefont {F.}~\bibnamefont {Aryasetiawan}}, \ and\
  \bibinfo {author} {\bibfnamefont {P.}~\bibnamefont {Werner}},\ }\href
  {\doibase 10.1103/PhysRevB.94.201106} {\bibfield  {journal} {\bibinfo
  {journal} {Physical Review B}\ }\textbf {\bibinfo {volume} {94}},\ \bibinfo
  {pages} {201106} (\bibinfo {year} {2016})}\BibitemShut {NoStop}%
\bibitem [{\citenamefont {Choi}\ \emph {et~al.}(2016)\citenamefont {Choi},
  \citenamefont {Kutepov}, \citenamefont {Haule}, \citenamefont {{van
  Schilfgaarde}},\ and\ \citenamefont {Kotliar}}]{choi2016}%
  \BibitemOpen
  \bibfield  {author} {\bibinfo {author} {\bibfnamefont {S.}~\bibnamefont
  {Choi}}, \bibinfo {author} {\bibfnamefont {A.}~\bibnamefont {Kutepov}},
  \bibinfo {author} {\bibfnamefont {K.}~\bibnamefont {Haule}}, \bibinfo
  {author} {\bibfnamefont {M.}~\bibnamefont {{van Schilfgaarde}}}, \ and\
  \bibinfo {author} {\bibfnamefont {G.}~\bibnamefont {Kotliar}},\ }\href
  {\doibase 10.1038/npjquantmats.2016.1} {\bibfield  {journal} {\bibinfo
  {journal} {npj Quantum Materials}\ }\textbf {\bibinfo {volume} {1}},\
  \bibinfo {pages} {1} (\bibinfo {year} {2016})}\BibitemShut {NoStop}%
\bibitem [{\citenamefont {Nilsson}\ \emph {et~al.}(2017)\citenamefont
  {Nilsson}, \citenamefont {Boehnke}, \citenamefont {Werner},\ and\
  \citenamefont {Aryasetiawan}}]{nilsson2017}%
  \BibitemOpen
  \bibfield  {author} {\bibinfo {author} {\bibfnamefont {F.}~\bibnamefont
  {Nilsson}}, \bibinfo {author} {\bibfnamefont {L.}~\bibnamefont {Boehnke}},
  \bibinfo {author} {\bibfnamefont {P.}~\bibnamefont {Werner}}, \ and\ \bibinfo
  {author} {\bibfnamefont {F.}~\bibnamefont {Aryasetiawan}},\ }\href {\doibase
  10.1103/PhysRevMaterials.1.043803} {\bibfield  {journal} {\bibinfo  {journal}
  {Physical Review Materials}\ }\textbf {\bibinfo {volume} {1}},\ \bibinfo
  {pages} {043803} (\bibinfo {year} {2017})}\BibitemShut {NoStop}%
\bibitem [{\citenamefont {Sun}\ and\ \citenamefont {Kotliar}(2002)}]{sun2002}%
  \BibitemOpen
  \bibfield  {author} {\bibinfo {author} {\bibfnamefont {P.}~\bibnamefont
  {Sun}}\ and\ \bibinfo {author} {\bibfnamefont {G.}~\bibnamefont {Kotliar}},\
  }\href {\doibase 10.1103/PhysRevB.66.085120} {\bibfield  {journal} {\bibinfo
  {journal} {Physical Review B}\ }\textbf {\bibinfo {volume} {66}},\ \bibinfo
  {pages} {085120} (\bibinfo {year} {2002})}\BibitemShut {NoStop}%
\bibitem [{\citenamefont {Lichtenstein}\ and\ \citenamefont
  {Katsnelson}(1998)}]{lichtenstein1998}%
  \BibitemOpen
  \bibfield  {author} {\bibinfo {author} {\bibfnamefont {A.~I.}\ \bibnamefont
  {Lichtenstein}}\ and\ \bibinfo {author} {\bibfnamefont {M.~I.}\ \bibnamefont
  {Katsnelson}},\ }\href {\doibase 10.1103/PhysRevB.57.6884} {\bibfield
  {journal} {\bibinfo  {journal} {Physical Review B}\ }\textbf {\bibinfo
  {volume} {57}},\ \bibinfo {pages} {6884} (\bibinfo {year}
  {1998})}\BibitemShut {NoStop}%
\bibitem [{\citenamefont {Dhawan}\ \emph {et~al.}(2021)\citenamefont {Dhawan},
  \citenamefont {Metcalf},\ and\ \citenamefont {Zgid}}]{dhawan2021}%
  \BibitemOpen
  \bibfield  {author} {\bibinfo {author} {\bibfnamefont {D.}~\bibnamefont
  {Dhawan}}, \bibinfo {author} {\bibfnamefont {M.}~\bibnamefont {Metcalf}}, \
  and\ \bibinfo {author} {\bibfnamefont {D.}~\bibnamefont {Zgid}},\ }\href
  {http://arxiv.org/abs/2010.05441} {\bibfield  {journal} {\bibinfo  {journal}
  {arXiv:2010.05441 [physics, physics:quant-ph]}\ } (\bibinfo {year}
  {2021})}\BibitemShut {NoStop}%
\bibitem [{\citenamefont {Otten}\ \emph {et~al.}(2021)\citenamefont {Otten},
  \citenamefont {Hermes}, \citenamefont {Pandharkar}, \citenamefont {Alexeev},
  \citenamefont {Gray},\ and\ \citenamefont {Gagliardi}}]{otten2021}%
  \BibitemOpen
  \bibfield  {author} {\bibinfo {author} {\bibfnamefont {M.}~\bibnamefont
  {Otten}}, \bibinfo {author} {\bibfnamefont {M.}~\bibnamefont {Hermes}},
  \bibinfo {author} {\bibfnamefont {R.}~\bibnamefont {Pandharkar}}, \bibinfo
  {author} {\bibfnamefont {Y.}~\bibnamefont {Alexeev}}, \bibinfo {author}
  {\bibfnamefont {S.}~\bibnamefont {Gray}}, \ and\ \bibinfo {author}
  {\bibfnamefont {L.}~\bibnamefont {Gagliardi}},\ }\href {\doibase
  10.33774/chemrxiv-2021-0nmwt} {\  (\bibinfo {year} {2021}),\
  10.33774/chemrxiv-2021-0nmwt}\BibitemShut {NoStop}%
\bibitem [{\citenamefont {Seo}\ \emph {et~al.}(2016)\citenamefont {Seo},
  \citenamefont {Govoni},\ and\ \citenamefont {Galli}}]{seo2016}%
  \BibitemOpen
  \bibfield  {author} {\bibinfo {author} {\bibfnamefont {H.}~\bibnamefont
  {Seo}}, \bibinfo {author} {\bibfnamefont {M.}~\bibnamefont {Govoni}}, \ and\
  \bibinfo {author} {\bibfnamefont {G.}~\bibnamefont {Galli}},\ }\href
  {\doibase 10.1038/srep20803} {\bibfield  {journal} {\bibinfo  {journal}
  {Scientific Reports}\ }\textbf {\bibinfo {volume} {6}},\ \bibinfo {pages}
  {20803} (\bibinfo {year} {2016})}\BibitemShut {NoStop}%
\bibitem [{\citenamefont {Seo}\ \emph {et~al.}(2017)\citenamefont {Seo},
  \citenamefont {Ma}, \citenamefont {Govoni},\ and\ \citenamefont
  {Galli}}]{seo2017}%
  \BibitemOpen
  \bibfield  {author} {\bibinfo {author} {\bibfnamefont {H.}~\bibnamefont
  {Seo}}, \bibinfo {author} {\bibfnamefont {H.}~\bibnamefont {Ma}}, \bibinfo
  {author} {\bibfnamefont {M.}~\bibnamefont {Govoni}}, \ and\ \bibinfo {author}
  {\bibfnamefont {G.}~\bibnamefont {Galli}},\ }\href {\doibase
  10.1103/PhysRevMaterials.1.075002} {\bibfield  {journal} {\bibinfo  {journal}
  {Physical Review Materials}\ }\textbf {\bibinfo {volume} {1}},\ \bibinfo
  {pages} {075002} (\bibinfo {year} {2017})}\BibitemShut {NoStop}%
\bibitem [{\citenamefont {Iv{\'a}dy}\ \emph {et~al.}(2018)\citenamefont
  {Iv{\'a}dy}, \citenamefont {Abrikosov},\ and\ \citenamefont
  {Gali}}]{ivady2018}%
  \BibitemOpen
  \bibfield  {author} {\bibinfo {author} {\bibfnamefont {V.}~\bibnamefont
  {Iv{\'a}dy}}, \bibinfo {author} {\bibfnamefont {I.~A.}\ \bibnamefont
  {Abrikosov}}, \ and\ \bibinfo {author} {\bibfnamefont {A.}~\bibnamefont
  {Gali}},\ }\href {\doibase 10.1038/s41524-018-0132-5} {\bibfield  {journal}
  {\bibinfo  {journal} {npj Computational Materials}\ }\textbf {\bibinfo
  {volume} {4}},\ \bibinfo {pages} {1} (\bibinfo {year} {2018})}\BibitemShut
  {NoStop}%
\bibitem [{\citenamefont {Anderson}\ \emph {et~al.}(2019)\citenamefont
  {Anderson}, \citenamefont {Bourassa}, \citenamefont {Miao}, \citenamefont
  {Wolfowicz}, \citenamefont {Mintun}, \citenamefont {Crook}, \citenamefont
  {Abe}, \citenamefont {Hassan}, \citenamefont {Son}, \citenamefont {Ohshima},\
  and\ \citenamefont {Awschalom}}]{anderson2019}%
  \BibitemOpen
  \bibfield  {author} {\bibinfo {author} {\bibfnamefont {C.~P.}\ \bibnamefont
  {Anderson}}, \bibinfo {author} {\bibfnamefont {A.}~\bibnamefont {Bourassa}},
  \bibinfo {author} {\bibfnamefont {K.~C.}\ \bibnamefont {Miao}}, \bibinfo
  {author} {\bibfnamefont {G.}~\bibnamefont {Wolfowicz}}, \bibinfo {author}
  {\bibfnamefont {P.~J.}\ \bibnamefont {Mintun}}, \bibinfo {author}
  {\bibfnamefont {A.~L.}\ \bibnamefont {Crook}}, \bibinfo {author}
  {\bibfnamefont {H.}~\bibnamefont {Abe}}, \bibinfo {author} {\bibfnamefont
  {J.~U.}\ \bibnamefont {Hassan}}, \bibinfo {author} {\bibfnamefont {N.~T.}\
  \bibnamefont {Son}}, \bibinfo {author} {\bibfnamefont {T.}~\bibnamefont
  {Ohshima}}, \ and\ \bibinfo {author} {\bibfnamefont {D.~D.}\ \bibnamefont
  {Awschalom}},\ }\href {\doibase 10.1126/science.aax9406} {\bibfield
  {journal} {\bibinfo  {journal} {Science}\ }\textbf {\bibinfo {volume}
  {366}},\ \bibinfo {pages} {1225} (\bibinfo {year} {2019})}\BibitemShut
  {NoStop}%
\bibitem [{\citenamefont {Sun}\ and\ \citenamefont {Chan}(2016)}]{sun2016}%
  \BibitemOpen
  \bibfield  {author} {\bibinfo {author} {\bibfnamefont {Q.}~\bibnamefont
  {Sun}}\ and\ \bibinfo {author} {\bibfnamefont {G.~K.-L.}\ \bibnamefont
  {Chan}},\ }\href {\doibase 10.1021/acs.accounts.6b00356} {\bibfield
  {journal} {\bibinfo  {journal} {Accounts of Chemical Research}\ }\textbf
  {\bibinfo {volume} {49}},\ \bibinfo {pages} {2705} (\bibinfo {year}
  {2016})}\BibitemShut {NoStop}%
\bibitem [{\citenamefont {Jones}\ \emph {et~al.}(2020)\citenamefont {Jones},
  \citenamefont {Mosquera}, \citenamefont {Schatz},\ and\ \citenamefont
  {Ratner}}]{jones2020}%
  \BibitemOpen
  \bibfield  {author} {\bibinfo {author} {\bibfnamefont {L.~O.}\ \bibnamefont
  {Jones}}, \bibinfo {author} {\bibfnamefont {M.~A.}\ \bibnamefont {Mosquera}},
  \bibinfo {author} {\bibfnamefont {G.~C.}\ \bibnamefont {Schatz}}, \ and\
  \bibinfo {author} {\bibfnamefont {M.~A.}\ \bibnamefont {Ratner}},\ }\href
  {\doibase 10.1021/jacs.9b10780} {\bibfield  {journal} {\bibinfo  {journal}
  {Journal of the American Chemical Society}\ }\textbf {\bibinfo {volume}
  {142}},\ \bibinfo {pages} {3281} (\bibinfo {year} {2020})}\BibitemShut
  {NoStop}%
\bibitem [{\citenamefont {Lin}\ and\ \citenamefont {Truhlar}(2006)}]{lin2006}%
  \BibitemOpen
  \bibfield  {author} {\bibinfo {author} {\bibfnamefont {H.}~\bibnamefont
  {Lin}}\ and\ \bibinfo {author} {\bibfnamefont {D.~G.}\ \bibnamefont
  {Truhlar}},\ }\href {\doibase 10.1007/s00214-006-0143-z} {\bibfield
  {journal} {\bibinfo  {journal} {Theoretical Chemistry Accounts}\ }\textbf
  {\bibinfo {volume} {117}},\ \bibinfo {pages} {185} (\bibinfo {year}
  {2006})}\BibitemShut {NoStop}%
\bibitem [{\citenamefont {Wang}\ \emph {et~al.}(2014)\citenamefont {Wang},
  \citenamefont {Yang}, \citenamefont {Xu}, \citenamefont {Isegawa},
  \citenamefont {Leverentz},\ and\ \citenamefont {Truhlar}}]{wang2014}%
  \BibitemOpen
  \bibfield  {author} {\bibinfo {author} {\bibfnamefont {B.}~\bibnamefont
  {Wang}}, \bibinfo {author} {\bibfnamefont {K.~R.}\ \bibnamefont {Yang}},
  \bibinfo {author} {\bibfnamefont {X.}~\bibnamefont {Xu}}, \bibinfo {author}
  {\bibfnamefont {M.}~\bibnamefont {Isegawa}}, \bibinfo {author} {\bibfnamefont
  {H.~R.}\ \bibnamefont {Leverentz}}, \ and\ \bibinfo {author} {\bibfnamefont
  {D.~G.}\ \bibnamefont {Truhlar}},\ }\href {\doibase 10.1021/ar500068a}
  {\bibfield  {journal} {\bibinfo  {journal} {Accounts of Chemical Research}\
  }\textbf {\bibinfo {volume} {47}},\ \bibinfo {pages} {2731} (\bibinfo {year}
  {2014})}\BibitemShut {NoStop}%
\bibitem [{\citenamefont {Pezeshki}\ and\ \citenamefont
  {Lin}(2015)}]{pezeshki2015}%
  \BibitemOpen
  \bibfield  {author} {\bibinfo {author} {\bibfnamefont {S.}~\bibnamefont
  {Pezeshki}}\ and\ \bibinfo {author} {\bibfnamefont {H.}~\bibnamefont {Lin}},\
  }\href {\doibase 10.1080/08927022.2014.911870} {\bibfield  {journal}
  {\bibinfo  {journal} {Molecular Simulation}\ }\textbf {\bibinfo {volume}
  {41}},\ \bibinfo {pages} {168} (\bibinfo {year} {2015})}\BibitemShut
  {NoStop}%
\bibitem [{\citenamefont {He}\ and\ \citenamefont
  {Evangelista}(2020)}]{he2020}%
  \BibitemOpen
  \bibfield  {author} {\bibinfo {author} {\bibfnamefont {N.}~\bibnamefont
  {He}}\ and\ \bibinfo {author} {\bibfnamefont {F.~A.}\ \bibnamefont
  {Evangelista}},\ }\href {\doibase 10.1063/1.5142481} {\bibfield  {journal}
  {\bibinfo  {journal} {The Journal of Chemical Physics}\ }\textbf {\bibinfo
  {volume} {152}},\ \bibinfo {pages} {094107} (\bibinfo {year}
  {2020})}\BibitemShut {NoStop}%
\bibitem [{\citenamefont {Gujarati}\ \emph {et~al.}(2022)\citenamefont
  {Gujarati}, \citenamefont {Motta}, \citenamefont {Friedhoff}, \citenamefont
  {Rice}, \citenamefont {Nguyen}, \citenamefont {Barkoutsos}, \citenamefont
  {Thompson}, \citenamefont {Smith}, \citenamefont {Kagele}, \citenamefont
  {Brei}, \citenamefont {Jones},\ and\ \citenamefont
  {Williams}}]{gujarati2022}%
  \BibitemOpen
  \bibfield  {author} {\bibinfo {author} {\bibfnamefont {T.~P.}\ \bibnamefont
  {Gujarati}}, \bibinfo {author} {\bibfnamefont {M.}~\bibnamefont {Motta}},
  \bibinfo {author} {\bibfnamefont {T.~N.}\ \bibnamefont {Friedhoff}}, \bibinfo
  {author} {\bibfnamefont {J.~E.}\ \bibnamefont {Rice}}, \bibinfo {author}
  {\bibfnamefont {N.}~\bibnamefont {Nguyen}}, \bibinfo {author} {\bibfnamefont
  {P.~K.}\ \bibnamefont {Barkoutsos}}, \bibinfo {author} {\bibfnamefont
  {R.~J.}\ \bibnamefont {Thompson}}, \bibinfo {author} {\bibfnamefont
  {T.}~\bibnamefont {Smith}}, \bibinfo {author} {\bibfnamefont
  {M.}~\bibnamefont {Kagele}}, \bibinfo {author} {\bibfnamefont
  {M.}~\bibnamefont {Brei}}, \bibinfo {author} {\bibfnamefont {B.~A.}\
  \bibnamefont {Jones}}, \ and\ \bibinfo {author} {\bibfnamefont
  {K.}~\bibnamefont {Williams}},\ }\href@noop {} {\bibfield  {journal}
  {\bibinfo  {journal} {arXiv:2203.07536 [quant-ph]}\ } (\bibinfo {year}
  {2022})},\ \Eprint {http://arxiv.org/abs/2203.07536} {arXiv:2203.07536
  [quant-ph]} \BibitemShut {NoStop}%
\bibitem [{\citenamefont {Lau}\ \emph {et~al.}(2021)\citenamefont {Lau},
  \citenamefont {Knizia},\ and\ \citenamefont {Berkelbach}}]{lau2021}%
  \BibitemOpen
  \bibfield  {author} {\bibinfo {author} {\bibfnamefont {B.~T.~G.}\
  \bibnamefont {Lau}}, \bibinfo {author} {\bibfnamefont {G.}~\bibnamefont
  {Knizia}}, \ and\ \bibinfo {author} {\bibfnamefont {T.~C.}\ \bibnamefont
  {Berkelbach}},\ }\href {\doibase 10.1021/acs.jpclett.0c03274} {\bibfield
  {journal} {\bibinfo  {journal} {The Journal of Physical Chemistry Letters}\
  }\textbf {\bibinfo {volume} {12}},\ \bibinfo {pages} {1104} (\bibinfo {year}
  {2021})}\BibitemShut {NoStop}%
\bibitem [{\citenamefont {Cui}\ \emph {et~al.}(2020)\citenamefont {Cui},
  \citenamefont {Zhu},\ and\ \citenamefont {Chan}}]{cui2020}%
  \BibitemOpen
  \bibfield  {author} {\bibinfo {author} {\bibfnamefont {Z.-H.}\ \bibnamefont
  {Cui}}, \bibinfo {author} {\bibfnamefont {T.}~\bibnamefont {Zhu}}, \ and\
  \bibinfo {author} {\bibfnamefont {G.~K.-L.}\ \bibnamefont {Chan}},\ }\href
  {\doibase 10.1021/acs.jctc.9b00933} {\bibfield  {journal} {\bibinfo
  {journal} {Journal of Chemical Theory and Computation}\ }\textbf {\bibinfo
  {volume} {16}},\ \bibinfo {pages} {119} (\bibinfo {year} {2020})}\BibitemShut
  {NoStop}%
\bibitem [{\citenamefont {Cui}\ \emph {et~al.}(2022)\citenamefont {Cui},
  \citenamefont {Zhai}, \citenamefont {Zhang},\ and\ \citenamefont
  {Chan}}]{cui2022}%
  \BibitemOpen
  \bibfield  {author} {\bibinfo {author} {\bibfnamefont {Z.-H.}\ \bibnamefont
  {Cui}}, \bibinfo {author} {\bibfnamefont {H.}~\bibnamefont {Zhai}}, \bibinfo
  {author} {\bibfnamefont {X.}~\bibnamefont {Zhang}}, \ and\ \bibinfo {author}
  {\bibfnamefont {G.~K.-L.}\ \bibnamefont {Chan}},\ }\href@noop {} {\bibfield
  {journal} {\bibinfo  {journal} {arXiv:2112.09735 [cond-mat]}\ } (\bibinfo
  {year} {2022})},\ \Eprint {http://arxiv.org/abs/2112.09735} {arXiv:2112.09735
  [cond-mat]} \BibitemShut {NoStop}%
\bibitem [{\citenamefont {Anderson}(1961)}]{anderson1961}%
  \BibitemOpen
  \bibfield  {author} {\bibinfo {author} {\bibfnamefont {P.~W.}\ \bibnamefont
  {Anderson}},\ }\href {\doibase 10.1103/PhysRev.124.41} {\bibfield  {journal}
  {\bibinfo  {journal} {Physical Review}\ }\textbf {\bibinfo {volume} {124}},\
  \bibinfo {pages} {41} (\bibinfo {year} {1961})}\BibitemShut {NoStop}%
\bibitem [{\citenamefont {Werner}\ and\ \citenamefont
  {Millis}(2007)}]{werner2007}%
  \BibitemOpen
  \bibfield  {author} {\bibinfo {author} {\bibfnamefont {P.}~\bibnamefont
  {Werner}}\ and\ \bibinfo {author} {\bibfnamefont {A.~J.}\ \bibnamefont
  {Millis}},\ }\href {\doibase 10.1103/PhysRevLett.99.146404} {\bibfield
  {journal} {\bibinfo  {journal} {Physical Review Letters}\ }\textbf {\bibinfo
  {volume} {99}},\ \bibinfo {pages} {146404} (\bibinfo {year}
  {2007})}\BibitemShut {NoStop}%
\bibitem [{\citenamefont {Nilsson}\ and\ \citenamefont
  {Aryasetiawan}(2018)}]{nilsson2018}%
  \BibitemOpen
  \bibfield  {author} {\bibinfo {author} {\bibfnamefont {F.}~\bibnamefont
  {Nilsson}}\ and\ \bibinfo {author} {\bibfnamefont {F.}~\bibnamefont
  {Aryasetiawan}},\ }\href {\doibase 10.3390/computation6010026} {\bibfield
  {journal} {\bibinfo  {journal} {Computation}\ }\textbf {\bibinfo {volume}
  {6}},\ \bibinfo {pages} {26} (\bibinfo {year} {2018})}\BibitemShut {NoStop}%
\bibitem [{\citenamefont {Sakuma}\ \emph {et~al.}(2013)\citenamefont {Sakuma},
  \citenamefont {Werner},\ and\ \citenamefont {Aryasetiawan}}]{sakuma2013}%
  \BibitemOpen
  \bibfield  {author} {\bibinfo {author} {\bibfnamefont {R.}~\bibnamefont
  {Sakuma}}, \bibinfo {author} {\bibfnamefont {P.}~\bibnamefont {Werner}}, \
  and\ \bibinfo {author} {\bibfnamefont {F.}~\bibnamefont {Aryasetiawan}},\
  }\href {\doibase 10.1103/PhysRevB.88.235110} {\bibfield  {journal} {\bibinfo
  {journal} {Physical Review B}\ }\textbf {\bibinfo {volume} {88}},\ \bibinfo
  {pages} {235110} (\bibinfo {year} {2013})}\BibitemShut {NoStop}%
\bibitem [{\citenamefont {Petocchi}\ \emph {et~al.}(2020)\citenamefont
  {Petocchi}, \citenamefont {Nilsson}, \citenamefont {Aryasetiawan},\ and\
  \citenamefont {Werner}}]{petocchi2020}%
  \BibitemOpen
  \bibfield  {author} {\bibinfo {author} {\bibfnamefont {F.}~\bibnamefont
  {Petocchi}}, \bibinfo {author} {\bibfnamefont {F.}~\bibnamefont {Nilsson}},
  \bibinfo {author} {\bibfnamefont {F.}~\bibnamefont {Aryasetiawan}}, \ and\
  \bibinfo {author} {\bibfnamefont {P.}~\bibnamefont {Werner}},\ }\href
  {\doibase 10.1103/PhysRevResearch.2.013191} {\bibfield  {journal} {\bibinfo
  {journal} {Physical Review Research}\ }\textbf {\bibinfo {volume} {2}},\
  \bibinfo {pages} {013191} (\bibinfo {year} {2020})}\BibitemShut {NoStop}%
\bibitem [{\citenamefont {Tomczak}\ \emph {et~al.}(2017)\citenamefont
  {Tomczak}, \citenamefont {Liu}, \citenamefont {Toschi}, \citenamefont
  {Kresse},\ and\ \citenamefont {Held}}]{tomczak2017}%
  \BibitemOpen
  \bibfield  {author} {\bibinfo {author} {\bibfnamefont {J.~M.}\ \bibnamefont
  {Tomczak}}, \bibinfo {author} {\bibfnamefont {P.}~\bibnamefont {Liu}},
  \bibinfo {author} {\bibfnamefont {A.}~\bibnamefont {Toschi}}, \bibinfo
  {author} {\bibfnamefont {G.}~\bibnamefont {Kresse}}, \ and\ \bibinfo {author}
  {\bibfnamefont {K.}~\bibnamefont {Held}},\ }\href {\doibase
  10.1140/epjst/e2017-70053-1} {\bibfield  {journal} {\bibinfo  {journal} {The
  European Physical Journal Special Topics}\ }\textbf {\bibinfo {volume}
  {226}},\ \bibinfo {pages} {2565} (\bibinfo {year} {2017})}\BibitemShut
  {NoStop}%
\bibitem [{\citenamefont {Reining}(2018)}]{reining2018}%
  \BibitemOpen
  \bibfield  {author} {\bibinfo {author} {\bibfnamefont {L.}~\bibnamefont
  {Reining}},\ }\href {\doibase 10.1002/wcms.1344} {\bibfield  {journal}
  {\bibinfo  {journal} {WIREs Computational Molecular Science}\ }\textbf
  {\bibinfo {volume} {8}},\ \bibinfo {pages} {e1344} (\bibinfo {year}
  {2018})}\BibitemShut {NoStop}%
\bibitem [{\citenamefont {Onida}\ \emph {et~al.}(2002)\citenamefont {Onida},
  \citenamefont {Reining},\ and\ \citenamefont {Rubio}}]{onida2002}%
  \BibitemOpen
  \bibfield  {author} {\bibinfo {author} {\bibfnamefont {G.}~\bibnamefont
  {Onida}}, \bibinfo {author} {\bibfnamefont {L.}~\bibnamefont {Reining}}, \
  and\ \bibinfo {author} {\bibfnamefont {A.}~\bibnamefont {Rubio}},\ }\href
  {\doibase 10.1103/RevModPhys.74.601} {\bibfield  {journal} {\bibinfo
  {journal} {Reviews of Modern Physics}\ }\textbf {\bibinfo {volume} {74}},\
  \bibinfo {pages} {601} (\bibinfo {year} {2002})}\BibitemShut {NoStop}%
\bibitem [{\citenamefont {Hedin}(1999)}]{hedin1999}%
  \BibitemOpen
  \bibfield  {author} {\bibinfo {author} {\bibfnamefont {L.}~\bibnamefont
  {Hedin}},\ }\href {\doibase 10.1088/0953-8984/11/42/201} {\bibfield
  {journal} {\bibinfo  {journal} {Journal of Physics: Condensed Matter}\
  }\textbf {\bibinfo {volume} {11}},\ \bibinfo {pages} {R489–R528} (\bibinfo
  {year} {1999})}\BibitemShut {NoStop}%
\bibitem [{\citenamefont {Aryasetiawan}\ and\ \citenamefont
  {Gunnarsson}(1998)}]{aryasetiawan1998}%
  \BibitemOpen
  \bibfield  {author} {\bibinfo {author} {\bibfnamefont {F.}~\bibnamefont
  {Aryasetiawan}}\ and\ \bibinfo {author} {\bibfnamefont {O.}~\bibnamefont
  {Gunnarsson}},\ }\href {\doibase 10.1088/0034-4885/61/3/002} {\bibfield
  {journal} {\bibinfo  {journal} {Reports on Progress in Physics}\ }\textbf
  {\bibinfo {volume} {61}},\ \bibinfo {pages} {237} (\bibinfo {year}
  {1998})}\BibitemShut {NoStop}%
\bibitem [{\citenamefont {Golze}\ \emph {et~al.}(2019)\citenamefont {Golze},
  \citenamefont {Dvorak},\ and\ \citenamefont {Rinke}}]{golze2019}%
  \BibitemOpen
  \bibfield  {author} {\bibinfo {author} {\bibfnamefont {D.}~\bibnamefont
  {Golze}}, \bibinfo {author} {\bibfnamefont {M.}~\bibnamefont {Dvorak}}, \
  and\ \bibinfo {author} {\bibfnamefont {P.}~\bibnamefont {Rinke}},\ }\href
  {\doibase 10.3389/fchem.2019.00377} {\bibfield  {journal} {\bibinfo
  {journal} {Frontiers in Chemistry}\ }\textbf {\bibinfo {volume} {7}},\
  \bibinfo {pages} {377} (\bibinfo {year} {2019})}\BibitemShut {NoStop}%
\bibitem [{\citenamefont {Choi}\ \emph {et~al.}(2019)\citenamefont {Choi},
  \citenamefont {Semon}, \citenamefont {Kang}, \citenamefont {Kutepov},\ and\
  \citenamefont {Kotliar}}]{choi2019}%
  \BibitemOpen
  \bibfield  {author} {\bibinfo {author} {\bibfnamefont {S.}~\bibnamefont
  {Choi}}, \bibinfo {author} {\bibfnamefont {P.}~\bibnamefont {Semon}},
  \bibinfo {author} {\bibfnamefont {B.}~\bibnamefont {Kang}}, \bibinfo {author}
  {\bibfnamefont {A.}~\bibnamefont {Kutepov}}, \ and\ \bibinfo {author}
  {\bibfnamefont {G.}~\bibnamefont {Kotliar}},\ }\href {\doibase
  10.1016/j.cpc.2019.07.003} {\bibfield  {journal} {\bibinfo  {journal}
  {Computer Physics Communications}\ }\textbf {\bibinfo {volume} {244}},\
  \bibinfo {pages} {277} (\bibinfo {year} {2019})}\BibitemShut {NoStop}%
\bibitem [{\citenamefont {Tomczak}\ \emph {et~al.}(2012)\citenamefont
  {Tomczak}, \citenamefont {Casula}, \citenamefont {Miyake}, \citenamefont
  {Aryasetiawan},\ and\ \citenamefont {Biermann}}]{tomczak2012}%
  \BibitemOpen
  \bibfield  {author} {\bibinfo {author} {\bibfnamefont {J.~M.}\ \bibnamefont
  {Tomczak}}, \bibinfo {author} {\bibfnamefont {M.}~\bibnamefont {Casula}},
  \bibinfo {author} {\bibfnamefont {T.}~\bibnamefont {Miyake}}, \bibinfo
  {author} {\bibfnamefont {F.}~\bibnamefont {Aryasetiawan}}, \ and\ \bibinfo
  {author} {\bibfnamefont {S.}~\bibnamefont {Biermann}},\ }\href {\doibase
  10.1209/0295-5075/100/67001} {\bibfield  {journal} {\bibinfo  {journal} {EPL
  (Europhysics Letters)}\ }\textbf {\bibinfo {volume} {100}},\ \bibinfo {pages}
  {67001} (\bibinfo {year} {2012})}\BibitemShut {NoStop}%
\bibitem [{\citenamefont {Aryasetiawan}\ \emph {et~al.}(2004)\citenamefont
  {Aryasetiawan}, \citenamefont {Imada}, \citenamefont {Georges}, \citenamefont
  {Kotliar}, \citenamefont {Biermann},\ and\ \citenamefont
  {Lichtenstein}}]{aryasetiawan2004}%
  \BibitemOpen
  \bibfield  {author} {\bibinfo {author} {\bibfnamefont {F.}~\bibnamefont
  {Aryasetiawan}}, \bibinfo {author} {\bibfnamefont {M.}~\bibnamefont {Imada}},
  \bibinfo {author} {\bibfnamefont {A.}~\bibnamefont {Georges}}, \bibinfo
  {author} {\bibfnamefont {G.}~\bibnamefont {Kotliar}}, \bibinfo {author}
  {\bibfnamefont {S.}~\bibnamefont {Biermann}}, \ and\ \bibinfo {author}
  {\bibfnamefont {A.~I.}\ \bibnamefont {Lichtenstein}},\ }\href {\doibase
  10.1103/PhysRevB.70.195104} {\bibfield  {journal} {\bibinfo  {journal}
  {Physical Review B}\ }\textbf {\bibinfo {volume} {70}},\ \bibinfo {pages}
  {195104} (\bibinfo {year} {2004})}\BibitemShut {NoStop}%
\bibitem [{\citenamefont {Aryasetiawan}\ \emph {et~al.}(2009)\citenamefont
  {Aryasetiawan}, \citenamefont {Tomczak}, \citenamefont {Miyake},\ and\
  \citenamefont {Sakuma}}]{aryasetiawan2009}%
  \BibitemOpen
  \bibfield  {author} {\bibinfo {author} {\bibfnamefont {F.}~\bibnamefont
  {Aryasetiawan}}, \bibinfo {author} {\bibfnamefont {J.~M.}\ \bibnamefont
  {Tomczak}}, \bibinfo {author} {\bibfnamefont {T.}~\bibnamefont {Miyake}}, \
  and\ \bibinfo {author} {\bibfnamefont {R.}~\bibnamefont {Sakuma}},\ }\href
  {\doibase 10.1103/PhysRevLett.102.176402} {\bibfield  {journal} {\bibinfo
  {journal} {Physical Review Letters}\ }\textbf {\bibinfo {volume} {102}},\
  \bibinfo {pages} {176402} (\bibinfo {year} {2009})}\BibitemShut {NoStop}%
\bibitem [{\citenamefont {Miyake}\ and\ \citenamefont
  {Aryasetiawan}(2008)}]{miyake2008}%
  \BibitemOpen
  \bibfield  {author} {\bibinfo {author} {\bibfnamefont {T.}~\bibnamefont
  {Miyake}}\ and\ \bibinfo {author} {\bibfnamefont {F.}~\bibnamefont
  {Aryasetiawan}},\ }\href {\doibase 10.1103/PhysRevB.77.085122} {\bibfield
  {journal} {\bibinfo  {journal} {Physical Review B}\ }\textbf {\bibinfo
  {volume} {77}},\ \bibinfo {pages} {085122} (\bibinfo {year}
  {2008})}\BibitemShut {NoStop}%
\bibitem [{\citenamefont {Hampel}\ \emph {et~al.}(2020)\citenamefont {Hampel},
  \citenamefont {Beck},\ and\ \citenamefont {Ederer}}]{hampel2020}%
  \BibitemOpen
  \bibfield  {author} {\bibinfo {author} {\bibfnamefont {A.}~\bibnamefont
  {Hampel}}, \bibinfo {author} {\bibfnamefont {S.}~\bibnamefont {Beck}}, \ and\
  \bibinfo {author} {\bibfnamefont {C.}~\bibnamefont {Ederer}},\ }\href
  {\doibase 10.1103/PhysRevResearch.2.033088} {\bibfield  {journal} {\bibinfo
  {journal} {Physical Review Research}\ }\textbf {\bibinfo {volume} {2}},\
  \bibinfo {pages} {033088} (\bibinfo {year} {2020})}\BibitemShut {NoStop}%
\bibitem [{\citenamefont {Bhandary}\ and\ \citenamefont
  {Held}(2021)}]{bhandary2021}%
  \BibitemOpen
  \bibfield  {author} {\bibinfo {author} {\bibfnamefont {S.}~\bibnamefont
  {Bhandary}}\ and\ \bibinfo {author} {\bibfnamefont {K.}~\bibnamefont
  {Held}},\ }\href {\doibase 10.1103/PhysRevB.103.245116} {\bibfield  {journal}
  {\bibinfo  {journal} {Physical Review B}\ }\textbf {\bibinfo {volume}
  {103}},\ \bibinfo {pages} {245116} (\bibinfo {year} {2021})}\BibitemShut
  {NoStop}%
\bibitem [{\citenamefont {Lee}\ and\ \citenamefont {Haule}(2017)}]{lee2017}%
  \BibitemOpen
  \bibfield  {author} {\bibinfo {author} {\bibfnamefont {J.}~\bibnamefont
  {Lee}}\ and\ \bibinfo {author} {\bibfnamefont {K.}~\bibnamefont {Haule}},\
  }\href {\doibase 10.1103/PhysRevB.95.155104} {\bibfield  {journal} {\bibinfo
  {journal} {Physical Review B}\ }\textbf {\bibinfo {volume} {95}},\ \bibinfo
  {pages} {155104} (\bibinfo {year} {2017})}\BibitemShut {NoStop}%
\bibitem [{\citenamefont {Eidelstein}\ \emph {et~al.}(2020)\citenamefont
  {Eidelstein}, \citenamefont {Gull},\ and\ \citenamefont
  {Cohen}}]{eidelstein2020}%
  \BibitemOpen
  \bibfield  {author} {\bibinfo {author} {\bibfnamefont {E.}~\bibnamefont
  {Eidelstein}}, \bibinfo {author} {\bibfnamefont {E.}~\bibnamefont {Gull}}, \
  and\ \bibinfo {author} {\bibfnamefont {G.}~\bibnamefont {Cohen}},\ }\href
  {\doibase 10.1103/PhysRevLett.124.206405} {\bibfield  {journal} {\bibinfo
  {journal} {Physical Review Letters}\ }\textbf {\bibinfo {volume} {124}},\
  \bibinfo {pages} {206405} (\bibinfo {year} {2020})}\BibitemShut {NoStop}%
\bibitem [{\citenamefont {Seth}\ \emph {et~al.}(2016)\citenamefont {Seth},
  \citenamefont {Krivenko}, \citenamefont {Ferrero},\ and\ \citenamefont
  {Parcollet}}]{seth2016}%
  \BibitemOpen
  \bibfield  {author} {\bibinfo {author} {\bibfnamefont {P.}~\bibnamefont
  {Seth}}, \bibinfo {author} {\bibfnamefont {I.}~\bibnamefont {Krivenko}},
  \bibinfo {author} {\bibfnamefont {M.}~\bibnamefont {Ferrero}}, \ and\
  \bibinfo {author} {\bibfnamefont {O.}~\bibnamefont {Parcollet}},\ }\href
  {\doibase 10.1016/j.cpc.2015.10.023} {\bibfield  {journal} {\bibinfo
  {journal} {Computer Physics Communications}\ }\textbf {\bibinfo {volume}
  {200}},\ \bibinfo {pages} {274} (\bibinfo {year} {2016})}\BibitemShut
  {NoStop}%
\bibitem [{\citenamefont {Werner}\ and\ \citenamefont
  {Millis}(2010)}]{werner2010}%
  \BibitemOpen
  \bibfield  {author} {\bibinfo {author} {\bibfnamefont {P.}~\bibnamefont
  {Werner}}\ and\ \bibinfo {author} {\bibfnamefont {A.~J.}\ \bibnamefont
  {Millis}},\ }\href {\doibase 10.1103/PhysRevLett.104.146401} {\bibfield
  {journal} {\bibinfo  {journal} {Physical Review Letters}\ }\textbf {\bibinfo
  {volume} {104}},\ \bibinfo {pages} {146401} (\bibinfo {year}
  {2010})}\BibitemShut {NoStop}%
\bibitem [{\citenamefont {Medvedeva}\ \emph {et~al.}(2017)\citenamefont
  {Medvedeva}, \citenamefont {Iskakov}, \citenamefont {Krien}, \citenamefont
  {Mazurenko},\ and\ \citenamefont {Lichtenstein}}]{medvedeva2017}%
  \BibitemOpen
  \bibfield  {author} {\bibinfo {author} {\bibfnamefont {D.}~\bibnamefont
  {Medvedeva}}, \bibinfo {author} {\bibfnamefont {S.}~\bibnamefont {Iskakov}},
  \bibinfo {author} {\bibfnamefont {F.}~\bibnamefont {Krien}}, \bibinfo
  {author} {\bibfnamefont {V.~V.}\ \bibnamefont {Mazurenko}}, \ and\ \bibinfo
  {author} {\bibfnamefont {A.~I.}\ \bibnamefont {Lichtenstein}},\ }\href
  {\doibase 10.1103/PhysRevB.96.235149} {\bibfield  {journal} {\bibinfo
  {journal} {Physical Review B}\ }\textbf {\bibinfo {volume} {96}},\ \bibinfo
  {pages} {235149} (\bibinfo {year} {2017})}\BibitemShut {NoStop}%
\bibitem [{\citenamefont {Werner}\ and\ \citenamefont
  {Casula}(2016)}]{werner2016}%
  \BibitemOpen
  \bibfield  {author} {\bibinfo {author} {\bibfnamefont {P.}~\bibnamefont
  {Werner}}\ and\ \bibinfo {author} {\bibfnamefont {M.}~\bibnamefont
  {Casula}},\ }\href {\doibase 10.1088/0953-8984/28/38/383001} {\bibfield
  {journal} {\bibinfo  {journal} {Journal of Physics: Condensed Matter}\
  }\textbf {\bibinfo {volume} {28}},\ \bibinfo {pages} {383001} (\bibinfo
  {year} {2016})}\BibitemShut {NoStop}%
\bibitem [{\citenamefont {Adler}\ \emph {et~al.}(2018)\citenamefont {Adler},
  \citenamefont {Kang}, \citenamefont {Yee},\ and\ \citenamefont
  {Kotliar}}]{adler2018}%
  \BibitemOpen
  \bibfield  {author} {\bibinfo {author} {\bibfnamefont {R.}~\bibnamefont
  {Adler}}, \bibinfo {author} {\bibfnamefont {C.-J.}\ \bibnamefont {Kang}},
  \bibinfo {author} {\bibfnamefont {C.-H.}\ \bibnamefont {Yee}}, \ and\
  \bibinfo {author} {\bibfnamefont {G.}~\bibnamefont {Kotliar}},\ }\href
  {\doibase 10.1088/1361-6633/aadca4} {\bibfield  {journal} {\bibinfo
  {journal} {Reports on Progress in Physics}\ }\textbf {\bibinfo {volume}
  {82}},\ \bibinfo {pages} {012504} (\bibinfo {year} {2018})}\BibitemShut
  {NoStop}%
\bibitem [{\citenamefont {Haule}(2015)}]{haule2015}%
  \BibitemOpen
  \bibfield  {author} {\bibinfo {author} {\bibfnamefont {K.}~\bibnamefont
  {Haule}},\ }\href {\doibase 10.1103/PhysRevLett.115.196403} {\bibfield
  {journal} {\bibinfo  {journal} {Physical Review Letters}\ }\textbf {\bibinfo
  {volume} {115}},\ \bibinfo {pages} {196403} (\bibinfo {year}
  {2015})}\BibitemShut {NoStop}%
\bibitem [{\citenamefont {Haule}\ \emph {et~al.}(2010)\citenamefont {Haule},
  \citenamefont {Yee},\ and\ \citenamefont {Kim}}]{haule2010}%
  \BibitemOpen
  \bibfield  {author} {\bibinfo {author} {\bibfnamefont {K.}~\bibnamefont
  {Haule}}, \bibinfo {author} {\bibfnamefont {C.-H.}\ \bibnamefont {Yee}}, \
  and\ \bibinfo {author} {\bibfnamefont {K.}~\bibnamefont {Kim}},\ }\href
  {\doibase 10.1103/PhysRevB.81.195107} {\bibfield  {journal} {\bibinfo
  {journal} {Physical Review B}\ }\textbf {\bibinfo {volume} {81}},\ \bibinfo
  {pages} {195107} (\bibinfo {year} {2010})}\BibitemShut {NoStop}%
\bibitem [{\citenamefont {Haule}\ \emph {et~al.}(2014)\citenamefont {Haule},
  \citenamefont {Birol},\ and\ \citenamefont {Kotliar}}]{haule2014}%
  \BibitemOpen
  \bibfield  {author} {\bibinfo {author} {\bibfnamefont {K.}~\bibnamefont
  {Haule}}, \bibinfo {author} {\bibfnamefont {T.}~\bibnamefont {Birol}}, \ and\
  \bibinfo {author} {\bibfnamefont {G.}~\bibnamefont {Kotliar}},\ }\href
  {\doibase 10.1103/PhysRevB.90.075136} {\bibfield  {journal} {\bibinfo
  {journal} {Physical Review B}\ }\textbf {\bibinfo {volume} {90}},\ \bibinfo
  {pages} {075136} (\bibinfo {year} {2014})}\BibitemShut {NoStop}%
\bibitem [{\citenamefont {{van Roekeghem}}\ \emph {et~al.}(2014)\citenamefont
  {{van Roekeghem}}, \citenamefont {Ayral}, \citenamefont {Tomczak},
  \citenamefont {Casula}, \citenamefont {Xu}, \citenamefont {Ding},
  \citenamefont {Ferrero}, \citenamefont {Parcollet}, \citenamefont {Jiang},\
  and\ \citenamefont {Biermann}}]{vanroekeghem2014}%
  \BibitemOpen
  \bibfield  {author} {\bibinfo {author} {\bibfnamefont {A.}~\bibnamefont {{van
  Roekeghem}}}, \bibinfo {author} {\bibfnamefont {T.}~\bibnamefont {Ayral}},
  \bibinfo {author} {\bibfnamefont {J.~M.}\ \bibnamefont {Tomczak}}, \bibinfo
  {author} {\bibfnamefont {M.}~\bibnamefont {Casula}}, \bibinfo {author}
  {\bibfnamefont {N.}~\bibnamefont {Xu}}, \bibinfo {author} {\bibfnamefont
  {H.}~\bibnamefont {Ding}}, \bibinfo {author} {\bibfnamefont {M.}~\bibnamefont
  {Ferrero}}, \bibinfo {author} {\bibfnamefont {O.}~\bibnamefont {Parcollet}},
  \bibinfo {author} {\bibfnamefont {H.}~\bibnamefont {Jiang}}, \ and\ \bibinfo
  {author} {\bibfnamefont {S.}~\bibnamefont {Biermann}},\ }\href {\doibase
  10.1103/PhysRevLett.113.266403} {\bibfield  {journal} {\bibinfo  {journal}
  {Physical Review Letters}\ }\textbf {\bibinfo {volume} {113}},\ \bibinfo
  {pages} {266403} (\bibinfo {year} {2014})}\BibitemShut {NoStop}%
\bibitem [{\citenamefont {Yeh}\ \emph {et~al.}(2021)\citenamefont {Yeh},
  \citenamefont {Iskakov}, \citenamefont {Zgid},\ and\ \citenamefont
  {Gull}}]{yeh_electron_2021}%
  \BibitemOpen
  \bibfield  {author} {\bibinfo {author} {\bibfnamefont {C.-N.}\ \bibnamefont
  {Yeh}}, \bibinfo {author} {\bibfnamefont {S.}~\bibnamefont {Iskakov}},
  \bibinfo {author} {\bibfnamefont {D.}~\bibnamefont {Zgid}}, \ and\ \bibinfo
  {author} {\bibfnamefont {E.}~\bibnamefont {Gull}},\ }\href {\doibase
  10.1103/PhysRevB.103.195149} {\bibfield  {journal} {\bibinfo  {journal}
  {Physical Review B}\ }\textbf {\bibinfo {volume} {103}},\ \bibinfo {pages}
  {195149} (\bibinfo {year} {2021})}\BibitemShut {NoStop}%
\bibitem [{\citenamefont {Iskakov}\ \emph {et~al.}(2021)\citenamefont
  {Iskakov}, \citenamefont {Yeh}, \citenamefont {Gull},\ and\ \citenamefont
  {Zgid}}]{iskakov_ab_2020}%
  \BibitemOpen
  \bibfield  {author} {\bibinfo {author} {\bibfnamefont {S.}~\bibnamefont
  {Iskakov}}, \bibinfo {author} {\bibfnamefont {C.-N.}\ \bibnamefont {Yeh}},
  \bibinfo {author} {\bibfnamefont {E.}~\bibnamefont {Gull}}, \ and\ \bibinfo
  {author} {\bibfnamefont {D.}~\bibnamefont {Zgid}},\ }\href {\doibase
  10.1103/PhysRevB.102.085105} {\bibfield  {journal} {\bibinfo  {journal}
  {Physical Review B}\ }\textbf {\bibinfo {volume} {102}},\ \bibinfo {pages}
  {085105} (\bibinfo {year} {2021})}\BibitemShut {NoStop}%
\bibitem [{\citenamefont {Kananenka}\ \emph {et~al.}(2015)\citenamefont
  {Kananenka}, \citenamefont {Gull},\ and\ \citenamefont
  {Zgid}}]{kananenka2015}%
  \BibitemOpen
  \bibfield  {author} {\bibinfo {author} {\bibfnamefont {A.~A.}\ \bibnamefont
  {Kananenka}}, \bibinfo {author} {\bibfnamefont {E.}~\bibnamefont {Gull}}, \
  and\ \bibinfo {author} {\bibfnamefont {D.}~\bibnamefont {Zgid}},\ }\href
  {\doibase 10.1103/PhysRevB.91.121111} {\bibfield  {journal} {\bibinfo
  {journal} {Physical Review B}\ }\textbf {\bibinfo {volume} {91}},\ \bibinfo
  {pages} {121111} (\bibinfo {year} {2015})}\BibitemShut {NoStop}%
\bibitem [{\citenamefont {Lan}\ \emph {et~al.}(2015)\citenamefont {Lan},
  \citenamefont {Kananenka},\ and\ \citenamefont {Zgid}}]{lan2015}%
  \BibitemOpen
  \bibfield  {author} {\bibinfo {author} {\bibfnamefont {T.~N.}\ \bibnamefont
  {Lan}}, \bibinfo {author} {\bibfnamefont {A.~A.}\ \bibnamefont {Kananenka}},
  \ and\ \bibinfo {author} {\bibfnamefont {D.}~\bibnamefont {Zgid}},\ }\href
  {\doibase 10.1063/1.4938562} {\bibfield  {journal} {\bibinfo  {journal} {The
  Journal of Chemical Physics}\ }\textbf {\bibinfo {volume} {143}},\ \bibinfo
  {pages} {241102} (\bibinfo {year} {2015})}\BibitemShut {NoStop}%
\bibitem [{\citenamefont {Lan}\ \emph {et~al.}(2017)\citenamefont {Lan},
  \citenamefont {Shee}, \citenamefont {Li}, \citenamefont {Gull},\ and\
  \citenamefont {Zgid}}]{lan2017a}%
  \BibitemOpen
  \bibfield  {author} {\bibinfo {author} {\bibfnamefont {T.~N.}\ \bibnamefont
  {Lan}}, \bibinfo {author} {\bibfnamefont {A.}~\bibnamefont {Shee}}, \bibinfo
  {author} {\bibfnamefont {J.}~\bibnamefont {Li}}, \bibinfo {author}
  {\bibfnamefont {E.}~\bibnamefont {Gull}}, \ and\ \bibinfo {author}
  {\bibfnamefont {D.}~\bibnamefont {Zgid}},\ }\href {\doibase
  10.1103/PhysRevB.96.155106} {\bibfield  {journal} {\bibinfo  {journal}
  {Physical Review B}\ }\textbf {\bibinfo {volume} {96}},\ \bibinfo {pages}
  {155106} (\bibinfo {year} {2017})}\BibitemShut {NoStop}%
\bibitem [{\citenamefont {Muechler}\ \emph {et~al.}(2021)\citenamefont
  {Muechler}, \citenamefont {Badrtdinov}, \citenamefont {Hampel}, \citenamefont
  {Cano}, \citenamefont {R{\"o}sner},\ and\ \citenamefont
  {Dreyer}}]{muechler2021}%
  \BibitemOpen
  \bibfield  {author} {\bibinfo {author} {\bibfnamefont {L.}~\bibnamefont
  {Muechler}}, \bibinfo {author} {\bibfnamefont {D.~I.}\ \bibnamefont
  {Badrtdinov}}, \bibinfo {author} {\bibfnamefont {A.}~\bibnamefont {Hampel}},
  \bibinfo {author} {\bibfnamefont {J.}~\bibnamefont {Cano}}, \bibinfo {author}
  {\bibfnamefont {M.}~\bibnamefont {R{\"o}sner}}, \ and\ \bibinfo {author}
  {\bibfnamefont {C.~E.}\ \bibnamefont {Dreyer}},\ }\href@noop {} {\bibfield
  {journal} {\bibinfo  {journal} {arXiv:2105.08705 [cond-mat]}\ } (\bibinfo
  {year} {2021})},\ \Eprint {http://arxiv.org/abs/2105.08705} {arXiv:2105.08705
  [cond-mat]} \BibitemShut {NoStop}%
\bibitem [{\citenamefont {Govoni}\ and\ \citenamefont
  {Galli}(2015)}]{govoni2015}%
  \BibitemOpen
  \bibfield  {author} {\bibinfo {author} {\bibfnamefont {M.}~\bibnamefont
  {Govoni}}\ and\ \bibinfo {author} {\bibfnamefont {G.}~\bibnamefont {Galli}},\
  }\href {\doibase 10.1021/ct500958p} {\bibfield  {journal} {\bibinfo
  {journal} {Journal of Chemical Theory and Computation}\ }\textbf {\bibinfo
  {volume} {11}},\ \bibinfo {pages} {2680} (\bibinfo {year}
  {2015})}\BibitemShut {NoStop}%
\bibitem [{\citenamefont {Scherpelz}\ \emph {et~al.}(2016)\citenamefont
  {Scherpelz}, \citenamefont {Govoni}, \citenamefont {Hamada},\ and\
  \citenamefont {Galli}}]{scherpelz2016}%
  \BibitemOpen
  \bibfield  {author} {\bibinfo {author} {\bibfnamefont {P.}~\bibnamefont
  {Scherpelz}}, \bibinfo {author} {\bibfnamefont {M.}~\bibnamefont {Govoni}},
  \bibinfo {author} {\bibfnamefont {I.}~\bibnamefont {Hamada}}, \ and\ \bibinfo
  {author} {\bibfnamefont {G.}~\bibnamefont {Galli}},\ }\href {\doibase
  10.1021/acs.jctc.6b00114} {\bibfield  {journal} {\bibinfo  {journal} {Journal
  of Chemical Theory and Computation}\ }\textbf {\bibinfo {volume} {12}},\
  \bibinfo {pages} {3523} (\bibinfo {year} {2016})}\BibitemShut {NoStop}%
\bibitem [{\citenamefont {Govoni}\ and\ \citenamefont
  {Galli}(2018)}]{govoni2018}%
  \BibitemOpen
  \bibfield  {author} {\bibinfo {author} {\bibfnamefont {M.}~\bibnamefont
  {Govoni}}\ and\ \bibinfo {author} {\bibfnamefont {G.}~\bibnamefont {Galli}},\
  }\href {\doibase 10.1021/acs.jctc.7b00952} {\bibfield  {journal} {\bibinfo
  {journal} {Journal of Chemical Theory and Computation}\ }\textbf {\bibinfo
  {volume} {14}},\ \bibinfo {pages} {1895} (\bibinfo {year}
  {2018})}\BibitemShut {NoStop}%
\bibitem [{\citenamefont {Govoni}\ \emph {et~al.}(2021)\citenamefont {Govoni},
  \citenamefont {Whitmer}, \citenamefont {{de Pablo}}, \citenamefont {Gygi},\
  and\ \citenamefont {Galli}}]{govoni2021}%
  \BibitemOpen
  \bibfield  {author} {\bibinfo {author} {\bibfnamefont {M.}~\bibnamefont
  {Govoni}}, \bibinfo {author} {\bibfnamefont {J.}~\bibnamefont {Whitmer}},
  \bibinfo {author} {\bibfnamefont {J.}~\bibnamefont {{de Pablo}}}, \bibinfo
  {author} {\bibfnamefont {F.}~\bibnamefont {Gygi}}, \ and\ \bibinfo {author}
  {\bibfnamefont {G.}~\bibnamefont {Galli}},\ }\href {\doibase
  10.1038/s41524-021-00501-z} {\bibfield  {journal} {\bibinfo  {journal} {npj
  Computational Materials}\ }\textbf {\bibinfo {volume} {7}},\ \bibinfo {pages}
  {1} (\bibinfo {year} {2021})}\BibitemShut {NoStop}%
\bibitem [{\citenamefont {Sheng}\ \emph {et~al.}(2022)\citenamefont {Sheng},
  \citenamefont {Vorwerk}, \citenamefont {Govoni},\ and\ \citenamefont
  {Galli}}]{sheng2022}%
  \BibitemOpen
  \bibfield  {author} {\bibinfo {author} {\bibfnamefont {N.}~\bibnamefont
  {Sheng}}, \bibinfo {author} {\bibfnamefont {C.}~\bibnamefont {Vorwerk}},
  \bibinfo {author} {\bibfnamefont {M.}~\bibnamefont {Govoni}}, \ and\ \bibinfo
  {author} {\bibfnamefont {G.}~\bibnamefont {Galli}},\ }\href@noop {}
  {\bibfield  {journal} {\bibinfo  {journal} {arXiv:2203.05493 [cond-mat,
  physics:physics, physics:quant-ph]}\ } (\bibinfo {year} {2022})},\ \Eprint
  {http://arxiv.org/abs/2203.05493} {arXiv:2203.05493 [cond-mat,
  physics:physics, physics:quant-ph]} \BibitemShut {NoStop}%
\bibitem [{\citenamefont {Casula}\ \emph {et~al.}(2012)\citenamefont {Casula},
  \citenamefont {Rubtsov},\ and\ \citenamefont {Biermann}}]{casula2012}%
  \BibitemOpen
  \bibfield  {author} {\bibinfo {author} {\bibfnamefont {M.}~\bibnamefont
  {Casula}}, \bibinfo {author} {\bibfnamefont {A.}~\bibnamefont {Rubtsov}}, \
  and\ \bibinfo {author} {\bibfnamefont {S.}~\bibnamefont {Biermann}},\ }\href
  {\doibase 10.1103/PhysRevB.85.035115} {\bibfield  {journal} {\bibinfo
  {journal} {Physical Review B}\ }\textbf {\bibinfo {volume} {85}},\ \bibinfo
  {pages} {035115} (\bibinfo {year} {2012})}\BibitemShut {NoStop}%
\bibitem [{\citenamefont {Krivenko}\ and\ \citenamefont
  {Biermann}(2015)}]{krivenko2015}%
  \BibitemOpen
  \bibfield  {author} {\bibinfo {author} {\bibfnamefont {I.~S.}\ \bibnamefont
  {Krivenko}}\ and\ \bibinfo {author} {\bibfnamefont {S.}~\bibnamefont
  {Biermann}},\ }\href {\doibase 10.1103/PhysRevB.91.155149} {\bibfield
  {journal} {\bibinfo  {journal} {Physical Review B}\ }\textbf {\bibinfo
  {volume} {91}},\ \bibinfo {pages} {155149} (\bibinfo {year}
  {2015})}\BibitemShut {NoStop}%
\bibitem [{\citenamefont {Nomura}\ \emph {et~al.}(2014)\citenamefont {Nomura},
  \citenamefont {Sakai},\ and\ \citenamefont {Arita}}]{nomura2014}%
  \BibitemOpen
  \bibfield  {author} {\bibinfo {author} {\bibfnamefont {Y.}~\bibnamefont
  {Nomura}}, \bibinfo {author} {\bibfnamefont {S.}~\bibnamefont {Sakai}}, \
  and\ \bibinfo {author} {\bibfnamefont {R.}~\bibnamefont {Arita}},\ }\href
  {\doibase 10.1103/PhysRevB.89.195146} {\bibfield  {journal} {\bibinfo
  {journal} {Physical Review B}\ }\textbf {\bibinfo {volume} {89}},\ \bibinfo
  {pages} {195146} (\bibinfo {year} {2014})}\BibitemShut {NoStop}%
\bibitem [{\citenamefont {Mizuno}\ \emph {et~al.}(2021)\citenamefont {Mizuno},
  \citenamefont {Ochi},\ and\ \citenamefont {Kuroki}}]{mizuno2021}%
  \BibitemOpen
  \bibfield  {author} {\bibinfo {author} {\bibfnamefont {R.}~\bibnamefont
  {Mizuno}}, \bibinfo {author} {\bibfnamefont {M.}~\bibnamefont {Ochi}}, \ and\
  \bibinfo {author} {\bibfnamefont {K.}~\bibnamefont {Kuroki}},\ }\href
  {\doibase 10.1103/PhysRevB.104.035160} {\bibfield  {journal} {\bibinfo
  {journal} {Physical Review B}\ }\textbf {\bibinfo {volume} {104}},\ \bibinfo
  {pages} {035160} (\bibinfo {year} {2021})}\BibitemShut {NoStop}%
\bibitem [{\citenamefont {Kotliar}\ \emph {et~al.}(2001)\citenamefont
  {Kotliar}, \citenamefont {Savrasov}, \citenamefont {P{\'a}lsson},\ and\
  \citenamefont {Biroli}}]{kotliar2001}%
  \BibitemOpen
  \bibfield  {author} {\bibinfo {author} {\bibfnamefont {G.}~\bibnamefont
  {Kotliar}}, \bibinfo {author} {\bibfnamefont {S.~Y.}\ \bibnamefont
  {Savrasov}}, \bibinfo {author} {\bibfnamefont {G.}~\bibnamefont
  {P{\'a}lsson}}, \ and\ \bibinfo {author} {\bibfnamefont {G.}~\bibnamefont
  {Biroli}},\ }\href {\doibase 10.1103/PhysRevLett.87.186401} {\bibfield
  {journal} {\bibinfo  {journal} {Physical Review Letters}\ }\textbf {\bibinfo
  {volume} {87}},\ \bibinfo {pages} {186401} (\bibinfo {year}
  {2001})}\BibitemShut {NoStop}%
\bibitem [{\citenamefont {De~Leo}\ \emph {et~al.}(2008)\citenamefont {De~Leo},
  \citenamefont {Civelli},\ and\ \citenamefont {Kotliar}}]{deleo2008}%
  \BibitemOpen
  \bibfield  {author} {\bibinfo {author} {\bibfnamefont {L.}~\bibnamefont
  {De~Leo}}, \bibinfo {author} {\bibfnamefont {M.}~\bibnamefont {Civelli}}, \
  and\ \bibinfo {author} {\bibfnamefont {G.}~\bibnamefont {Kotliar}},\ }\href
  {\doibase 10.1103/PhysRevB.77.075107} {\bibfield  {journal} {\bibinfo
  {journal} {Physical Review B}\ }\textbf {\bibinfo {volume} {77}},\ \bibinfo
  {pages} {075107} (\bibinfo {year} {2008})}\BibitemShut {NoStop}%
\bibitem [{\citenamefont {Gull}\ \emph {et~al.}(2011)\citenamefont {Gull},
  \citenamefont {Staar}, \citenamefont {Fuchs}, \citenamefont {Nukala},
  \citenamefont {Summers}, \citenamefont {Pruschke}, \citenamefont
  {Schulthess},\ and\ \citenamefont {Maier}}]{gull_submatrix_2011}%
  \BibitemOpen
  \bibfield  {author} {\bibinfo {author} {\bibfnamefont {E.}~\bibnamefont
  {Gull}}, \bibinfo {author} {\bibfnamefont {P.}~\bibnamefont {Staar}},
  \bibinfo {author} {\bibfnamefont {S.}~\bibnamefont {Fuchs}}, \bibinfo
  {author} {\bibfnamefont {P.}~\bibnamefont {Nukala}}, \bibinfo {author}
  {\bibfnamefont {M.~S.}\ \bibnamefont {Summers}}, \bibinfo {author}
  {\bibfnamefont {T.}~\bibnamefont {Pruschke}}, \bibinfo {author}
  {\bibfnamefont {T.~C.}\ \bibnamefont {Schulthess}}, \ and\ \bibinfo {author}
  {\bibfnamefont {T.}~\bibnamefont {Maier}},\ }\href {\doibase
  10.1103/PhysRevB.83.075122} {\bibfield  {journal} {\bibinfo  {journal}
  {Physical Review B}\ }\textbf {\bibinfo {volume} {83}},\ \bibinfo {pages}
  {075122} (\bibinfo {year} {2011})}\BibitemShut {NoStop}%
\bibitem [{\citenamefont {LeBlanc}\ \emph {et~al.}()\citenamefont {LeBlanc},
  \citenamefont {Antipov}, \citenamefont {Becca}, \citenamefont {Bulik},
  \citenamefont {Chan}, \citenamefont {Chung}, \citenamefont {Deng},
  \citenamefont {Ferrero}, \citenamefont {Henderson}, \citenamefont
  {Jiménez-Hoyos}, \citenamefont {Kozik}, \citenamefont {Liu}, \citenamefont
  {Millis}, \citenamefont {Prokof’ev}, \citenamefont {Qin}, \citenamefont
  {Scuseria}, \citenamefont {Shi}, \citenamefont {Svistunov}, \citenamefont
  {Tocchio}, \citenamefont {Tupitsyn}, \citenamefont {White}, \citenamefont
  {Zhang}, \citenamefont {Zheng}, \citenamefont {Zhu}, \citenamefont {Gull},\
  and\ \citenamefont {{Simons Collaboration on the Many-Electron
  Problem}}}]{leblanc_solutions_2015}%
  \BibitemOpen
  \bibfield  {author} {\bibinfo {author} {\bibfnamefont {J.}~\bibnamefont
  {LeBlanc}}, \bibinfo {author} {\bibfnamefont {A.~E.}\ \bibnamefont
  {Antipov}}, \bibinfo {author} {\bibfnamefont {F.}~\bibnamefont {Becca}},
  \bibinfo {author} {\bibfnamefont {I.~W.}\ \bibnamefont {Bulik}}, \bibinfo
  {author} {\bibfnamefont {G.~K.-L.}\ \bibnamefont {Chan}}, \bibinfo {author}
  {\bibfnamefont {C.-M.}\ \bibnamefont {Chung}}, \bibinfo {author}
  {\bibfnamefont {Y.}~\bibnamefont {Deng}}, \bibinfo {author} {\bibfnamefont
  {M.}~\bibnamefont {Ferrero}}, \bibinfo {author} {\bibfnamefont {T.~M.}\
  \bibnamefont {Henderson}}, \bibinfo {author} {\bibfnamefont {C.~A.}\
  \bibnamefont {Jiménez-Hoyos}}, \bibinfo {author} {\bibfnamefont
  {E.}~\bibnamefont {Kozik}}, \bibinfo {author} {\bibfnamefont {X.-W.}\
  \bibnamefont {Liu}}, \bibinfo {author} {\bibfnamefont {A.~J.}\ \bibnamefont
  {Millis}}, \bibinfo {author} {\bibfnamefont {N.}~\bibnamefont {Prokof’ev}},
  \bibinfo {author} {\bibfnamefont {M.}~\bibnamefont {Qin}}, \bibinfo {author}
  {\bibfnamefont {G.~E.}\ \bibnamefont {Scuseria}}, \bibinfo {author}
  {\bibfnamefont {H.}~\bibnamefont {Shi}}, \bibinfo {author} {\bibfnamefont
  {B.}~\bibnamefont {Svistunov}}, \bibinfo {author} {\bibfnamefont {L.~F.}\
  \bibnamefont {Tocchio}}, \bibinfo {author} {\bibfnamefont {I.}~\bibnamefont
  {Tupitsyn}}, \bibinfo {author} {\bibfnamefont {S.~R.}\ \bibnamefont {White}},
  \bibinfo {author} {\bibfnamefont {S.}~\bibnamefont {Zhang}}, \bibinfo
  {author} {\bibfnamefont {B.-X.}\ \bibnamefont {Zheng}}, \bibinfo {author}
  {\bibfnamefont {Z.}~\bibnamefont {Zhu}}, \bibinfo {author} {\bibfnamefont
  {E.}~\bibnamefont {Gull}}, \ and\ \bibinfo {author} {\bibnamefont {{Simons
  Collaboration on the Many-Electron Problem}}},\ }\href {\doibase
  10.1103/PhysRevX.5.041041} {\ \textbf {\bibinfo {volume} {5}},\ \bibinfo
  {pages} {041041}}\BibitemShut {NoStop}%
\bibitem [{\citenamefont {Jamet}\ \emph {et~al.}(2021)\citenamefont {Jamet},
  \citenamefont {Agarwal}, \citenamefont {Lupo}, \citenamefont {Browne},
  \citenamefont {Weber},\ and\ \citenamefont {Rungger}}]{jamet2021}%
  \BibitemOpen
  \bibfield  {author} {\bibinfo {author} {\bibfnamefont {F.}~\bibnamefont
  {Jamet}}, \bibinfo {author} {\bibfnamefont {A.}~\bibnamefont {Agarwal}},
  \bibinfo {author} {\bibfnamefont {C.}~\bibnamefont {Lupo}}, \bibinfo {author}
  {\bibfnamefont {D.~E.}\ \bibnamefont {Browne}}, \bibinfo {author}
  {\bibfnamefont {C.}~\bibnamefont {Weber}}, \ and\ \bibinfo {author}
  {\bibfnamefont {I.}~\bibnamefont {Rungger}},\ }\href@noop {} {\bibfield
  {journal} {\bibinfo  {journal} {arXiv:2105.13298 [quant-ph]}\ } (\bibinfo
  {year} {2021})},\ \Eprint {http://arxiv.org/abs/2105.13298} {arXiv:2105.13298
  [quant-ph]} \BibitemShut {NoStop}%
\bibitem [{\citenamefont {Wecker}\ \emph {et~al.}(2015)\citenamefont {Wecker},
  \citenamefont {Hastings}, \citenamefont {Wiebe}, \citenamefont {Clark},
  \citenamefont {Nayak},\ and\ \citenamefont {Troyer}}]{wecker2015}%
  \BibitemOpen
  \bibfield  {author} {\bibinfo {author} {\bibfnamefont {D.}~\bibnamefont
  {Wecker}}, \bibinfo {author} {\bibfnamefont {M.~B.}\ \bibnamefont
  {Hastings}}, \bibinfo {author} {\bibfnamefont {N.}~\bibnamefont {Wiebe}},
  \bibinfo {author} {\bibfnamefont {B.~K.}\ \bibnamefont {Clark}}, \bibinfo
  {author} {\bibfnamefont {C.}~\bibnamefont {Nayak}}, \ and\ \bibinfo {author}
  {\bibfnamefont {M.}~\bibnamefont {Troyer}},\ }\href {\doibase
  10.1103/PhysRevA.92.062318} {\bibfield  {journal} {\bibinfo  {journal}
  {Physical Review A}\ }\textbf {\bibinfo {volume} {92}},\ \bibinfo {pages}
  {062318} (\bibinfo {year} {2015})}\BibitemShut {NoStop}%
\bibitem [{\citenamefont {Huang}\ \emph {et~al.}(2022)\citenamefont {Huang},
  \citenamefont {Govoni},\ and\ \citenamefont {Galli}}]{huang2022}%
  \BibitemOpen
  \bibfield  {author} {\bibinfo {author} {\bibfnamefont {B.}~\bibnamefont
  {Huang}}, \bibinfo {author} {\bibfnamefont {M.}~\bibnamefont {Govoni}}, \
  and\ \bibinfo {author} {\bibfnamefont {G.}~\bibnamefont {Galli}},\ }\href
  {\doibase 10.1103/PRXQuantum.3.010339} {\bibfield  {journal} {\bibinfo
  {journal} {PRX Quantum}\ }\textbf {\bibinfo {volume} {3}},\ \bibinfo {pages}
  {010339} (\bibinfo {year} {2022})}\BibitemShut {NoStop}%
\bibitem [{\citenamefont {McClean}\ \emph {et~al.}(2017)\citenamefont
  {McClean}, \citenamefont {{Kimchi-Schwartz}}, \citenamefont {Carter},\ and\
  \citenamefont {{de Jong}}}]{mcclean2017}%
  \BibitemOpen
  \bibfield  {author} {\bibinfo {author} {\bibfnamefont {J.~R.}\ \bibnamefont
  {McClean}}, \bibinfo {author} {\bibfnamefont {M.~E.}\ \bibnamefont
  {{Kimchi-Schwartz}}}, \bibinfo {author} {\bibfnamefont {J.}~\bibnamefont
  {Carter}}, \ and\ \bibinfo {author} {\bibfnamefont {W.~A.}\ \bibnamefont {{de
  Jong}}},\ }\href {\doibase 10.1103/PhysRevA.95.042308} {\bibfield  {journal}
  {\bibinfo  {journal} {Physical Review A}\ }\textbf {\bibinfo {volume} {95}},\
  \bibinfo {pages} {042308} (\bibinfo {year} {2017})}\BibitemShut {NoStop}%
\bibitem [{\citenamefont {Endo}\ \emph {et~al.}(2021)\citenamefont {Endo},
  \citenamefont {Cai}, \citenamefont {Benjamin},\ and\ \citenamefont
  {Yuan}}]{endo2021}%
  \BibitemOpen
  \bibfield  {author} {\bibinfo {author} {\bibfnamefont {S.}~\bibnamefont
  {Endo}}, \bibinfo {author} {\bibfnamefont {Z.}~\bibnamefont {Cai}}, \bibinfo
  {author} {\bibfnamefont {S.~C.}\ \bibnamefont {Benjamin}}, \ and\ \bibinfo
  {author} {\bibfnamefont {X.}~\bibnamefont {Yuan}},\ }\href {\doibase
  10.7566/JPSJ.90.032001} {\bibfield  {journal} {\bibinfo  {journal} {Journal
  of the Physical Society of Japan}\ }\textbf {\bibinfo {volume} {90}},\
  \bibinfo {pages} {032001} (\bibinfo {year} {2021})}\BibitemShut {NoStop}%
\bibitem [{\citenamefont {Bauer}\ \emph {et~al.}(2020)\citenamefont {Bauer},
  \citenamefont {Bravyi}, \citenamefont {Motta},\ and\ \citenamefont {{Kin-Lic
  Chan}}}]{bauer2020}%
  \BibitemOpen
  \bibfield  {author} {\bibinfo {author} {\bibfnamefont {B.}~\bibnamefont
  {Bauer}}, \bibinfo {author} {\bibfnamefont {S.}~\bibnamefont {Bravyi}},
  \bibinfo {author} {\bibfnamefont {M.}~\bibnamefont {Motta}}, \ and\ \bibinfo
  {author} {\bibfnamefont {G.}~\bibnamefont {{Kin-Lic Chan}}},\ }\href
  {\doibase 10.1021/acs.chemrev.9b00829} {\bibfield  {journal} {\bibinfo
  {journal} {{Chemical Reviews}}\ }\textbf {\bibinfo {volume} {120}},\ \bibinfo
  {pages} {12685} (\bibinfo {year} {2020})}\BibitemShut {NoStop}%
\bibitem [{\citenamefont {Korol}\ \emph {et~al.}(2021)\citenamefont {Korol},
  \citenamefont {Choo},\ and\ \citenamefont {Mezzacapo}}]{korol2021}%
  \BibitemOpen
  \bibfield  {author} {\bibinfo {author} {\bibfnamefont {K.~J.~M.}\
  \bibnamefont {Korol}}, \bibinfo {author} {\bibfnamefont {K.}~\bibnamefont
  {Choo}}, \ and\ \bibinfo {author} {\bibfnamefont {A.}~\bibnamefont
  {Mezzacapo}},\ }\href@noop {} {\bibfield  {journal} {\bibinfo  {journal}
  {arXiv:2111.08090 [cond-mat, physics:quant-ph]}\ } (\bibinfo {year}
  {2021})},\ \Eprint {http://arxiv.org/abs/2111.08090} {arXiv:2111.08090
  [cond-mat, physics:quant-ph]} \BibitemShut {NoStop}%
\bibitem [{\citenamefont {Wecker}\ \emph {et~al.}(2014)\citenamefont {Wecker},
  \citenamefont {Bauer}, \citenamefont {Clark}, \citenamefont {Hastings},\ and\
  \citenamefont {Troyer}}]{wecker2014}%
  \BibitemOpen
  \bibfield  {author} {\bibinfo {author} {\bibfnamefont {D.}~\bibnamefont
  {Wecker}}, \bibinfo {author} {\bibfnamefont {B.}~\bibnamefont {Bauer}},
  \bibinfo {author} {\bibfnamefont {B.~K.}\ \bibnamefont {Clark}}, \bibinfo
  {author} {\bibfnamefont {M.~B.}\ \bibnamefont {Hastings}}, \ and\ \bibinfo
  {author} {\bibfnamefont {M.}~\bibnamefont {Troyer}},\ }\href {\doibase
  10.1103/PhysRevA.90.022305} {\bibfield  {journal} {\bibinfo  {journal}
  {Physical Review A}\ }\textbf {\bibinfo {volume} {90}},\ \bibinfo {pages}
  {022305} (\bibinfo {year} {2014})}\BibitemShut {NoStop}%
\bibitem [{\citenamefont {Preskill}(2018)}]{preskill2018}%
  \BibitemOpen
  \bibfield  {author} {\bibinfo {author} {\bibfnamefont {J.}~\bibnamefont
  {Preskill}},\ }\href {\doibase 10.22331/q-2018-08-06-79} {\bibfield
  {journal} {\bibinfo  {journal} {Quantum}\ }\textbf {\bibinfo {volume} {2}},\
  \bibinfo {pages} {79} (\bibinfo {year} {2018})}\BibitemShut {NoStop}%
\bibitem [{\citenamefont {Lebreuilly}\ \emph {et~al.}(2021)\citenamefont
  {Lebreuilly}, \citenamefont {Noh}, \citenamefont {Wang}, \citenamefont
  {Girvin},\ and\ \citenamefont {Jiang}}]{lebreuilly_autonomous_2021}%
  \BibitemOpen
  \bibfield  {author} {\bibinfo {author} {\bibfnamefont {J.}~\bibnamefont
  {Lebreuilly}}, \bibinfo {author} {\bibfnamefont {K.}~\bibnamefont {Noh}},
  \bibinfo {author} {\bibfnamefont {C.-H.}\ \bibnamefont {Wang}}, \bibinfo
  {author} {\bibfnamefont {S.~M.}\ \bibnamefont {Girvin}}, \ and\ \bibinfo
  {author} {\bibfnamefont {L.}~\bibnamefont {Jiang}},\ }\href
  {http://arxiv.org/abs/2103.05007} {\bibfield  {journal} {\bibinfo  {journal}
  {arXiv:2103.05007 [quant-ph]}\ } (\bibinfo {year} {2021})}\BibitemShut
  {NoStop}%
\bibitem [{\citenamefont {Fedorov}\ \emph {et~al.}(2021)\citenamefont
  {Fedorov}, \citenamefont {Otten}, \citenamefont {Gray},\ and\ \citenamefont
  {Alexeev}}]{fedorov2021}%
  \BibitemOpen
  \bibfield  {author} {\bibinfo {author} {\bibfnamefont {D.~A.}\ \bibnamefont
  {Fedorov}}, \bibinfo {author} {\bibfnamefont {M.~J.}\ \bibnamefont {Otten}},
  \bibinfo {author} {\bibfnamefont {S.~K.}\ \bibnamefont {Gray}}, \ and\
  \bibinfo {author} {\bibfnamefont {Y.}~\bibnamefont {Alexeev}},\ }\href
  {\doibase 10.1063/5.0046930} {\bibfield  {journal} {\bibinfo  {journal} {The
  Journal of Chemical Physics}\ }\textbf {\bibinfo {volume} {154}},\ \bibinfo
  {pages} {164103} (\bibinfo {year} {2021})}\BibitemShut {NoStop}%
\bibitem [{\citenamefont {Macridin}\ \emph {et~al.}(2018)\citenamefont
  {Macridin}, \citenamefont {Spentzouris}, \citenamefont {Amundson},\ and\
  \citenamefont {Harnik}}]{macridin_electron-phonon_2018}%
  \BibitemOpen
  \bibfield  {author} {\bibinfo {author} {\bibfnamefont {A.}~\bibnamefont
  {Macridin}}, \bibinfo {author} {\bibfnamefont {P.}~\bibnamefont
  {Spentzouris}}, \bibinfo {author} {\bibfnamefont {J.}~\bibnamefont
  {Amundson}}, \ and\ \bibinfo {author} {\bibfnamefont {R.}~\bibnamefont
  {Harnik}},\ }\href {\doibase 10.1103/PhysRevLett.121.110504} {\bibfield
  {journal} {\bibinfo  {journal} {Physical Review Letters}\ }\textbf {\bibinfo
  {volume} {121}},\ \bibinfo {pages} {110504} (\bibinfo {year}
  {2018})}\BibitemShut {NoStop}%
\bibitem [{\citenamefont {Powers}\ \emph {et~al.}(2021)\citenamefont {Powers},
  \citenamefont {Bassman},\ and\ \citenamefont
  {de~Jong}}]{powers_exploring_2021}%
  \BibitemOpen
  \bibfield  {author} {\bibinfo {author} {\bibfnamefont {C.}~\bibnamefont
  {Powers}}, \bibinfo {author} {\bibfnamefont {L.}~\bibnamefont {Bassman}}, \
  and\ \bibinfo {author} {\bibfnamefont {W.~A.}\ \bibnamefont {de~Jong}},\
  }\href {http://arxiv.org/abs/2109.01619} {\bibfield  {journal} {\bibinfo
  {journal} {{arXiv}:2109.01619 [quant-ph]}\ } (\bibinfo {year}
  {2021})}\BibitemShut {NoStop}%
\bibitem [{\citenamefont {Wu}\ and\ \citenamefont
  {Hsieh}(2019)}]{wu_variational_2019}%
  \BibitemOpen
  \bibfield  {author} {\bibinfo {author} {\bibfnamefont {J.}~\bibnamefont
  {Wu}}\ and\ \bibinfo {author} {\bibfnamefont {T.~H.}\ \bibnamefont {Hsieh}},\
  }\href {\doibase 10.1103/PhysRevLett.123.220502} {\bibfield  {journal}
  {\bibinfo  {journal} {Physical Review Letters}\ }\textbf {\bibinfo {volume}
  {123}},\ \bibinfo {pages} {220502} (\bibinfo {year} {2019})}\BibitemShut
  {NoStop}%
\end{thebibliography}%

\end{document}